\documentclass[11pt,reqno]{amsart}
\usepackage[letterpaper,left=35mm,right=35mm,top=30mm,bottom=32mm]{geometry}
\usepackage{amsmath,amssymb,amsfonts,mathtools,mathrsfs}
\usepackage[OT1]{fontenc}
\usepackage{microtype}
\usepackage{cite}
\usepackage[hidelinks]{hyperref}
\numberwithin{equation}{section}
\newtheorem{theorem}{Theorem}[section]
\newtheorem{proposition}{Proposition}[section]
\newtheorem{lemma}{Lemma}[section]

\theoremstyle{remark}

\newcommand{\T}{\mathbb T}
\newcommand{\R}{\mathbb R}
\newcommand{\C}{\mathbb C}
\newcommand{\Z}{\mathbb Z}
\newcommand{\N}{\mathbb N}
\usepackage{xcolor}

\newcommand{\Q}{\mathbb Q}
\newcommand{\U}{\mathrm U}
\newcommand{\uu}{\mathfrak u}

\DeclareMathOperator{\diag}{diag}
\DeclareMathOperator{\spec}{spec}
\DeclareMathOperator{\ran}{ran}
\DeclareMathOperator{\tr}{tr}
\DeclareMathOperator{\rank}{rank}
\newcommand{\HS}{\mathrm{HS}}
\newcommand{\coc}[1]{C^{\omega}_{#1}}

\allowdisplaybreaks[2]
\title[The Supercritical Ten Martini Problem]{The Supercritical Ten Martini Problem}
\hypersetup{pdftitle={The Supercritical Ten Martini Problem}}
\date{}

\author {Lingrui Ge}
\address{Beijing International Center for Mathematical Research, Peking University, Beijing, China
	} \email{gelingrui@bicmr.pku.edu.cn}
\author{Svetlana Jitomirskaya}
\address{
	Department of Mathematics, University of California, Berkeley CA, 94703} \email{sjitomi@berkeley.edu}
\author{Jiangong You}
\address{
	Chern Institute of Mathematics and LPMC, Nankai University, Tianjin 300071, China} \email{jyou@nankai.edu.cn}

\begin{document}

\begin{abstract}
We solve the ten martini problem for the
positive-Lyapunov part of the spectrum of any nonconstant real analytic
one-frequency Schr\"odinger operator.
The proof uses the intrinsic symplectic structure of the dual center of Ge--Jitomirskaya \cite{GJmanuscript} and extends
the Kotani--almost-reducibility strategy of Ge--Jitomirskaya--You \cite{GJY} from
two-dimensional centers to arbitrary finite center dimension. 
\end{abstract}
\maketitle
\section{Introduction}\label{sec:introduction}
The Ten Martini Problem for a one-frequency Schr\"odinger family
\begin{equation}\label{eq:operator-intro}
 (H_{v,\alpha,\theta}u)_n=u_{n+1}+u_{n-1}+v(\theta+n\alpha)u_n,
 \qquad n\in\Z,
\end{equation}
on $\ell^2(\Z)$ is to prove that its spectrum is a Cantor set for
\emph{all} irrational frequencies $\alpha$. Here  $v\in C^\omega(\T,\R)$, where $\T=\R/\Z,$ is nonconstant.
 For irrational $\alpha$, the spectrum is independent of
$\theta\in\T$ and will be denoted by $\Sigma_{v,\alpha}$.

The problem was originally posed for the almost Mathieu operator,
$v(x)=2\lambda\cos 2\pi x$, $\lambda\ne0$. Its name was coined by
Barry Simon after a challenge of Mark Kac, and Simon popularized the
problem through his lists~\cite{sim15,simXXI}. After its solution
by Avila--Jitomirskaya~\cite{AJTen}, interest turned to { its robustness/universality}:
whether the same conclusion holds for other sampling functions, in
particular for  analytic  perturbations of the cosine.
 In the two-dimensional magnetic interpretation of the almost Mathieu
operator, $\alpha$ is the magnetic flux through a lattice cell in units
of the flux quantum~\cite{Peierls}, and other $v$ correspond to different allowed hoppings of the tight-binding electron on a 2D lattice, making {robustness/universality} question particularly natural. The \emph{all-$\alpha$}  statement is important for physics, e.g.  quantum Hall effect applications \cite{ntw}. 

The robustness question led us 
to introduce the concept of \emph{Type~I} energies and operators,
defined through a new acceleration-type quantity, the
\emph{T-acceleration} \cite{GJY2023}. The Type~I condition requires T-acceleration
one. It holds for the almost Mathieu operator in all three regimes
of Avila's global theory~\cite{AvilaGlobal} and is stable under analytic
perturbations on a sufficiently wide strip. The resulting global class extends
well beyond perturbations of the almost Mathieu family and requires
neither evenness nor self-duality. We proved Cantor
spectrum for every irrational $\alpha$ throughout this class,
first for even trigonometric polynomial $v$~\cite{GJY2023} and then for
general analytic potentials~\cite{GJY}. The Type~I concept also underlies
the subsequent dry Ten Martini results for analytic operators \cite{LXZ} discussed
below. At supercritical spectral energies, Type~I is equivalent to
acceleration one; the present paper removes this restriction.

The corresponding statement for arbitrary nonconstant analytic $v$
cannot be expected without further assumptions. Indeed,
Argentieri--Avila~\cite{ArgentieriAvila} have announced examples of small
real analytic potentials whose spectra contain intervals. These examples
are {\it subcritical}: the Lyapunov exponent remains zero under small complex
shifts of the phase. In the supercritical regime, however, Cantor spectrum
is expected to hold for every nonconstant $v\in C^\omega(\T,\R)$ and
every irrational frequency. This was formulated as a conjecture by
You~\cite[Conjecture~1]{YouKAM2024}; see also
\cite{MarxJitomirskaya2017}. We prove this conjecture using the intrinsic symplectic center
of~\cite{GJmanuscript} to extend the Kotani and almost reducibility
arguments of~\cite{GJY} from a two-dimensional center to one of
arbitrary finite dimension. Write
\[
 S_E^v(x)=\begin{pmatrix}E-v(x)&-1\\1&0\end{pmatrix},
 \qquad
 L(E)=\lim_{n\to\infty}\frac1n
 \int_\T\log\|(S_E^v)_n(x)\|\,dx,
\]
and set $\mathcal O=\{E:L(E)>0\}$.\footnote{The set $\mathcal O$
is open by continuity of the Lyapunov exponent
\cite{BJContinuity,AJS}.}

\begin{theorem}\label{thm:main}
For every nonconstant $v\in C^\omega(\mathbb T,\mathbb R)$ and
every $\alpha\in\mathbb R\setminus\mathbb Q$,
$\Sigma_{v,\alpha}\cap\overline{\mathcal O}$ is a Cantor set.
\footnote{Provided, of course, it is nonempty.}
\end{theorem}

In particular, the whole spectrum is Cantor whenever $L(E)>0$
throughout the spectrum. By Sorets--Spencer
\cite{SoretsSpencer,bourg}, this applies to $\lambda v$ for every
nonconstant real analytic $v$ and all sufficiently large $|\lambda|$.
\subsection{Main ideas}
The proof extends the Kotani--almost-reducibility argument
of~\cite{GJY2023,GJY} from two-dimensional dual centers to arbitrary
finite dimension. To explain the role of the center, recall that Aubry
duality~\cite{AubryAndre,AvronSimon} associates to
\eqref{eq:operator-intro} the operator
\[
 (L_{v,\alpha,x}w)_n
 =\sum_{s\in\Z}\widehat v_s w_{n+s}
       +2\cos 2\pi(x+n\alpha)w_n,
\]
where $\widehat v_s$ are the Fourier coefficients of $v$.
For a trigonometric polynomial of degree $d$, the dual eigenvalue
equation $L_{v,\alpha,x}w=Ew$ gives rise to $2d$-dimensional transfer dynamics, and a key discovery  in \cite{GJYZ} is that their multiplicative Jensen formula implies this dynamics is partially hyperbolic with $2k$-dimensional center, where $k$ is the T-acceleration in Avila's global theory \cite{AvilaGlobal}.  For general
analytic $v$, the dual operator has infinite range and there is no such
finite-dimensional transfer matrix. Nevertheless,
it was shown in \cite{GJmanuscript}  that the dual eigenvalue
equation admits an intrinsic $2k$-dimensional symplectic center.
 The center is
therefore two-dimensional at Type~I energies. Higher acceleration requires working
with a higher-dimensional center, which the constructions of \cite{GJYZ,GJmanuscript} both allow. A part of the argument of \cite{GJY}, however, significantly used the two-dimensionaity of the intrinsic center, and this is what we extend here
 to arbitrary finite center dimension.

 Suppose that a nondegenerate interval
$I\subset\Sigma_{v,\alpha}\cap\mathcal O$ exists. After restricting { and shrinking}
$I$, the intrinsic center has constant dimension $2k$ and admits an
analytic injective solution frame (Proposition~\ref{prop:actual-center}).
We extend the Kotani argument of~\cite{GJY} to this center,
obtaining a reflectionless identity and a compact reduction analytic in
both energy and phase (Theorem~\ref{thm:center-kotani}). The issue here
is that only the center exponents vanish, while the complementary
directions are hyperbolic. The partial-vanishing form of Kotani--Simon
theory~\cite{KS} for the intrinsic {\it long-range} center in the presence of
complementary hyperbolic directions was developed in \cite{GJY} for the case of two-dimensional center. Here  we extend it to the center of arbitrary dimension. For partially hyperbolic finite-range
operators over general ergodic bases, a related extension was developed
by Ge--Xu~\cite{GeXu}. 

The reduced center cocycle takes values in
$\mathrm U(m)\times\mathrm U(m)$,
\[
 U_E=U_{+,E}\oplus U_{-,E},
\]
and its complexified Lyapunov exponents vanish on a common strip.
Moreover, the two blocks have strict energy monotonicity with {\it opposite
signs} (Proposition~\ref{prop:compact-energy}). This allows us to vary
the energy to produce a resonance between two eigenphases, one from each
block. Our use of energy monotonicity is related to the monotone cocycle
theory of Avila--Krikorian~\cite{AKMonotonic} and its extension
to higher dimensions by Xu~\cite{XuJEMS}.

We first fix an interior energy $E_c$ and prove almost reducibility
on a fixed strip (Proposition~\ref{prop:center-preparation}).
We use Krikorian's renormalization method~\cite{KrCompact,Kr99,Kr01},
expressed in terms of commuting returns as in~\cite{AK,HPZ}.
Subcriticality of the center, namely the vanishing of its
complexified Lyapunov exponents on a strip, gives the required
analytic estimates. We then apply the analytic embedding theorem
of You--Zhou~\cite{YZ} and Lemma~\ref{lem:monotone-suspension}
to obtain a {\it continuous-time} family whose two blocks are
monotone in energy with opposite signs.

To choose the energy, we follow the spectral argument
of~\cite{AYZ,GJY}, with additional estimates needed in higher
dimension. At each step we fix the resonant subspaces at a
reference energy. The opposite monotonicity of the two blocks
allows us to vary the energy until two selected eigenfrequencies,
one from each block, have the required difference. We evaluate
the conjugating matrix at this energy and use the same matrix
for the whole energy family. This preserves the monotonicity
bounds without differentiating the conjugating matrices in energy.

In Section~\ref{sec:local-marked}, we prove that for every $\mu>0$
there are an energy $E_*$, an integer $k$ and conjugacies
\[
 B_j(x+\alpha)U_{E_*}(x)B_j(x)^{-1}=C_j+F_j(x),
 \qquad C_j=C_{+,j}\oplus C_{-,j},
\]
such that the eigenphases $\theta_{\pm,j}$ of $C_{\pm,j}$ satisfy
\[
 \theta_{+,j}-\theta_{-,j}=k\alpha\pmod\Z,
\]
and
\[
 S_j\longrightarrow\infty,\qquad
 \delta_j:=\|F_j\|_0\le Ce^{-S_j},\qquad
 \|B_j'\|_0+\|(B_j^{-1})'\|_0\le C_{\rm fixed}+\mu S_j,
\]
where primes denote phase derivatives
(Proposition~\ref{prop:marked-resonance}).
We will choose $\mu$ sufficiently small in the final duality argument.

Indeed, the resonance gives two approximate solutions of the original
scalar equation at the same phase. Injectivity of the center frame
 at finitely many lattice sites, gives a uniform lower bound
on their $L^2$ Gram matrix. The phase derivative estimates control
their Fourier first moments, so a fixed amount of mass lies in a
window of radius $R_j\le C+A\mu S_j$, with $A$ fixed by the
observation map. On the other hand, the scalar Wronskian and
propagation estimates give
\[
 \delta_jR_jK^{2R_j}\ge c>0,
\]
where $K>1$ bounds the scalar transfer matrices. Choosing
$4A\mu\log K<1$, we obtain
\[
 0<c\le\delta_jR_jK^{2R_j}
 \le C(1+S_j)e^{-S_j/2}\longrightarrow0,
\]
a contradiction. This excludes intervals in
$\Sigma_{v,\alpha}\cap\mathcal O$.

\subsection{History and related results}
The history of the almost Mathieu Ten Martini Problem includes the early results of
Bellissard--Simon~\cite{BS82}, and
Helffer--Sj\"ostrand~\cite{HelfferSjostrand}. Last~\cite{Last94} proved zero measure, thus Cantor, spectrum at $\lambda=1$ for a.e. $\alpha.$ A proof for a.e. $\alpha$ containing the complementary to Last's set was given by Avila-Krikorian \cite{AK}.\footnote{See also \cite{BJK} for a simple all frequency argument.}
Outside the critical point, Choi--Elliott--Yui~\cite{CEY90} and Puig~\cite{Puig} obtained
breakthroughs on the Liouvillean and Diophantine sides, respectively, thus proving Cantor spectrum for both topologically generic and full measure explicit set of frequencies, leaving however an arithmetically critical region untreated.
The proof
of~\cite{AJTen} covered all but the very Liouville frequencies, for which it referred to the argument of \cite{CEY90} to establish the full ten martini.

For other continuous $v,$ there were a number of Cantor spectrum results, all requiring to exclude parameters. Avila--Bochi--Damanik~\cite{ABD09,ABD12} proved that (dry) Cantor spectrum holds Baire generically for continuous $v$. Perturbative results of Sinai~\cite{Sinai1987} (at large coupling for $C^2$ cosine-type $v$) and
Eliasson~\cite{Eliasson} (at small coupling for {generic} analytic $v$) %based on Moser--P\"oschel~\cite{MP} argument and KAM reducibility) 
proved Cantor spectrum in corresponding perturbative\footnote{Meaning, frequency-dependent.} regimes
at Diophantine frequencies.\footnote{Sinai's theorem has further frequency restrictions besides the Diophantine condition, removed in later proofs \cite{WangZhang,FormanVandenBoom}} Different proofs of Sinai's result were later given by Wang--Zhang~\cite{WangZhang} and
Forman--VandenBoom~\cite{FormanVandenBoom}.  For general analytic potentials, Goldstein--Schlag~\cite{GS} proved Cantor spectrum in the positive Lyapunov exponent regime under a strong Diophantine condition on the frequency, excluding also a further, implicit, potential-dependent set of frequencies.

Thus, before \cite{GJY2023, GJY} there were no ten martini results for continuous $v$ other than $\lambda\cos 2\pi x.$

If allowing discontinuity,  the ten martini was proved for many piecewise constant $v$ by  Damanik-Lenz \cite{dl} by establishing measure zero spectrum. For $v$ that are monotone on the period,  the spectrum is necessarily of positive measure \cite{jk,jls}. It was a surprise in \cite{kps} that the spectrum is also Cantor in the localization regime, proved by a simple argument using stability
of localization under rank-one perturbations. Jitomirskaya--Stoiber~\cite{JS}
recently proved ten martini for a large
class of such $v$  in particular, any $v\in C^1([0,1])$ with $v'\ge 0,\;\max v'>0.$ Their proof
introduces a completely different mechanism for the all-frequency problem:
monotonicity and interlacing yield gaps of size at least $c_J/q$
in the periodic union spectrum inside each interval $J$ with
positive Lyapunov exponent, leading to a result by certain gap continuity. Monotonicity  is crucial for their argument.
 
The dry Ten Martini Problem asks whether every gap permitted by gap
labelling is open. Avila--You--Zhou~\cite{AYZ} proved this for the
noncritical almost Mathieu operator, for the critical case, see very recent work of
Borgnia--Faust~\cite{BF} and Avila--You--Zhou~\cite{AYZc}. Band--Beckus--Loewy~\cite{BBL} proved the  dry theorem for
Sturmian Hamiltonians at every irrational frequency and every
nonzero coupling.\footnote{Technically, this does not fall under the definition of dry ten martini, since $v=v_\alpha$ in the Sturmian case, but morally it does. For gap labelling and examples of
gap closing on decorated graphs, see
Band--Sofer~\cite{BandSofer}.}

For general analytic potentials,
Ge--Jitomirskaya--You~\cite{GJY2023} formulated the corresponding
question for the Type~I class they introduced. This was resolved for supercritical
Type~I energies in the work of Li, Xu, and Zhou~\cite{LXZ} and at subcritical Type~I
energies in ~\cite{GeDry,LiDry}.

The role of Type~I in the latter results is crucial: it identifies the
regime of a {\it two-dimensional} dual center. The supercritical proof
of Li--Xu--Zhou~\cite{LXZ} builds on this concept, the duality approach
to global theory of~\cite{GJYZ}, and the
intrinsic center and its subcriticality established in~\cite{GJmanuscript}. In the subcritical Type I regime,
Ge~\cite{GeDry} gives the details of the dry statement announced
in~\cite{GJY}, combining the all-frequency Puig argument developed
there with quantitative almost reducibility. Li~\cite{LiDry} gives a different proof. Moreover, Ge--Xu--Wang--Zhao \cite{GXWZ}
have constructed a supercritical Type~II counterexample to
the dry Ten Martini statement. Thus our Cantor spectrum
 theorem cannot in general be strengthened to a dry result for {\it all} supercritical potentials. In particular, the route through full gap opening
used in~\cite{AYZ,LXZ} cannot apply in this generality.

\subsection{Structure of the paper}
Section~\ref{sec:preliminaries} contains the preliminaries.
Sections~\ref{sec:center}--\ref{sec:global-to-local} construct 
the monotone compact center. Sections~\ref{sec:local-marked}
and~\ref{sec:duality-wronskian} give the local reduction and scalar
contradiction. 
Appendix~\ref{app:elementary} contains some elementary calculations.

\section{Preliminaries}\label{sec:preliminaries}\label{sec:notation}
Let $\T=\R/\Z$ with the normalized Lebesgue measure. 
Throughout, we assume $\alpha\in(0,1)\setminus\Q$. Constants $C,c>0$ may change
from line to line.

\subsection{Notation and analytic norms}\label{subsec:analytic-notation}
We use the operator norm on matrices. For $B\in M_m(\C)$, we use the standard notations
\[
 \|B\|_{\mathrm{HS}}=(\tr B^*B)^{1/2},\qquad \mathbb H=\{z:\Im z>0\},\qquad\mathbb D=\{z:|z|<1\},
\]
\[
 \U(m)=\{B:B^*B=I\},\qquad
 \mathfrak u(m)=\{B:B^*=-B\}.
\]
For $h>0$, put
\[
 S_h^d=\{z\in\C^d:|\Im z_j|<h\},\qquad
 \T_h^d=S_h^d/\Z^d,\qquad S_h=S_h^1.
\]
We write $C_h^\omega$ for the bounded periodic holomorphic functions
on $S_h^d$ and use the norm $\|F\|_h=\sup_{S_h^d}\|F\|$.
Group and Lie algebra conditions are imposed on the real torus.
For a family on a conjugation-invariant energy domain $\mathcal E$, define
\begin{equation}\label{eq:reflected-adjoint}
 F^\natural(z,x)=F(\bar z,\bar x)^*,\qquad
 \|F\|_{\mathcal E,h}=\sup_{\mathcal E\times S_h^d}\|F\|.
\end{equation}
Then we have
\[
 (FG)^\natural=G^\natural F^\natural,\qquad
 (\partial F)^\natural=\partial F^\natural.
\]

Clearly,
\begin{equation}\label{eq:unitaryinverse}
 F\in C_h^\omega(\T^d,\U(m))\quad\Longrightarrow\quad F^{-1}=F^\natural.
\end{equation}
For
$
 F(x)=\sum_{k\in\Z^d}\widehat F(k)e^{2\pi i\langle k,x\rangle},
\; |k|_1=\sum_{j=1}^d|k_j|,
$
define
\[
 |F|_h=\sum_{k\in\Z^d}\|\widehat F(k)\|_{\mathrm{HS}}e^{2\pi h|k|_1}.
\]
For $0<r<h$, Cauchy's estimate gives
\begin{equation}\label{eq:cauchy}
 \|\partial_{x_j}^{\ell}F\|_r
 \le \ell!(h-r)^{-\ell}\|F\|_h.
\end{equation}
Shifting contours in the Fourier coefficients gives
\begin{equation}\label{eq:fouriernorm}
 \|F\|_r\le |F|_r
 \le C_{m,d}\bigl(1+(h-r)^{-d}\bigr)\|F\|_h.
\end{equation}
We also have
\begin{equation}\label{eq:wiener}
 |FG|_h\le |F|_h|G|_h,\qquad
 |VFV^{-1}|_h=|F|_h\quad(V\in\U(m)).
\end{equation}

Note that keeping only selected Fourier coefficients in each matrix entry
does not increase $|\cdot|_h$. This also preserves
$F(x)^*=\pm F(x)$ on the real torus if the retained indices
are invariant under $(a,b,k)\mapsto(b,a,-k)$.

\subsection{Cocycles and Lyapunov exponents}\label{subsec:cocycle-preparations}
For $A\in C(\T,\mathrm{GL}(m,\C))$ and $n\ge 1$, define the iterates by \begin{equation}\label{eq:iterates}
 \begin{gathered}
 A_n(x)=A(x+(n-1)\alpha)\cdots A(x),\qquad n\ge1,\\
 A_0=I,\qquad A_{-n}(x)=A_n(x-n\alpha)^{-1}.
 \end{gathered}
 \end{equation}
We abbreviate $F(x+\alpha)$ by $F^+(x)$. Under a change of coordinates,
\begin{equation}\label{eq:gauge}
 A^B=B^+AB^{-1},\qquad
 (A^B)_n(x)=B(x+n\alpha)A_n(x)B(x)^{-1}.
\end{equation}
The Lyapunov exponents $L_1\ge\cdots\ge L_m$, counted with multiplicity,
are characterized by
\begin{equation}\label{eq:LE}
 \sum_{a=1}^jL_a(\alpha,A)
 =\lim_{n\to\infty}\frac1n
 \int_\T\log\|\textstyle\bigwedge^j A_n(x)\|\,dx.
\end{equation}
An analytic cocycle is regular at index $j$ if
$y\mapsto\sum_{a\le j}L_a(\alpha,A(\cdot+iy))$ is affine near zero
\cite{AJS}.

We call $(\alpha,A)$ almost reducible on a fixed strip $S_r$ if
there are periodic holomorphic invertible $B_j$ and constants $C_j$
such that
\[
 \|B_j^+AB_j^{-1}-C_j\|_r\longrightarrow0.
\]
In the unitary case, $B_j\in C^\omega(\T,\U(m))$ and $C_j\in\U(m)$.

\subsection{The Schr\"odinger operator and its spectrum}\label{subsec:scalar-preparations}
The operator \eqref{eq:operator-intro} is self-adjoint and
$\|H_{v,\alpha,\theta}\|\le2+\|v\|_0$. It is well known \cite{avronsimon} that for continuous real $v$ and irrational $\alpha$, the spectrum of
$H_{v,\alpha,\theta}$ is independent of $\theta$ and perfect.

For the Schr\"odinger cocycle, write
\begin{gather}
 A_z(x)=\begin{pmatrix}z-v(x)&-1\\1&0\end{pmatrix},\qquad
 L_y(z)=L_1(\alpha,A_z(\cdot+iy)),\qquad
 L(E)=L_0(E),\label{eq:schLE}
 \end{gather}
 Avila \cite{AvilaGlobal} defined  the {\it acceleration} $\omega(E)$ by
 \begin{equation}
     \omega(E)=\frac1{2\pi}\partial_y^+L_y(E)\big|_{y=0}.
 \label{eq:acceleration}
\end{equation}
 and proved that the acceleration is quantized: $\omega(E)\in \mathbb{Z}.$ The same formula \eqref{eq:acceleration} defines $\omega(z)$ for complex energy $z$.

\subsection{Aubry duality}\label{subsec:aubry-preparations}
For $f\in L^2(\T)$, put
$\widehat f_n=\int_\T f(x)e^{-2\pi inx}\,dx$. Fix $\rho>0$ such that
\begin{equation}\label{eq:potential}
 v(x)=\sum_{s\in\Z}\widehat v_s e^{2\pi isx},\qquad
 \widehat v_{-s}=\overline{\widehat v_s},\qquad
 A_\rho=\sum_{s\in\Z}|\widehat v_s|e^{\rho|s|}<\infty.
\end{equation}
 
Define the Aubry fual operator by
\begin{equation}\label{eq:scalar-dual-maps}
 (\mathcal L_\theta f)(x)
 =\sum_{s\in\Z}\widehat v_s e^{2\pi is\theta}f(x+s\alpha)
   +2\cos(2\pi x)f(x).
\end{equation}
 Then
\[
 \widehat{\mathcal L_\theta f}_n
=\widehat f_{n-1}+\widehat f_{n+1}
   +v(\theta+n\alpha)\widehat f_n.
\]

\subsection{Solution frames and generalized Wronskians}\label{subsec:frame-preparations}
For $0<\beta<\rho$, set
\begin{equation}\label{eq:weightedspace}
 X_\beta=\{u:\|u\|_\beta=\sup_{n\in\Z}e^{-\beta|n|}|u_n|<\infty\},
 \qquad T:X_\beta\to X_\beta, \;(Tu)_n=u_{n+1}.
\end{equation}
Then
\[
 \|T^{\pm1}\|\le e^\beta,\qquad
 \Bigl\|\sum_s\widehat v_sT^s\Bigr\|
 \le\sum_s|\widehat v_s|e^{\beta|s|}\le A_\rho.
\]
For $Z:S_h\to\mathcal L(\C^q,X_\beta)$, put
$\|Z\|_{h,\beta}=\sup_{x\in S_h,\,|a|=1}\|Z(x)a\|_\beta$.
For an energy family $\mathcal E$, the same norm includes the supremum over
$z\in\mathcal E$.
If
\[
 (L_{v,\alpha,x}-z)Z_z=0,\qquad TZ_z=Z_z^+M_z,
\]
then the site rows satisfy
\begin{equation}\label{eq:rows}
 P_n(z,x):=Z_z(x;n)=P_0(z,x+n\alpha)M_{z,n}(x),\qquad P_1=P_0^+M_z.
\end{equation}
A change of frame gives
\begin{equation}\label{eq:framegauge}
 \widetilde Z=ZQ
 \quad\Longrightarrow\quad
 T\widetilde Z=\widetilde Z^+\widetilde M,\qquad
 \widetilde M=(Q^+)^{-1}MQ.
\end{equation}
At the cut $n$, define the generalized Wronskian
\begin{equation}\label{eq:green-pairing}
 W_n(u,w)=\sum_{i<n\le j}
 \bigl(\overline{u_i}\widehat v_{j-i}w_j
       -\overline{u_j}\widehat v_{i-j}w_i\bigr).
\end{equation}
For the corresponding finite-range definition, see, e.g.,~\cite{Novikov98}. The generalized Wronskian has the following elementary properties.
\begin{lemma}\label{lem:green-pairing}
For $u,w\in X_\beta$, the series in \eqref{eq:green-pairing}
converges absolutely. Moreover,
\begin{gather}
 |W_0(u,w)|\le2\|u\|_\beta\|w\|_\beta
       \sum_{s\ge1}s|\widehat v_s|e^{\beta s},\label{eq:green-bound}\\
 W_{n+1}-W_n
 =\overline{u_n}(L_{v,\alpha,x}w)_n
  -\overline{(L_{v,\alpha,x}u)_n}w_n
 \quad(x\in\R).\label{eq:green-jump}
\end{gather}
\end{lemma}
\begin{proof}
Write $j=i+s$. For $s\ge1$,
\[
 \#\{i\in\Z:i<n\le i+s\}=s,\qquad
 i<n\le i+s\ \Longrightarrow\ |i|+|i+s|\le2|n|+s.
\]
Since $\beta<\rho$,
\[
 \sum_{s\ge1}s|\widehat v_s|e^{\beta s}
 \le A_\rho\sup_{s\ge1}se^{-(\rho-\beta)s}<\infty.
\]
Thus
\begin{align*}
 &\sum_{i<n\le j}
 \bigl(|u_i|\,|\widehat v_{j-i}|\,|w_j|
       +|u_j|\,|\widehat v_{i-j}|\,|w_i|\bigr)\\
 &\qquad\le2e^{2\beta|n|}\|u\|_\beta\|w\|_\beta
 \sum_{s\ge1}s|\widehat v_s|e^{\beta s}<\infty.
\end{align*}
Taking $n=0$ gives \eqref{eq:green-bound}. By absolute convergence we can
cancel the terms crossing both cuts. Using
$\widehat v_{-s}=\overline{\widehat v_s}$ and the real diagonal, we get
\begin{align*}
 W_{n+1}-W_n
 &=\overline{u_n}\sum_{s\ne0}\widehat v_s w_{n+s}
   -\sum_{s\ne0}\overline{\widehat v_su_{n+s}}w_n\\
 &=\overline{u_n}(L_{v,\alpha,x}w)_n
   -\overline{(L_{v,\alpha,x}u)_n}w_n.
\end{align*}
\end{proof}

\begin{lemma}\label{lem:end-decay}
Suppose $u,w\in X_\beta$ and
$|u_{\sigma n}|+|w_{\sigma n}|\le Ce^{-\delta n}$ for $n\ge0$,
where $\delta>0$ and $\sigma\in\{-1,1\}$. Then
$W_{\sigma n}(u,w)\to0$, uniformly when
$C,\delta,\|u\|_\beta,\|w\|_\beta$ are fixed.
\end{lemma}
\begin{proof}
Decrease $\delta$ so that $0<\delta<\rho$. For $\sigma=1$,
\[
 |u_i|+|w_i|\le Ce^{\beta|i|},\qquad
 |u_j|+|w_j|\le Ce^{-\delta j}\quad(j\ge0).
\]
For $0\le i<n\le j$,
\[
 e^{-\delta(i+j)}=e^{-2\delta j}e^{\delta(j-i)}
 \le e^{-2\delta n}e^{\delta(j-i)}.
\]
Split the sum at $i=0$ and use
$|\widehat v_s|\le A_\rho e^{-\rho|s|}$ to obtain
\begin{align*}
 |W_n(u,w)|
 &\le C\sum_{i<0}e^{(\rho-\beta)i}
             \sum_{j\ge n}e^{-(\rho+\delta)j}
      +Ce^{-2\delta n}\sum_{s\ge1}s|\widehat v_s|e^{\delta s}\\
 &\le C_1e^{-(\rho+\delta)n}+C_2e^{-2\delta n}\longrightarrow0.
\end{align*}
For $\sigma=-1$, set
\[
 u'_k=u_{-k},\qquad w'_k=w_{-k},\qquad
 \widehat v'_s=\widehat v_{-s}.
\]
The change $(i,j)\mapsto(-j,-i)$ gives
\[
 i<-n\le j\ \Longleftrightarrow\ -j<n+1\le-i,
 \qquad W_{-n}(u,w)=-W'_{n+1}(u',w')\longrightarrow0.
\]
The constants depend only on $\rho,\beta,\delta,A_\rho$ and the stated
sequence bounds. Hence the convergence is uniform.
\end{proof}

Write $u^*G_vw=W_0(u,w)$.
For a solution frame $Z_z(x):\C^q\to X_\beta$, the matrix
$Z^\natural G_vZ$ is defined entrywise by
\[
(Z^\natural G_vZ){ab}(z,x)
=W_0\bigl(Z{\bar z}(\bar x)e_a,Z_z(x)e_b\bigr),
\qquad 1\le a,b\le q,
\]
where $e_1,\ldots,e_q$ are the standard basis vectors of $\C^q$. 
By \eqref{eq:green-bound}, their defining series converge.
Set
\begin{equation}\label{eq:J}
 J_{2m}=\begin{pmatrix}0&I_m\\-I_m&0\end{pmatrix},\qquad
 \mathcal J_m=\diag(-iI_m,iI_m),\qquad
 V=\frac1{\sqrt2}\begin{pmatrix}I_m&I_m\\-iI_m&iI_m\end{pmatrix}.
\end{equation}
Since $V^*V=I$ and $V^*J_{2m}V=\mathcal J_m$,
\[
 Z^*G_vZ=J_{2m}
 \quad\Longrightarrow\quad
 (ZV)^*G_v(ZV)=\mathcal J_m,\qquad
 T(ZV)=(ZV)^+(V^{-1}MV).
\]

For $E,x\in\R$, $y>0$, let
$(L_{v,\alpha,x}-E-iy)u=0$, $u\in X_\beta$, with exponential decay
at $+\infty$. By \eqref{eq:green-jump},
\[
 W_{n+1}(u,u)-W_n(u,u)
 =\bigl((E+iy)-(E-iy)\bigr)|u_n|^2=2iy|u_n|^2.
\]
Summing from $0$ to $N$ and applying Lemma~\ref{lem:end-decay},
\[
 W_{N+1}(u,u)-W_0(u,u)=2iy\sum_{n=0}^N|u_n|^2,
 \qquad W_{N+1}(u,u)\longrightarrow0.
\]
Hence
\begin{equation}\label{eq:halflinemass}
 \frac i2W_0(u,u)=y\sum_{n\ge0}|u_n|^2.
\end{equation}
For energy $E-iy$ and exponential decay at $-\infty$,
\[
 W_0(u,u)-W_{-N}(u,u)=-2iy\sum_{n=-N}^{-1}|u_n|^2,
 \qquad W_{-N}(u,u)\longrightarrow0,
\]
and therefore
\[
 \frac i2W_0(u,u)=y\sum_{n\le-1}|u_n|^2.
\]

\section{The intrinsic symplectic center dynamics}\label{sec:center}\label{sec:dual-frames}
The following proposition follows from the construction of finite-range solution frames and their convergence on a common domain in~\cite[Theorem~6.3, Lemma~6.6, (6.35), (6.38)--(6.46)]{GJmanuscript}, together with the multiplicative Jensen formulas in~\cite{GJYZ}. The proof is given in Section~\ref{subsec:center-construction}.

\begin{proposition}\label{prop:actual-center}
Let $I\subset\Sigma_{v,\alpha}\cap\mathcal O$ be an interval with $|I|>0$.
There exist $m\ge1$, a compact interval $J\subset\operatorname{int}I$
with $|J|>0$, a bounded connected conjugation-invariant neighborhood
$\mathcal E$ of $J$, $h>0$, and $0<\beta<\rho$ such that the following
hold. There are bounded periodic holomorphic maps
\[
 Z:\mathcal E\times S_h\longrightarrow\mathcal L(\C^{2m},X_\beta),
 \qquad
 M:\mathcal E\times S_h\longrightarrow\mathrm{GL}(2m,\C)
\]
satisfying
\begin{equation}\label{eq:center-solution}
 (L_{v,\alpha,x}-z)Z_z(x)=0,\qquad
 TZ_z(x)=Z_z(x+\alpha)M_z(x),
\end{equation}
\begin{equation}\label{eq:center-green-form}
 Z^\natural G_vZ=\mathcal J_m,\qquad
 M^\natural\mathcal J_mM=\mathcal J_m,\qquad
 \ker Z=\{0\}.
\end{equation}
The first $m$ nonnegative dual exponents of the Fourier truncations of
$v$ have limits $0\le\gamma_1(z)\le\cdots\le\gamma_m(z)$ on $\mathcal E$,
and
\begin{equation}\label{eq:center-spectral-data}
 \begin{split}
 L_j(\alpha,M_z)&=\gamma_{m+1-j}(z),\\
 L_{2m+1-j}(\alpha,M_z)&=-\gamma_{m+1-j}(\bar z),
 \qquad 1\le j\le m,
 \end{split}
\end{equation}
and
\begin{equation}\label{eq:center-nonreal-regularity}
 L_j(\alpha,M_z(\cdot+it))=L_j(\alpha,M_z),
 \qquad z\in\mathcal E,\quad |t|<h,\quad1\le j\le2m.
\end{equation}
Moreover,
\[
 L_j(\alpha,M_E(\cdot+it))=0,
 \qquad E\in J,\quad |t|<h,\quad1\le j\le2m,
\]
while, for $z\in\mathcal E\cap\mathbb H$,
\begin{equation}\label{eq:center-nonreal-gap}
 L_m(\alpha,M_z)=\gamma_1(z)>0>
 -\gamma_1(\bar z)=L_{m+1}(\alpha,M_z).
\end{equation}
\end{proposition}

\subsection{Complex-energy flatness}
The following lemma is essentially proved in \cite[Theorem~5.1]{GJmanuscript}, however, we only state the result for real E there, so we give a sketch of  the proof for all $z\in\C$ here for completeness.
\begin{lemma}\label{lem:complex-dual-flatness}
Let $w$ be a nonconstant real trigonometric polynomial of degree $d$,
and let $\widehat A_z$ be its dual transfer matrix. Write
$L^w(z)=L_1(\alpha,A_z^w)$. If $L^w(z)>0$, then
\begin{equation}\label{eq:finite-dual-complex-flatness}
 L_j(\alpha,\widehat A_z(\cdot+it))
 =L_j(\alpha,\widehat A_z),
 \qquad1\le j\le2d,\quad2\pi|t|<L^w(z).
\end{equation}
\end{lemma}
\begin{proof}
Put
\[
 \ell_j(z,t)=L_j(\alpha,\widehat A_z(\cdot+it)),\qquad
 S_k(z,t)=\sum_{j=1}^k\ell_j(z,t),\qquad S_0=0.
\]
By \cite[proof of Theorem~5.1, (5.2)--(5.6)]{GJmanuscript},
\begin{gather}\label{slo}
 |\partial_t^\pm S_d(z,t)|\le2\pi,\qquad
 S_d(z,\sigma t)=2\pi t-\log|\widehat w_d|\quad(t\gg1,\ \sigma=\pm1),\\
 S_d(z,0)+\log|\widehat w_d|=L^w(z)=L^w(\bar z),\\
 \ell_j(z,t)=-\ell_{2d+1-j}(\bar z,-t).
\end{gather}
First suppose $z\notin\R$. The self-adjoint resolvent criterion gives
\[
 \ell_d(z,0)>0>\ell_{d+1}(z,0).
\]
For $\sigma=\pm1$, let
\[
 t_k^\sigma(z)=\sup\{b\ge0:\ell_k(z,\sigma t)=\ell_k(z,0)
                                  \text{ for }0\le t\le b\},
\]
\[
 b_k^\sigma(z)=\sup\{b\ge0:S_k(z,\sigma t)=S_k(z,0)
                                  \text{ for }0\le t\le b\}.
\]
By \cite[proof of Theorem~5.1, claims~(1)--(3)]{GJmanuscript},
\begin{equation}\label{eq:complex-turning-order}
 0<b_d^\sigma(z)\le b_k^\sigma(z),\qquad1\le k\le d.
\end{equation}
Indeed, the nested flat intervals at the turning points $t_j^\sigma$
give $b_k^\sigma\ge t_d^\sigma>0$. For $d>1$, claim~(3) and convexity
of $S_{d-1}(z,\sigma\cdot)$ give
\[
 \left.\partial_t^+S_d(z,\sigma t)\right|_{t=t_d^\sigma}>0,
 \qquad b_d^\sigma=t_d^\sigma.
\]
For $d=1$, use $S_1=\ell_1$; for $\sigma=-1$, reflect the phase and
replace $\alpha$ by $-\alpha$. Since $\ell_j=S_j-S_{j-1}$,
\[
 \ell_j(z,\sigma t)=\ell_j(z,0),
 \qquad1\le j\le d,\quad0\le t\le b_d^\sigma(z).
\]
Since $S_d(z,\sigma t)$ is constant for $0\le t\le b_d^\sigma(z)$ and its slope does not exceed $2\pi$, for sufficiently large $t$ we have
\[
2\pi t-\log|\widehat w_d|
=S_d(z,\sigma t)
\le S_d(z,0)+2\pi(t-b_d^\sigma(z)).
\]

and hence
\[
 0<b_d^\sigma(z)\le\frac{L^w(z)}{2\pi}<\infty.
\]
Set
\[
 b_*=\min_{\zeta\in\{z,\bar z\},\ \sigma=\pm1}b_d^\sigma(\zeta).
\]
By \eqref{eq:complex-turning-order} and reflection,
\[
 \ell_d(\zeta,\sigma b_*)=\ell_d(\zeta,0)>0>
 -\ell_d(\bar\zeta,0)=\ell_{d+1}(\zeta,\sigma b_*).
\]
Choose a minimizing pair $(\zeta,\sigma)$ and put
$f(t)=S_d(\zeta,\sigma t)$. By Lyapunov continuity, the middle gap
persists for $|t-b_*|<\delta$. Thus \cite[Remark~4.1]{GJmanuscript} gives
\[
 f'_+(t)\in2\pi\Z,\qquad |t-b_*|<\delta.
\]
By convexity, $f'_+(b_*)\ge0$. Right continuity and quantization give
\[
 f'_+(b_*)=0
 \ \Longrightarrow\ f'_+=0\text{ on }[b_*,b_*+\delta']
 \ \Longrightarrow\ b_d^\sigma(\zeta)\ge b_*+\delta'
\]
for some $\delta'>0$, a contradiction. Thus
\[
 f'_+(b_*)\in(0,2\pi]\cap2\pi\Z=\{2\pi\}.
\]
By convexity and the slope bound,
\[
 f(t)=f(0)+2\pi(t-b_*),\qquad t\ge b_*.
\]
By the large-height formula,
\[
 2\pi b_*=S_d(\zeta,0)+\log|\widehat w_d|
 =L^w(\zeta)=L^w(z).
\]
Thus
\[
 \frac{L^w(z)}{2\pi}=b_*\le b_d^\sigma(\zeta)
 \le\frac{L^w(z)}{2\pi},
 \qquad \zeta\in\{z,\bar z\},\quad\sigma=\pm1.
\]
Using $\ell_j=S_j-S_{j-1}$ and reflection gives
\eqref{eq:finite-dual-complex-flatness}.

For $E\in\R$ and $2\pi|t|<L^w(E)$, set $z_n=E+i/n$.
By continuity \cite{AJS},
$2\pi|t|<L^w(z_n)$ for large $n$, and hence
\[
 \ell_j(E,t)=\lim_{n\to\infty}\ell_j(z_n,t)
 =\lim_{n\to\infty}\ell_j(z_n,0)=\ell_j(E,0).
\qedhere
\]
\end{proof}

\subsection{Construction of the center}\label{subsec:center-construction}
\begin{proof}[Proof of Proposition~\ref{prop:actual-center}]
\par\smallskip\noindent\emph{Step 1. Constant center dimension.}\par\nobreak
We replace $I$ by its interior, still denoted by $I.$  Since $L(E)>0$
there, global theory \cite{AvilaGlobal,AJS} gives
\[
 \omega(E)=0
 \ \Longrightarrow\ (\alpha,A_E)\text{ is regular}
 \ \Longrightarrow\ (\alpha,A_E)\text{ is uniformly hyperbolic}
 \ \Longrightarrow\ E\notin\Sigma_{v,\alpha}
\]
Thus $\omega(I)\subset\Z_{>0}$. Choose
\[
 m=\min_{E\in I}\omega(E)=\omega(E_0)\ge1,
 \qquad E_0\in I.
\]
Fix $0<s_0<\rho/(2\pi)$ before the next positive turning point of
$L_t(E_0)$. Regularity and positivity at $t=\pm s_0$ give uniform
hyperbolicity and locally constant accelerations. By convexity,
on a smaller interval $E_0\in I'\subset I$,
\[
 m\le\omega(E)\le\frac1{2\pi}\partial_tL_t(E)|_{t=s_0}=m.
\]
Thus $\omega|_{I'}=m$. 
Let $y_1(E)\ge0$ be the first turning point of
$y\mapsto L_y(E)$. When such a turning point exists, the
$T$-acceleration is defined \cite {GJY} by
\[
 \overline{\omega}(E)
 =\lim_{y\downarrow y_1(E)}
 \frac{L_y(E)-L_{y_1(E)}(E)}{2\pi(y-y_1(E))}.
\]
In the supercritical regime, T-acceleration coincides with acceleration thus  is equal to $m$.
Hence
\cite[Theorem~6.3]{GJmanuscript} gives a center of dimension $2m$.

\par\smallskip\noindent\emph{Step 2. A common domain for the finite-range models.}\par\nobreak
Let
\[
 v_D(x)=\sum_{|s|\le D}\widehat v_s e^{2\pi i sx},\qquad
 d=l(D)=\max\{|s|\le D:\widehat v_s\ne0\}.
\]
Since $m\ge1$, take $D$ large enough that $v_D$ is nonconstant.
If $v$ is a trigonometric polynomial, we take $v_D=v$ and keep its
frame and cocycle fixed thereafter.
By \cite[Theorem~6.3, Lemma~6.6]{GJmanuscript}, there are a compact
interval $E_0\in\operatorname{int}J\subset J\subset I'$, a bounded
connected conjugation-invariant neighborhood $\mathcal E$ of $J$,
and $h>0$ on which the finite-range frames satisfy
\[
 O_D^\natural S_DO_D=\mathcal J_m,\qquad
 \widehat A_DO_D=O_D^+M_D.
\]
Here $\widehat A_D$ is the transfer matrix for $v_D$, $O_D$ is a
$2d\times2m$ center frame, and $S_D$  is the matrix of the generalized Wronskian,
defined in Step~7 below. Let us make the constant change of coordinates \eqref{eq:J}.
Order the site rows from $d-1$ to $-d$, and let $P_D=O_D(\cdot;0)$.
We therefore get bounded periodic holomorphic limits
$M,P$ satisfying
\begin{equation}\label{eq:center-approximation}
 \|M_D-M\|_{\mathcal E,h}+\|P_D-P\|_{\mathcal E,h}\le Ce^{-cD}.
\end{equation}
The domain and the constants in \eqref{eq:center-approximation} are
independent of the cutoff $D$.   

By normalization, invariance and holomorphic uniqueness we have $
 M_D^\natural\mathcal J_mM_D=\mathcal J_m.$ Thus, by \eqref{eq:center-approximation} we also have
 $M^\natural\mathcal J_mM=\mathcal J_m.
$
Hence
\[
 M_D^{-1}=\mathcal J_m^{-1}M_D^\natural\mathcal J_m,
 \qquad M^{-1}=\mathcal J_m^{-1}M^\natural\mathcal J_m,
\]
and, since the domain is conjugation-invariant,
\[
 \|M_D^{-1}-M^{-1}\|_{\mathcal E,h}
 =\|M_D-M\|_{\mathcal E,h}\le Ce^{-cD}.
\]
All subsequent shrinkings of
$J,\mathcal E,S_h$ are common to sufficiently large $D$.

The solution rows satisfy
\[
 Z_D(z,x;n)=P_D(z,x+n\alpha)M_{D,n}(z,x).
\]
For fixed $n\in\Z$, \eqref{eq:center-approximation} and
\cite[(6.39)--(6.46)]{GJmanuscript} give
\[
 M_{D,n}\longrightarrow M_n,
 \qquad Z_D(z,x;n)\longrightarrow
 Z(z,x;n):=P(z,x+n\alpha)M_n(z,x)
\]
uniformly on $\mathcal E\times S_h$. 

\par\smallskip\noindent\emph{Step 3. Uniform growth and spectral identification.}\par\nobreak
For $v_D$, let
$0\le\gamma_1^{v_D}(z)\le\cdots\le\gamma_d^{v_D}(z)$ be the
nonnegative dual exponents, and write $a_+=\max\{a,0\}$ for $a\in\R$.
In the normalization \eqref{eq:LE}, 
the complex energy version of the multiplicative Jensen formula (Theorem 7.1 in \cite{GJYZ}) gives,
for $\varepsilon>0$,
\begin{equation}\label{eq:finite-complex-jensen}
\begin{aligned}
 L_{\varepsilon}^{v_D}(z)-L_0^{v_D}(z)
 &=\sum_{i=1}^{d}
       \bigl(2\pi\varepsilon-\gamma_i^{v_D}(\bar z)\bigr)_+,\\
 L_{-\varepsilon}^{v_D}(z)-L_0^{v_D}(z)
 &=\sum_{i=1}^{d}
       \bigl(2\pi\varepsilon-\gamma_i^{v_D}(z)\bigr)_+.
\end{aligned}
\end{equation}
The first identity follows from the
second with $z$ replaced
by $\bar z$, using the reality of $v_D$. 

By uniform hyperbolicity at $\pm s_0$ and local constancy of acceleration
\cite{AJS}, on the common domain,
\begin{gather*}
 \partial_tL_t^{v_D}(z)|_{t=s_0}=2\pi m,\qquad
 \partial_tL_t^{v_D}(z)|_{t=-s_0}=-2\pi m,\\
 \gamma_m^{v_D}(\zeta)<2\pi s_0<\gamma_{m+1}^{v_D}(\zeta),
 \qquad \zeta\in\{z,\bar z\}.
\end{gather*}
When $d=m$, we use the convention $\gamma_{d+1}^{v_D}=+\infty$.
The inequalities are strict since uniform hyperbolicity persists near
both heights, excluding turning points in \eqref{eq:finite-complex-jensen}.

Since $\omega(E_0)=m$, by the real-energy Jensen formula 
\cite{GJYZ}, arguing as in the proof of Theorem~6.3 in \cite{GJmanuscript} we obtain
\[
 L_j(\alpha,M_{E_0})=0,\qquad1\le j\le2m.
\]
Fix
\[
 0<\beta<\rho,\qquad0<\eta<\min\{\beta,\pi s_0\}.
\]
By uniform subadditivity 
there exists $N\ge1$ such that
\[
 \max_\pm\sup_{x\in\T}\|M_{\pm N}(E_0,x)\|<e^{\eta N/2}.
\]
By continuity and \eqref{eq:center-approximation}, after shrinking
the domain and taking $D_0$ large,
\[
 \max_\pm\sup_{\mathcal E\times S_h}\|M_{D,\pm N}\|
 <e^{\eta N},\qquad D\ge D_0.
\]
Convergence of the finitely many remaining products gives
\[
 C_N:=\sup_{D\ge D_0}\max_{|r|<N}
             \|M_{D,r}\|_{\mathcal E,h}<\infty.
\]
Writing $n=qN+r$  we obtain
\begin{gather*}
 \max_\pm\|M_{D,\pm(qN+r)}\|_{\mathcal E,h}
 \le C_Ne^{\eta qN},
 \qquad q\ge0,\quad0\le r<N,\\
 -\eta\le L_j(\alpha,M_{D,z}(\cdot+it))\le\eta,
 \qquad z\in\mathcal E,\quad |t|<h,\quad1\le j\le2m.
\end{gather*}
Passing to the limit we obtain the same bounds for $M$.
After increasing $D_0$ and shrinking the common domain, by the Lyapunov
continuity we have
\[
 \inf_{D\ge D_0,\ z\in\mathcal E}L_0^{v_D}(z)>2\pi h.
\]
By Lemma~\ref{lem:complex-dual-flatness}, all dual exponents are
constant for $|t|<h$.

At $t=0$, the $2m$ dual exponents in $(-2\pi s_0,2\pi s_0)$ are
\[
 \gamma_m^{v_D}(z),\ldots,\gamma_1^{v_D}(z),
 -\gamma_1^{v_D}(\bar z),\ldots,-\gamma_m^{v_D}(\bar z).
\]
For real $z$, this is the usual symplectic $\pm$ symmetry.
For real $x$, the finite-range dual operator $L_{v_D,\alpha,x}$
is self-adjoint. Hence every nonreal $z$ belongs to its resolvent
set, and the associated transfer cocycle is uniformly hyperbolic,
with $d$ positive and $d$ negative Lyapunov exponents.  By the generalized Wronskian identity ~\eqref{eq:green-jump} 
$\ell_j(z,0)=-\ell_{2d+1-j}(\bar z,0)$.

By \cite[Lemma~6.6, (6.35), (6.38)--(6.46)]{GJmanuscript},
\[
 \rank O_D(z,x)=2m,\qquad(z,x)\in\mathcal E\times S_h.
\]
For fixed $D,z,t$, the frame $O_D(z,\cdot+it)$ and its inverse on
its range are bounded. Thus the restricted dual cocycle and
$M_{D,z}(\cdot+it)$ have the same $2m$ exponents, counted with
multiplicity. Since
\[
 \{L_j(\alpha,M_{D,z}(\cdot+it)):1\le j\le2m\}
 \subset[-\eta,\eta]\subset(-2\pi s_0,2\pi s_0),
\]
these are exactly the exponents listed above. Hence
\begin{equation}\label{eq:uniform-center-input}
 \begin{gathered}
 L_j(\alpha,M_{D,z}(\cdot+it))=\gamma_{m+1-j}^{v_D}(z),
 \qquad1\le j\le m,\\
 D\ge D_0,\qquad z\in\mathcal E,\qquad |t|<h.
 \end{gathered}
\end{equation}
The domain and the constants in the product bounds are independent of
$D\ge D_0$.

\par\smallskip\noindent\emph{Step 4. The limiting spectrum and Jensen formula.}\par\nobreak
Since the dimension is fixed at $2m$, \eqref{eq:center-approximation}
and Lyapunov continuity \cite{AJS} give
\[
 \gamma_i(z):=\lim_{D\to\infty}\gamma_i^{v_D}(z)
 =L_{m+1-i}(\alpha,M_z),\qquad1\le i\le m,
\]
and
\[
 L_j(\alpha,M_z(\cdot+it))
 =\gamma_{m+1-j}(z),\qquad
 z\in\mathcal E,\quad |t|<h,\quad1\le j\le m.
\]
Since $\gamma_{m+1}^{v_D}(\zeta)>2\pi s_0$ for $\zeta=z,\bar z$, the
terms $i>m$ in \eqref{eq:finite-complex-jensen} vanish for
$0<\varepsilon<s_0$.
By  Lyapunov continuity
\cite{AJS} and continuity of $a\mapsto a_+$,
\begin{equation}\label{eq:local-complex-jensen}
\begin{aligned}
 L_{\varepsilon}(z)-L_0(z)
 &=\sum_{i=1}^{m}\bigl(2\pi\varepsilon-\gamma_i(\bar z)\bigr)_+,\\
 L_{-\varepsilon}(z)-L_0(z)
 &=\sum_{i=1}^{m}\bigl(2\pi\varepsilon-\gamma_i(z)\bigr)_+,
 \qquad z\in\mathcal E.
\end{aligned}
\end{equation}
This includes the turning points. Since
$M^\natural\mathcal J_mM=\mathcal J_m$ and $\mathcal J_m$ is unitary,
\begin{gather*}
 M_n(z,x+it)^{-1}
 =\mathcal J_m^{-1}M_n(\bar z,x-it)^*\mathcal J_m,\\
 L_j(\alpha,M_z(\cdot+it))
 =-L_{2m+1-j}(\alpha,M_{\bar z}(\cdot-it)).
\end{gather*}
Hence \eqref{eq:center-spectral-data} and
\eqref{eq:center-nonreal-regularity} hold. For $E\in J$, Step~1 and
$\gamma_i(E)\ge0$ give
\[
 m=\omega(E)
 =\lim_{\varepsilon\downarrow0}
 \sum_{i=1}^m\frac{(2\pi\varepsilon-\gamma_i(E))_+}
                         {2\pi\varepsilon}
 =\#\{i\le m:\gamma_i(E)=0\}.
\]
Thus
\[
 \gamma_1(E)=\cdots=\gamma_m(E)=0,
 \qquad L_j(\alpha,M_E(\cdot+it))=0
 \quad(1\le j\le2m,\ |t|<h).
\]

\par\smallskip\noindent\emph{Step 5. The strict gap at nonreal energy.}\par\nobreak
Fix $z\in\mathcal E\setminus\R$. 
Since
$
 z\notin\sigma(H_{\tau v,\alpha,\theta})$ we have $
  (\alpha,A_z^{\tau v})$ is uniformly hyperbolic.
By local constancy of acceleration \cite{AJS} and
phase-independence at $\tau=0$, we have
$
 \omega(z)=\omega(\bar z)=0.
$
Taking the right derivative in \eqref{eq:local-complex-jensen} we obtain
\[
 0=\omega(z)=\lim_{\varepsilon\downarrow0}\sum_{i=1}^m
   \frac{(2\pi\varepsilon-\gamma_i(\bar z))_+}{2\pi\varepsilon}=\#\{i\le m:\gamma_i(\bar z)=0\}.
\]
Thus $\gamma_1(\bar z)>0$ and we obtain 
\eqref{eq:center-nonreal-gap}.

\par\smallskip\noindent\emph{Step 6. The weighted solution frame.}\par\nobreak
The product bounds from Step~3 and the boundedness of $P_D,P$ give
\begin{equation}\label{eq:frame-growth}
 \|M_{D,n}\|_{\mathcal E,h}+\|M_n\|_{\mathcal E,h}
 +\|Z_D(\cdot;n)\|_{\mathcal E,h}+\|Z(\cdot;n)\|_{\mathcal E,h}
 \le C_\eta e^{\eta|n|},\qquad n\in\Z.
\end{equation}
For $q\ge0$,
\[
 \|Z_D-Z\|_{\mathcal E,h,\beta}
 \le\max_{|n|\le q}\|Z_D(n)-Z(n)\|_{\mathcal E,h}
     +C_\eta e^{-(\beta-\eta)q}.
\]
Letting $D\to\infty$ first,
\[
 \limsup_{D\to\infty}\|Z_D-Z\|_{\mathcal E,h,\beta}
 \le C_\eta e^{-(\beta-\eta)q}\longrightarrow0
 \qquad(q\to\infty).
\]
By the same tail bound we obtain norm convergence of the holomorphic
finite-row truncations to $Z$. Thus $Z$ is a bounded holomorphic map
into $\mathcal L(\C^{2m},X_\beta)$.

Since $\beta<\rho$,
\begin{align*}
 \|L_{v,\alpha,x}-L_{v_D,\alpha,x}\|_{X_\beta\to X_\beta}
 &\le\sum_{|s|>D}|\widehat v_s|e^{\beta|s|}\\
 &\le A_\rho e^{-(\rho-\beta)D}\longrightarrow0.
\end{align*}
By $(L_{v_D,\alpha,x}-z)Z_{D,z}=0$,
\begin{align*}
 \|(L_{v,\alpha,x}-z)Z_z\|_\beta
 &\le C\|Z_z-Z_{D,z}\|_\beta\\
 &\quad+\|L_{v,\alpha,x}-L_{v_D,\alpha,x}\|
                  \|Z_{D,z}\|_\beta\longrightarrow0.
\end{align*}
Finally, by \eqref{eq:iterates} we have
\[
 M_{n+1}(x)=M_n(x+\alpha)M(x)\qquad(n\in\Z),
\]
so \eqref{eq:rows} yields
\[
 Z(x;n+1)=P(x+(n+1)\alpha)M_{n+1}(x)
          =Z(x+\alpha;n)M(x).
\]
This proves \eqref{eq:center-solution}.

\par\smallskip\noindent\emph{Step 7. Wronskian normalization.}\par\nobreak
For $d=l(D)$, put
\[
 \Phi_d(u)=(u_{d-1},\ldots,u_0,u_{-1},\ldots,u_{-d})^T,
 \qquad S_D=\begin{pmatrix}0&-C^*\\C&0\end{pmatrix},
\]
where
\[
 C_{sr}=\begin{cases}
 \widehat v_{d+s-r},&0\le s\le r\le d-1,\\
 0,&0\le r<s\le d-1.
 \end{cases}
\]
Setting $i=-1-s$, $j=d-1-r$, we obtain
\begin{align*}
 \Phi_d(u)^*S_D\Phi_d(w)
 &=\sum_{0\le s\le r\le d-1}
 \bigl(\overline{u_{-1-s}}\widehat v_{d+s-r}w_{d-1-r}
       -\overline{u_{d-1-r}}\widehat v_{-(d+s-r)}w_{-1-s}\bigr)\\
 &=u^*G_{v_D}w.
\end{align*}
Since $\Phi_d(Z_Da)=O_Da$, for real $(E,x)$,
\[
 Z_D^*G_{v_D}Z_D=O_D^*S_DO_D=\mathcal J_m.
\]
Since $i<0\le j$ implies $|i|+|j|=j-i$, \eqref{eq:green-bound} gives
\begin{align*}
 \|Z^*G_vZ-Z_D^*G_{v_D}Z_D\|
 &\le C(\|Z\|_\beta+\|Z_D\|_\beta)\|Z-Z_D\|_\beta\\
 &\quad+2\|Z_D\|_\beta^2
           \sum_{s>D}s|\widehat v_s|e^{\beta s}
 \longrightarrow0.
\end{align*}
The convergence is uniform for real parameters since $\beta<\rho$.
By normal convergence and holomorphic uniqueness,
\[
 Z^\natural G_vZ=\mathcal J_m
 \quad\text{on }\mathcal E\times S_h.
\]
Hence
\[
 Za=0\ \Longrightarrow\ \mathcal J_ma=0
 \ \Longrightarrow\ a=0.
\qedhere
\]
\end{proof}

\subsection{Finite observation}
The following lemma shows that a center solution is determined by
its values on a finite interval, with a quantitative bound uniform
on compact parameter sets.
\begin{lemma}\label{lem:finite-observation}
For every compact $K\subset(\mathcal E\cap\R)\times\T$, there are
$p\ge1$ and $c>0$ such that
\begin{equation}\label{eq:positive-window}
 \sum_{n=0}^{p-1}Z_E(x;n)^*Z_E(x;n)\ge cI_{2m},
 \qquad(E,x)\in K.
\end{equation}
\end{lemma}
\begin{proof}
Assume $K\ne\varnothing$ and set
$K_0=(\operatorname{pr}_E K)\times\T$, $S=\{a\in\C^{2m}:|a|=1\}$.
By injectivity, there is an increasing open cover
\[
 K_0\times S
 =\bigcup_{q\ge0}\Bigl\{(E,x,a):
        \sum_{n=-q}^q\|Z_E(x;n)a\|^2>0\Bigr\}.
\]
By compactness and continuity we obtain $q\ge0$ and $c_0>0$ such that
\[
 \min_{K_0\times S}\sum_{n=-q}^q\|Z_E(x;n)a\|^2=c_0>0.
\]
By homogeneity,
\[
 \sum_{n=-q}^q\|Z_E(x;n)a\|^2\ge c_0\|a\|^2,
 \qquad(E,x)\in K_0.
\]
By \eqref{eq:rows},
\[
 Z_E(x;n+q)=Z_E(x+q\alpha;n)M_{E,q}(x).
\]
Since $K_0$ is invariant under phase translations,
\begin{align*}
 &\sum_{n=0}^{2q}\|Z_E(x;n)a\|^2=\sum_{n=-q}^q
   \|Z_E(x+q\alpha;n)M_{E,q}(x)a\|^2\\
 &\ge c_0\|M_{E,q}(x)a\|^2\ge\frac{c_0\|a\|^2}
 {\displaystyle\sup_{K_0}\|M_{E,q}^{-1}\|^2}.
\end{align*}
The denominator is finite by continuity and invertibility. Take
\[
 p=2q+1,\qquad
 c=\frac{c_0}{\displaystyle\sup_{K_0}\|M_{E,q}^{-1}\|^2}.
\]
\end{proof}

\section{Kotani theory for the intrinsic center}\label{sec:kotani}
As we explained at the beginning, our main route to obtain all-frequency Cantor spectrum is through Kotani--almost reducibility for the intrinsic center. In \cite{GJY}, we established Kotani theory for the two dimensional intrinsic center of the duals of type I operators. Later, in \cite{GeXu}, the Kotani theory was extended to arbitrary {\it finite-range} partially hyperbolic operators with general ergodic underlying dynamical systems. In this section, we further extend \cite{GeXu} and the Green-form argument of
and \cite[Sections~7 and~10]{GJY} to the {\it infinite-range} dual of any Type $k$ operators. See also \cite{KotaniFloquet,KS,XuJEMS} for other related results.
All phase integrals below are over $\T$.

\begin{theorem}\label{thm:center-kotani}
Let $Z,M$ be as in Proposition~\ref{prop:actual-center}, and
$I\subset\operatorname{int}J$ be open.
For every $E_0\in I$, there exist an open $\mathcal E\ni E_0$, $h>0$,
and a periodic holomorphic $C:\mathcal E\times S_h\to \mathrm{GL}(2m,\C)$ such that
\[
 C_E^*(x)\mathcal J_mC_E(x)=\mathcal J_m,\qquad
 C_E(x+\alpha)^{-1}M_E(x)C_E(x)\in\U(m)\times\U(m)
\]
for $(E,x)\in(\mathcal E\cap\R)\times\T$.
\end{theorem}

Fix $E_0\in I$, and choose a nondegenerate compact interval
$I'\Subset I$ with $E_0\in\operatorname{int}I'$.
Let $\mathcal E$ be a bounded conjugation-invariant neighborhood of
$I'$ such that $\mathcal E\cap\R\subset I$.
Until Lemma~\ref{lem:analytic-continuation}, all shrinkings of
$\mathcal E$ will retain the whole interval $I'$ with
constants  uniform on $I'\times\T$.
For a Hermitian matrix $H$, we write $H=O(y^2R^*R)$ if
$
 |a^*Ha|\le Cy^2\|Ra\|^2
$ for every vector $a,
$
with $C$ uniform in the parameters. 

\subsection{Normalization of the solution frame}\label{subsec:canonical-gauge}
Set $Z\leftarrow ZV^{-1}$ and $N=VMV^{-1}$, with $V$ as in
\eqref{eq:J}. Then
\begin{equation}\label{eq:canonical-center}
 TZ=Z^+N,\qquad Z^\natural G_vZ=J_{2m},\qquad
 N^\natural J_{2m}N=J_{2m}.
\end{equation}
Write $P_n(E,x)=Z(E,x;n)$. By \eqref{eq:rows},
\[
 P_n=P_0(E,x+n\alpha)N_n,\qquad P_1=P_0^+N.
\]
\begin{lemma}\label{lem:midpoint-gauge}
There exists a periodic holomorphic $Q$ with $Q^\natural J_{2m}Q=J_{2m}$
such that, after
\[
 Z\longmapsto ZQ,\qquad N\longmapsto(Q^+)^{-1}NQ,
\]
\begin{equation}\label{eq:midpoint-horizontal}
 Z^\natural G_v\partial_EZ=-\tfrac12P_0^\natural P_0,
\end{equation}
and
\begin{equation}\label{eq:midpoint-energy}
 N^\natural J_{2m}\partial_EN
 =\tfrac12(P_0^\natural P_0+P_1^\natural P_1).
\end{equation}
Moreover, uniformly on compact  sets,
\begin{equation}\label{eq:height-expansion}
 Z(E+iy,x)^*G_vZ(E+iy,x)
 =J_{2m}-iyP_0(E+iy,x)^*P_0(E+iy,x)+O(y^2).
\end{equation}
\end{lemma}
\begin{proof}
Differentiating $Z^\natural G_vZ=J_{2m}$ gives
\[
 (Z^\natural G_v\partial_EZ)^\natural=Z^\natural G_v\partial_EZ.
\]
Put
\[
 H=Z^\natural G_v\partial_EZ+\tfrac12P_0^\natural P_0,
 \qquad H^\natural=H.
\]
On a simply connected symmetric neighborhood of $I'$, solve
\[
 \partial_EQ=-J_{2m}^{-1}HQ,\qquad Q(E_0,x)=I.
\]
By~\cite[Theorems~4.2 and~4.7]{TeschlODE}, $Q$ extends
analytically to a product neighborhood of $I'\times\T$.
Uniqueness implies that $Q$ is periodic in $x$. Since $J_{2m}^*=-J_{2m}$,
\[
 \partial_E(Q^\natural J_{2m}Q)
 =Q^\natural HQ-Q^\natural HQ=0,
 \qquad Q^\natural J_{2m}Q=J_{2m}.
\]
Thus
\[
 (ZQ)^\natural G_v\partial_E(ZQ)
 =Q^\natural(Z^\natural G_v\partial_EZ)Q+Q^\natural J_{2m}\partial_EQ
 =-\tfrac12(P_0Q)^\natural(P_0Q).
\]
For the remainder of this proof, we write $(Z,N,P_n)$ for
$(ZQ,(Q^+)^{-1}NQ,P_nQ)$. The generalized Wronskian identity~\eqref{eq:green-jump},
applied to $Z$ and $\partial_E Z$, and $TZ=Z^+N$ give
\begin{align*}
 P_0^\natural P_0
 &=(TZ)^\natural G_v(T\partial_EZ)-Z^\natural G_v\partial_EZ\\
 &=N^\natural(Z^\natural G_v\partial_EZ)^+N
   +N^\natural J_{2m}\partial_EN-Z^\natural G_v\partial_EZ.
\end{align*}
By \eqref{eq:midpoint-horizontal} and $P_1=P_0^+N$,
\[
 N^\natural J_{2m}\partial_EN
 =\tfrac12(P_0^\natural P_0+P_1^\natural P_1).
\]

For real $E,x$,
$(\partial_EZ)^*G_vZ=-Z^*G_v\partial_EZ$. By Taylor expansion,
\begin{align*}
 Z(E+iy,x)^*G_vZ(E+iy,x)
 &=J_{2m}+iy\bigl(Z^*G_v\partial_EZ-(\partial_EZ)^*G_vZ\bigr)+O(y^2)\\
 &=J_{2m}+2iyZ^*G_v\partial_EZ+O(y^2)\\
 &=J_{2m}-iyP_0(E,x)^*P_0(E,x)+O(y^2).
\end{align*}
The same expansion gives
\begin{gather*}
 P_0(E+iy,x)=P_0(E,x)+O(y),\\
 y\bigl(P_0(E+iy,x)^*P_0(E+iy,x)-P_0(E,x)^*P_0(E,x)\bigr)=O(y^2).
\end{gather*}
This proves \eqref{eq:height-expansion}. Its remainder is skew-Hermitian,
so we write it as $iR_y$. Then
\[
 R_y=R_y^*,\qquad \|R_y\|\le Cy^2,\qquad
 -Cy^2F^*F\le F^*R_yF\le Cy^2F^*F
\]
for every matrix $F$ of compatible size.
\end{proof}

\subsection{Half-line coordinates}
We now make the change of frame in Lemma~\ref{lem:midpoint-gauge}
and denote its matrix by $Q_{\rm mid}$. We continue to write
$Z,N,P_n$ for the new frame, cocycle, and site rows.
 
For $z=E+iy$, write
$\Delta_z=\partial_E^2+\partial_y^2$.

By Lemma~\ref{lem:finite-observation} and boundedness of
$Q_{\rm mid}^{\pm1}$, choose $p\ge1$ such that, for $(E,x)\in I'\times\T$,
\[
 \sum_{n=0}^{p-1}P_n(E,x)^*P_n(E,x)\ge cI.
\]
By covariance,
\begin{align*}
 \sum_{n=1}^{p}P_n^*P_n
 &=N^*\left(\sum_{n=0}^{p-1}P_n^*P_n\right)^+N
 \ge c\|N^{-1}\|^{-2}I,\\
 \sum_{n=-p}^{-1}P_n^*P_n
 &=N_{-p}^*\left(\sum_{n=0}^{p-1}P_n^*P_n\right)(x-p\alpha)N_{-p}
 \ge c\|N_{-p}^{-1}\|^{-2}I.
\end{align*}
The inverse products are uniformly bounded on $I'\times\T$.
By continuity, after shrinking $\mathcal E$,
\begin{gather*}
 \mathcal W_+=\{1,\ldots,p\},\qquad\mathcal W_-=\{-p,\ldots,-1\},\\
 \sum_{n\in\mathcal W_\pm}P_n(z,x)^*P_n(z,x)\ge cI
 \qquad((z,x)\in\mathcal E\times\T).
\end{gather*}

\begin{lemma}\label{lem:nonreal-dichotomy}
After shrinking the complex neighborhood $\mathcal E$ of $I'$, the
normalized cocycle has invariant planes
\[
 \C^{2m}=E^s(z,x)\oplus E^u(z,x),\qquad
 \dim E^s(z,x)=\dim E^u(z,x)=m\quad(z\in\mathcal E\setminus\R).
\]
For every compact $K\subset(\mathcal E\setminus\R)\times\T$, there exist
$C_K,c_K>0$ such that
\[
 \|N_n(z,x)|_{E^s(z,x)}\|+\|N_{-n}(z,x)|_{E^u(z,x)}\|
 \le C_Ke^{-c_Kn}\quad((z,x)\in K,\ n\ge0).
\]
Writing $L_1(z)\ge\cdots\ge L_{2m}(z)$ for the exponents of
$(\alpha,N_z)$, put $\gamma(z)=-\sum_{j=m+1}^{2m}L_j(z)$. Then
\[
 \Delta_z\gamma=0\quad(z\in\mathcal E,\ \Im z>0),
 \qquad\lim_{y\downarrow0}\sup_{E\in I'}\gamma(E+iy)=0
 \quad(I'\Subset I).
\]
\end{lemma}
\begin{proof}
By Proposition~\ref{prop:actual-center},
\[
 L_m(z)>0>L_{m+1}(z),\qquad
 \sum_{j=1}^mL_j(\alpha,N_z(\cdot+i\varepsilon))
 =\sum_{j=1}^mL_j(z)
\]
for small $|\varepsilon|$. By \cite{AJS}, there is an
$m$-dominated splitting $E^u\oplus E^s$. Uniform subadditivity

gives
\begin{gather*}
 \limsup_{n\to\infty}\frac1n\sup_x
       \log\|N_n(z,x)|_{E^s(z,x)}\|\le L_{m+1}(z)<0,\\
 \limsup_{n\to\infty}\frac1n\sup_x
       \log\|N_{-n}(z,x)|_{E^u(z,x)}\|\le-L_m(z)<0.
\end{gather*}
Contracting iterates persist locally; a finite cover of $K$ gives $C_K,c_K$.

By \cite{AJS}, $E^u$ is locally jointly
holomorphic in energy and phase. Apply this to
$(-\alpha,N_z(\cdot-\alpha)^{-1})$ for $E^s$.
The projections $\Pi_s$ onto $E^s$ along $E^u$ and $\Pi_u=I-\Pi_s$
are locally jointly holomorphic. Fix $z_0\in\mathcal E\cap\mathbb H$.
The bundles over the phase annulus are topologically trivial, hence
admit periodic holomorphic frames $V_s^0,V_u^0$ \cite{OkaGrauert}.
To extend the frames at $z_0$ holomorphically in energy, choose
a sufficiently small disk about $z_0$ and solve the following
linear ODE in $z$, with $x$ as a parameter
using e.g. \cite[Section~4.1]{TeschlODE}:

\[
 \partial_z\mathcal U=[\partial_z\Pi_s,\Pi_s]\mathcal U,
 \qquad\mathcal U(z_0,x)=I.
\]
Since $\tr[\partial_z\Pi_s,\Pi_s]=0$, $\det\mathcal U=1$.
Since $\Pi_s^2=\Pi_s$,
\[
 \Pi_s(\partial_z\Pi_s)\Pi_s=0,\qquad
 [[\partial_z\Pi_s,\Pi_s],\Pi_s]=\partial_z\Pi_s.
\]
Hence
\begin{align*}
 \partial_z(\mathcal U^{-1}\Pi_s\mathcal U)
 &=\mathcal U^{-1}
 \bigl(\partial_z\Pi_s-[[\partial_z\Pi_s,\Pi_s],\Pi_s]\bigr)\mathcal U=0,\\
 \mathcal U^{-1}\Pi_s\mathcal U&=\Pi_s(z_0,x).
\end{align*}
The same identity holds for $\Pi_u=I-\Pi_s$. Thus
$V_\sigma=\mathcal UV_\sigma^0$ frames $E^\sigma$, $\sigma=s,u$.

Write $NV_s=V_s^+T_s$. Then
\[
 \left\|\bigwedge^mN_n\bigm|_{\bigwedge^mE^s}\right\|
 =|\det T_{s,n}|\left(
 \frac{\det(V_s^*V_s)(x+n\alpha)}{\det(V_s^*V_s)(x)}\right)^{1/2}.
\]
The Gram ratio is bounded above and below uniformly in $x,n$. Take
logarithms in the identity, divide by $n$, integrate and let $n\to\infty$:
\[
 \gamma(z)=-\int_\T\log|\det T_s(z,x)|\,dx.
\]
Under $V_s\mapsto V_sD$,
\[
 T_s\mapsto(D^+)^{-1}T_sD,\qquad
 \int_\T\log|\det D|\,dx=\int_\T\log|\det D^+|\,dx.
\]
Thus the integrals agree on overlapping energy neighborhoods.
Since $\det T_s\ne0$,
\[
 \Delta_z\log|\det T_s(z,x)|=0
 \quad\Longrightarrow\quad\Delta_z\gamma(z)=0.
\]

Moreover,
\[
 \gamma(z)=\inf_{n\ge1}\frac1n\int_\T
 \log\|\bigwedge^mN_{-n}(z,x)\|\,dx\ge0.
\]
By continuity of finite products and vanishing of the real-energy exponents,
\[
 0\le\limsup_{z\to E}\gamma(z)
 \le\inf_{n\ge1}\frac1n\int_\T
 \log\|\bigwedge^mN_{-n}(E,x)\|\,dx=0.
\]
For $\varepsilon>0$, these continuous finite-product upper bounds give
 neighborhoods on which $\gamma<\varepsilon$; 
so by compactness we obtain the uniform limit.
\end{proof}

Take $\mathcal E$ to be rectangular and put
$\mathcal E_+=\mathcal E\cap\mathbb H$. Locally in energy, write
\[
 V_\sigma=\binom{V_{\sigma,1}}{V_{\sigma,2}},\qquad
 E^\sigma=\ran V_\sigma\quad(\sigma=s,u),\qquad NV_s=V_s^+T_s.
\]

For $z=E+iy\in\mathcal E_+$, we will use
\begin{equation}\label{eq:weyl-graph-definition}
 E^s(z,x)=\operatorname{ran}\binom{m_+(z,x)}{I_m},\qquad
 E^u(z,x)=\operatorname{ran}\binom{-m_-(z,x)}{I_m}.
\end{equation}

With $(z,x)$ suppressed, use the notation
\[
 F_+=\binom{m_+}{I_m},\quad \Phi_-=(-m_-,I_m),\quad
 A_n=P_nF_+,\quad C_n=\Phi_-P_n^\natural.
\]
\begin{lemma}\label{lem:graph-coordinates}
After shrinking $\mathcal E$, \eqref{eq:weyl-graph-definition} defines
periodic $m_\pm$, jointly holomorphic locally near
$\mathcal E_+\times\T$, with
\[
 E^u(z,x)=\ker(\Phi_-J_{2m}),\qquad \Im m_\pm>0.
\]
Set
\[
 K_+=y\left(\tfrac12A_0^*A_0+\sum_{n\ge1}A_n^*A_n\right),\qquad
 K_-=y\left(\sum_{n\le-1}C_nC_n^*+\tfrac12C_0C_0^*\right).
\]
Uniformly for $E\in I'\Subset I$ and small $y>0$,
\begin{equation}\label{eq:midpoint-comparison}
 (1-Cy)K_\pm\le \Im m_\pm\le(1+Cy)K_\pm.
\end{equation}
\end{lemma}
\begin{proof}
By \eqref{eq:halflinemass}, exponential decay and
\eqref{eq:height-expansion},
\begin{align*}
 V_s^*\tfrac{iJ_{2m}}2V_s
 &=y\left(\tfrac12V_s^*P_0^*P_0V_s+\sum_{n\ge1}V_s^*P_n^*P_nV_s\right)
   +O(y^2V_s^*V_s),\\
 -V_u^*\tfrac{iJ_{2m}}2V_u
 &=y\left(\sum_{n\le-1}V_u^*P_n^*P_nV_u+\tfrac12V_u^*P_0^*P_0V_u\right)
   +O(y^2V_u^*V_u).
\end{align*}
Each half-line sum contains a fixed observation window. Uniformly for
$E\in I'$ and small $y>0$,
\begin{align*}
 y\sum_{n\in\mathcal W_+}V_s^*P_n^*P_nV_s&\ge cyV_s^*V_s,\\
 y\sum_{n\in\mathcal W_-}V_u^*P_n^*P_nV_u&\ge cyV_u^*V_u.
\end{align*}
Consequently, for $Cy<c$,
\begin{align*}
 V_s^*\tfrac{iJ_{2m}}2V_s&\ge(cy-Cy^2)V_s^*V_s>0,\\
 -V_u^*\tfrac{iJ_{2m}}2V_u&\ge(cy-Cy^2)V_u^*V_u>0.
\end{align*}
Since $\C^m\times\{0\}$ is $iJ_{2m}$-isotropic,
\[
 \ker V_{\sigma,2}=\{0\},\qquad\det V_{\sigma,2}\ne0
 \quad(\sigma=s,u).
\]
Set
\begin{gather*}
 m_+=V_{s,1}V_{s,2}^{-1},\qquad
 m_-=-V_{u,1}V_{u,2}^{-1},\\
 F_+^*\tfrac{iJ_{2m}}2F_+=\Im m_+>0,\qquad
 -\binom{-m_-}{I_m}^*\tfrac{iJ_{2m}}2\binom{-m_-}{I_m}
 =\Im m_->0.
\end{gather*}
The quotients are frame-independent, periodic and locally jointly holomorphic.

For real $x$, pair the unstable solutions at $z$ and $\bar z$.
Green's identity and decay at $-\infty$ give
\[
 E^u(\overline z,x)^*J_{2m}E^u(z,x)=0.
\]
Moreover,
\[
 \Phi_-J_{2m}\binom{-m_-}{I_m}=(-m_-,I_m)\binom{I_m}{m_-}=0.
\]
Both $E^u(\bar z,x)$ and $\operatorname{ran}\Phi_-(z,x)^*$ have dimension
$m$ and lie in the $m$-dimensional $J_{2m}$-orthogonal complement of
$E^u(z,x)$. Hence
\[
 E^u(\overline z,x)=\operatorname{ran}\Phi_-(z,x)^*,\qquad
 P_n(\overline z,x)\Phi_-(z,x)^*=C_n(z,x)^*.
\]
By the half-line Green identities,
\begin{align*}
 y\sum_{n\ge0}A_n^*A_n
 &=\tfrac i2F_+^*Z(z,x)^*G_vZ(z,x)F_+,\\
 y\sum_{n\le-1}C_nC_n^*
 &=\tfrac i2\Phi_-Z(\overline z,x)^*G_vZ(\overline z,x)\Phi_-^*.
\end{align*}
By \eqref{eq:height-expansion} at $y$ and $-y$,
\[
 \Im m_+=K_++O(y^2F_+^*F_+),\qquad
 \Im m_-=K_-+O(y^2\Phi_-\Phi_-^*).
\]
By finite observation,
\[
 K_+\ge cyF_+^*F_+,\qquad K_-\ge cy\Phi_-\Phi_-^*.
\]
Hence
\[
 -CyK_\pm\le\Im m_\pm-K_\pm\le CyK_\pm.\qedhere
\]
\end{proof}

\subsection{The Johnson--Moser identity and Kotani estimates}
By Lemma~\ref{lem:graph-coordinates}, put
\[
 T_+=V_{s,2}^+T_sV_{s,2}^{-1}.
\]
Then
\[
 \int_\T\log|\det T_+|\,dx
 =\int_\T\log|\det T_s|\,dx=-\gamma(z).
\]
Following \cite{JM,GJYZ}, write
\[
 NF_+=F_+^+T_+,\qquad \Phi_-N^\natural=T_-\Phi_-^+,
\]
\[
 \Phi_-J_{2m}F_+=-(m_++m_-).
\]
For $n\in\Z$, define
\[
 N_nF_+=F_+(x+n\alpha)T_{+,n},\qquad
 \Phi_-N_n^\natural=T_{-,n}\Phi_-(x+n\alpha).
\]
In particular, the row transport has the order
$T_{-,n}=T_-(x)\cdots T_-(x+(n-1)\alpha)$ for $n>0$.

\begin{lemma}\label{lem:johnson-moser}
$\det(m_++m_-)\ne0$, and
\begin{equation}\label{eq:johnson-moser}
 \partial_y\gamma(z)=-\Im\int_\T\tr\bigl((m_++m_-)^{-1}C_0A_0\bigr)\,dx.
\end{equation}
\end{lemma}
\begin{proof}
Work locally in $z\in\mathcal E_+$. Since $\Im(m_++m_-)>0$ on
$\T$, shrink the product domain so that all factors below are holomorphic
and $m_++m_-$ is invertible. Put
\[
 \Lambda=-(m_++m_-)^{-1}\Phi_-J_{2m}.
\]
By the graph identities and invariance,
\[
 \Lambda F_+=I_m,\qquad \Lambda|_{E^u}=0,\qquad
 \Lambda^+N=T_+\Lambda.
\]
The last equality holds on $E^s$ by transport and on $E^u$ because
both sides vanish there. Also, $NE^s=(E^s)^+$ makes $T_+$ invertible.
Differentiate $NF_+=F_+^+T_+$:
\[
 (\partial_zN)F_++N\partial_zF_+
 =(\partial_zF_+)^+T_++F_+^+\partial_zT_+.
\]
Apply $\Lambda^+$:
\[
 \partial_zT_+
 =\Lambda^+(\partial_zN)F_++T_+\Lambda\partial_zF_+
   -\Lambda^+(\partial_zF_+)^+T_+.
\]
Using $T_+^{-1}\Lambda^+=\Lambda N^{-1}$ and trace cyclicity,
\begin{align*}
 \tr(T_+^{-1}\partial_zT_+)
 &=\tr\bigl(\Lambda N^{-1}(\partial_zN)F_+\bigr)\\
 &\quad+\tr(\Lambda\partial_zF_+)
       -\bigl[\tr(\Lambda\partial_zF_+)\bigr]^+.
\end{align*}
By the row and column transports,
\[
 m_++m_-=T_-(m_+^++m_-^+)T_+,\qquad
 C_1A_1=T_-(C_0A_0)^+T_+.
\]

By $J_{2m}N^{-1}=N^\natural J_{2m}$ and \eqref{eq:midpoint-energy},
\begin{align*}
 \Phi_-J_{2m}N^{-1}(\partial_zN)F_+
 &=\Phi_-N^\natural J_{2m}(\partial_zN)F_+\\
 &=\tfrac12\Phi_-(P_0^\natural P_0+P_1^\natural P_1)F_+
 =\tfrac12(C_0A_0+C_1A_1),\\
 (m_++m_-)^{-1}C_1A_1
 &=T_+^{-1}(m_+^++m_-^+)^{-1}(C_0A_0)^+T_+.
\end{align*}
Taking traces and using $\int_\T f^+=\int_\T f$, we obtain
\[
 \int_\T\tr(T_+^{-1}\partial_zT_+)\,dx
 =-\int_\T\tr\bigl((m_++m_-)^{-1}C_0A_0\bigr)\,dx.
\]
The holomorphic factors and their inverses are locally bounded on $\T$.
Thus differentiation under the integral and Cauchy--Riemann give
\begin{align*}
 \partial_y\gamma(z)
 &=-\int_\T\partial_y\log|\det T_+|\,dx\\
 &=\Im\int_\T\tr(T_+^{-1}\partial_zT_+)\,dx\\
 &=-\Im\int_\T\tr\bigl((m_++m_-)^{-1}C_0A_0\bigr)\,dx.\qedhere
\end{align*}
\end{proof}

\begin{lemma}\label{lem:kotani-estimates}
Uniformly for $E\in I'\Subset I$ and small $y>0$,
\begin{equation}\label{eq:kotani-estimate}
\begin{split}
 \max\left\{\int_\T\tr\bigl(A_0^*A_0(\Im m_+)^{-1}\bigr)\,dx,
 \int_\T\tr\bigl(C_0C_0^*(\Im m_-)^{-1}\bigr)\,dx\right\}
 &\le(1+Cy)\frac{2\gamma(z)}y,\\
 \max\left\{\int_\T\tr(A_0^*A_0K_+^{-1})\,dx,
 \int_\T\tr(C_0C_0^*K_-^{-1})\,dx\right\}
 &\le\frac{2\gamma(z)}y.
\end{split}
\end{equation}
\end{lemma}
\begin{proof}
Since
\[
 A_{n+1}(x)=A_n(x+\alpha)T_+(x),\qquad
 C_{n+1}(x)=T_-(x)C_n(x+\alpha),
\]
we have
\begin{align*}
 K_+-\tfrac y2A_0^*A_0
 &=T_+^*(K_+^++\tfrac y2(A_0^*A_0)^+)T_+,\\
 K_-+\tfrac y2C_0C_0^*
 &=T_-(K_-^+-\tfrac y2(C_0C_0^*)^+)T_-^*.
\end{align*}
Also, $m_++m_-=T_-(m_+^++m_-^+)T_+$ gives
\[
 \int_\T\log|\det T_-|\,dx
 =-\int_\T\log|\det T_+|\,dx=\gamma(z).
\]
By finite observation on the strict half-lines,
\[
 K_+-\tfrac y2A_0^*A_0=y\sum_{n\ge1}A_n^*A_n>0,\qquad
 K_--\tfrac y2C_0C_0^*=y\sum_{n\le-1}C_nC_n^*>0.
\]
Taking logarithmic determinants and phase integrals,
\[
 \begin{gathered}
 \int_\T\log\frac{\det(K_++\tfrac y2A_0^*A_0)}
 {\det(K_+-\tfrac y2A_0^*A_0)}\,dx=2\gamma(z),\\
 \int_\T\log\frac{\det(K_-+\tfrac y2C_0C_0^*)}
 {\det(K_--\tfrac y2C_0C_0^*)}\,dx=2\gamma(z).
 \end{gathered}
\]

Let $K>0$, $H\ge0$, $K-yH/2>0$, and let $\lambda_j$ be the
eigenvalues of $(y/2)K^{-1/2}HK^{-1/2}$. Since
\[
 0\le\lambda_j<1,\qquad
 \log\frac{1+t}{1-t}=2\int_0^t\frac{ds}{1-s^2}\ge2t\quad(0\le t<1),
\]
we have
\[
 \log\frac{\det(K+yH/2)}{\det(K-yH/2)}
 =\sum_j\log\frac{1+\lambda_j}{1-\lambda_j}
 \ge2\sum_j\lambda_j=y\tr(HK^{-1}).
\]
Apply this to $(K,H)=(K_+,A_0^*A_0),(K_-,C_0C_0^*)$ and integrate:
\[
 \int_\T\tr(A_0^*A_0K_+^{-1})\,dx\le\frac{2\gamma}y,\qquad
 \int_\T\tr(C_0C_0^*K_-^{-1})\,dx\le\frac{2\gamma}y.
\]
This is the second estimate. For the first, \eqref{eq:midpoint-comparison}
gives
\[
 (\Im m_\pm)^{-1}\le(1-Cy)^{-1}K_\pm^{-1}
 \le(1+2Cy)K_\pm^{-1}.
\]
Taking traces against $A_0^*A_0,C_0C_0^*\ge0$ and enlarging $C$
gives \eqref{eq:kotani-estimate}.
\end{proof}

\subsection{Boundary matching}
\begin{lemma}\label{lem:matrix-defect}
For $M,N\in M_m(\C)$ with $\Im M,\Im N>0$ and
$A\in M_{p\times m}(\C)$, $C\in M_{m\times p}(\C)$,
\begin{align*}
 &\tr\bigl(A^*A(\Im M)^{-1}\bigr)
 +\tr\bigl(CC^*(\Im N)^{-1}\bigr)
 +4\Im\tr\bigl((M+N)^{-1}CA\bigr)\\
 &=\bigl\|A(\Im M)^{-1/2}
 +2iC^*(M+N)^{-*}(\Im M)^{1/2}\bigr\|_{\HS}^2\\
 &\quad+\bigl\|C^*(M+N)^{-*}(M^*+N)(\Im N)^{-1/2}\bigr\|_{\HS}^2.
\end{align*}
\end{lemma}
\noindent The algebraic verification is in Appendix~\ref{app:matrix-identity}.

\begin{theorem}\label{thm:boundary-matching}
For a.e.\ $(E,x)\in\operatorname{int}I'\times\T$,
\begin{equation}\label{eq:boundary-matching}
 m_+(E+i0,x)=-m_-(E+i0,x)^*.
\end{equation}
\end{theorem}
\begin{proof}
Fix a compact interval $I_0\Subset\operatorname{int}I'$.
\par\smallskip\noindent\emph{Step 1. Finite boundary values.}\par\nobreak
Lemma~\ref{lem:graph-coordinates} gives the Cayley transforms
\[
 W_\pm=(m_\pm-iI)(m_\pm+iI)^{-1}.
\]
Since
\begin{gather*}
 I-W_\pm^*W_\pm=4(m_\pm+iI)^{-*}(\Im m_\pm)(m_\pm+iI)^{-1}>0,\\
 I-W_\pm=2i(m_\pm+iI)^{-1},
\end{gather*}
we have $\|W_\pm\|<1$ and $\det(I-W_\pm)\ne0$.
For each fixed phase, Fatou's theorem and boundary uniqueness for the
bounded, nonzero holomorphic function $\det(I-W_\pm)$ give, respectively:
\[
 W_\pm(E+i0,x)\text{ exists},\qquad
 \det(I-W_\pm(E+i0,x))\ne0\quad\text{for a.e. }E.
\]
Thus
\[
 m_\pm(E+i0,x)=i(I+W_\pm(E+i0,x))(I-W_\pm(E+i0,x))^{-1}
\]
for a.e. $(E,x)\in\operatorname{int}I'\times\T$ by Fubini.

\par\smallskip\noindent\emph{Step 2. Strict boundary positivity.}\par\nobreak
By Lemma~\ref{lem:nonreal-dichotomy}, the odd reflection of $\gamma$
is harmonic across $\operatorname{int}I'$. Taylor expansion gives,
uniformly for $E\in I_0$,
\begin{gather*}
 \gamma(E+iy)=y\partial_y\gamma(E+i0)+O(y^3),\\
 \gamma(E+iy)/y=O(1),\qquad
 \gamma(E+iy)/y-\partial_y\gamma(E+iy)=O(y^2).
\end{gather*}
Splitting the half-line sums gives
\begin{align*}
 &K_+-T_{+,n}^*K_+(x+n\alpha)T_{+,n}\\
 &\quad=y\left(\tfrac12A_0^*A_0+\sum_{\ell=1}^{n-1}A_\ell^*A_\ell
                  +\tfrac12A_n^*A_n\right)\ge0
 \quad(n\ge1),\\
 &K_--T_{-,n}K_-(x+n\alpha)T_{-,n}^*\\
 &\quad=y\left(\tfrac12C_nC_n^*+\sum_{\ell=n+1}^{-1}C_\ell C_\ell^*
                  +\tfrac12C_0C_0^*\right)\ge0
 \quad(n\le-1).
\end{align*}
By inversion and \eqref{eq:midpoint-comparison}, for $Cy<1/2$,
\begin{align*}
 T_{+,n}(\Im m_+)^{-1}T_{+,n}^*
 &\le\frac{1+Cy}{1-Cy}(\Im m_+(x+n\alpha))^{-1},\\
 T_{-,n}^*(\Im m_-)^{-1}T_{-,n}
 &\le\frac{1+Cy}{1-Cy}(\Im m_-(x+n\alpha))^{-1}.
\end{align*}
Since $A_n=A_0(x+n\alpha)T_{+,n}$ and
$C_n=T_{-,n}C_0(x+n\alpha)$, trace cyclicity, phase translation and
\eqref{eq:kotani-estimate} give
\begin{gather*}
 \int_\T\tr\bigl(A_n^*A_n(\Im m_+)^{-1}\bigr)\,dx
 \le C\gamma/y
 \quad(n\ge0),\\
 \int_\T\tr\bigl(C_nC_n^*(\Im m_-)^{-1}\bigr)\,dx
 \le C\gamma/y
 \quad(n\le-1).
\end{gather*}
For the fixed windows above,
\begin{align*}
 \sum_{n\in\mathcal W_+}A_n^*A_n
 &=F_+^*\left(\sum_{n\in\mathcal W_+}P_n(z,x)^*P_n(z,x)\right)F_+\\
 &\ge cF_+^*F_+\ge cI,\\
 \sum_{n\in\mathcal W_-}C_nC_n^*
 &=\Phi_-\left(\sum_{n\in\mathcal W_-}
       P_n(\bar z,x)^*P_n(\bar z,x)\right)\Phi_-^*\\
 &\ge c\Phi_-\Phi_-^*\ge cI.
\end{align*}
Summing over the fixed windows, for small $y>0$,
\[
 \int_\T\tr((\Im m_\pm)^{-1})\,dx\le C\gamma/y\le C'.
\]
For a.e. $E\in I_0$, Fatou's lemma gives
\[
 \int_\T\liminf_{y\downarrow0}
    \tr((\Im m_\pm(E+iy,x))^{-1})\,dx\le C'.
\]
If the boundary imaginary part were singular, then
\[
 \tr((\Im m_\pm(E+iy,x))^{-1})
 \ge\lambda_{\min}(\Im m_\pm(E+iy,x))^{-1}\longrightarrow\infty.
\]
Hence
\[
 \Im m_\pm(E+i0,x)>0\quad\text{a.e.}
\]
\par\smallskip\noindent\emph{Step 3. Vanishing of the boundary defect.}\par\nobreak
Fix an energy $E\in I_0$ for which Steps~1--2 hold for a.e. $x$.
Let $D_y(E,x)$ be the sum of squares in Lemma~\ref{lem:matrix-defect}
for $(M,N,A,C)=(m_+,m_-,A_0,C_0)$. The Johnson--Moser identity
\eqref{eq:johnson-moser} and the estimates
\eqref{eq:kotani-estimate} give
\begin{align*}
 0\le\int_\T D_y\,dx
 &=\int_\T\tr\bigl(A_0^*A_0(\Im m_+)^{-1}
                   +C_0C_0^*(\Im m_-)^{-1}\bigr)\,dx
       -4\partial_y\gamma(z)\\
 &\le4\left(\frac{\gamma(z)}y-\partial_y\gamma(z)\right)+4C\gamma(z)
 =O(y)\longrightarrow0.
\end{align*}
Since
\begin{gather*}
 \Im m_\pm(E+i0,x)>0
 \ \Longrightarrow\
 \Im(m_++m_-)(E+i0,x)>0,\\
 \det(m_++m_-)(E+i0,x)\ne0,
\end{gather*}
all factors in the defect have finite boundary limits, and
\[
 D_y(E,x)\longrightarrow D_0(E,x)\ge0\quad\text{a.e. }x.
\]
By Fatou's lemma,
\[
 0\le\int_\T D_0\,dx\le\liminf_{y\downarrow0}\int_\T D_y\,dx=0.
\]

\par\smallskip\noindent\emph{Step 4. Matching on the full center.}\par\nobreak
Let $\mathcal X_E^0$ be the full-measure phase set from Steps~1--3. Put
\[
 \mathcal X_E=\bigcap_{n\in\Z}(\mathcal X_E^0-n\alpha),\qquad
 |\mathcal X_E|=1,\qquad \mathcal X_E+n\alpha=\mathcal X_E\quad(n\in\Z).
\]
For $x\in\mathcal X_E$ and each fixed $n$, the graph normalizations give
\begin{align*}
 T_{+,n}(z,x)&=(0,I_m)N_n(z,x)F_+(z,x),\\
 T_{-,n}(z,x)&=\Phi_-(z,x)N_n^\natural(z,x)\binom0{I_m}.
\end{align*}
The right sides have finite nontangential limits at $E$.
Passing to these limits in the transport identities, and suppressing
$(E+i0,x)$, gives
\[
 N_nF_+=F_+(x+n\alpha)T_{+,n},\qquad
 \Phi_-N_n^*=T_{-,n}\Phi_-(x+n\alpha).
\]
Since $N_n(E,x)$ is invertible and $F_+,\Phi_-$ have rank $m$,
\begin{align*}
 m&=\rank(N_nF_+)
 =\rank(F_+(x+n\alpha)T_{+,n})\le\rank T_{+,n}\le m,\\
 m&=\rank(\Phi_-N_n^*)
 =\rank(T_{-,n}\Phi_-(x+n\alpha))\le\rank T_{-,n}\le m.
\end{align*}
Thus
\[
 \det T_{+,n}\ne0,\qquad\det T_{-,n}\ne0.
\]
On this invariant phase set, $N_n^*J_{2m}N_n=J_{2m}$ gives
\[
 \begin{aligned}
 \Im m_+&=T_{+,n}^*(\Im m_+(x+n\alpha))T_{+,n},\\
 \Im m_-&=T_{-,n}(\Im m_-(x+n\alpha))T_{-,n}^*,\\
 m_++m_-&=T_{-,n}(m_+(x+n\alpha)+m_-(x+n\alpha))T_{+,n}.
 \end{aligned}
\]
Invert these identities and use the row and column transports:
\begin{align*}
 T_{+,n}(\Im m_+)^{-1}T_{+,n}^*
 &=(\Im m_+(x+n\alpha))^{-1},\\
 T_{-,n}^*(\Im m_-)^{-1}T_{-,n}
 &=(\Im m_-(x+n\alpha))^{-1},\\
 (m_++m_-)^{-1}C_nA_n
 &=T_{+,n}^{-1}(m_+(x+n\alpha)+m_-(x+n\alpha))^{-1}\\
 &\quad\cdot(C_0A_0)(x+n\alpha)T_{+,n}.
\end{align*}
By trace cyclicity,
\begin{align*}
 &\tr(A_n^*A_n(\Im m_+)^{-1})+\tr(C_nC_n^*(\Im m_-)^{-1})
       +4\Im\tr((m_++m_-)^{-1}C_nA_n)\\
 &=D_0(E,x+n\alpha)=0
\end{align*}
for every $n\in\Z$. Take $\mathcal W=\{0,\ldots,p-1\}$, the observation
window for the full center chosen above. The second square in
Lemma~\ref{lem:matrix-defect} and invertibility of $(\Im m_-)^{-1/2}$ give
\[
 C_n^*(m_++m_-)^{-*}(m_+^*+m_-)=0\qquad(n\in\mathcal W).
\]
At real energy, $C_n^*=P_n\Phi_-^*$ and
$\Phi_-\Phi_-^*=I+m_-m_-^*\ge I$.
By finite observation,
\begin{align*}
 0&=\sum_{n\in\mathcal W}
 \bigl\|P_n\Phi_-^*(m_++m_-)^{-*}(m_+^*+m_-)\bigr\|_{\HS}^2\\
 &\ge c\bigl\|\Phi_-^*(m_++m_-)^{-*}(m_+^*+m_-)\bigr\|_{\HS}^2\\
 &\ge c\bigl\|(m_++m_-)^{-*}(m_+^*+m_-)\bigr\|_{\HS}^2.
\end{align*}
Since $\det(m_++m_-)\ne0$,
\[
 m_+^*+m_-=0.
\]
Exhausting $\operatorname{int}I'$ by a sequence of such intervals $I_0$ gives
\eqref{eq:boundary-matching}.
\end{proof}

\subsection{Analytic continuation and compact reduction}
Set
\[
 B=2^{-1/2}\begin{pmatrix}I&-iI\\I&iI\end{pmatrix}.
\]
Then $B^*J_0B=-iJ_{2m}$, where $J_0=\diag(I_m,-I_m)$.
By the Cayley definitions,
\begin{gather*}
 BF_+=2^{-1/2}\binom{W_+}{I_m}(m_++iI_m),\\
 B\binom{-m_-}{I_m}
 =-2^{-1/2}\binom{I_m}{W_-}(m_-+iI_m).
\end{gather*}
Thus the two images are the $J_0$-negative graph of $W_+$ and the
$J_0$-positive graph of $W_-$. By \eqref{eq:boundary-matching},
\[
 W_+(E+i0,x)=W_-(E+i0,x)^*.
\]
By $N^\natural J_{2m}N=J_{2m}$, the reflected $J_0$-orthogonal
complement, the negative graph of $W_-^\natural$, is invariant below
$\operatorname{int}I'$.

\begin{lemma}\label{lem:analytic-continuation}
For the fixed $E_0\in\operatorname{int}I'$, there exist an open
$\mathcal E_0\ni E_0$, $h>0$, and a periodic holomorphic
$W:\mathcal E_0\times S_h\to M_m(\C)$ such that
\[
 W(z,x)=\begin{cases}
 W_+(z,x),&\Im z>0,\\
 W_-^\natural(z,x),&\Im z<0.
 \end{cases}
\]
For every compact $K\subset(\mathcal E_0\cap\R)\times\T$, there exists
$c_K>0$ such that
\begin{equation}\label{eq:strict-interiority}
 I_m-W(E,x)^*W(E,x)\ge c_KI_m\qquad((E,x)\in K).
\end{equation}
\end{lemma}
\begin{proof}
\par\smallskip\noindent\emph{Step 1. Gluing across the real interval.}\par\nobreak
Choose a symmetric disk $D\Subset\mathcal E$ centered at $E_0$ with
$D\cap\R\subset\operatorname{int}I'$. By
\eqref{eq:boundary-matching} and Fubini, for a.e. $x$ the bounded
functions $W_+$ and $W_-^\natural$ have equal boundary values a.e. on
$D\cap\R$.
Define $W$ on the two half-disks as in the statement, and put
\[
 dA=d(\Re z)\,d(\Im z),\qquad
 \overline\partial=\tfrac12(\partial_{\Re z}+i\partial_{\Im z}).
\]
Fix a phase for which the boundary values agree. For
$\varphi\in C_c^\infty(D)$, Cauchy--Green on
$\{\Im z>t\}$ and $\{\Im z<-t\}$ gives
\begin{align*}
 \int_DW\overline\partial\varphi\,dA
 &=\frac1{2i}\lim_{t\downarrow0}\int_{D\cap\R}
 \bigl[W(E+it,x)\varphi(E+it)
       -W(E-it,x)\varphi(E-it)\bigr]\,dE\\
 &=0.
\end{align*}
The outer boundary terms vanish, and the integrands are bounded by
$2\|\varphi\|_\infty$. By dominated convergence and Weyl's lemma,
\[
 \overline\partial W=0,\qquad W\in\operatorname{Hol}(D),\qquad\|W\|\le1.
\]

\par\smallskip\noindent\emph{Step 2. Strict contraction at every phase.}\par\nobreak
Fix $z_*\in D\cap\mathbb H$, and choose a conformal map
$\chi:D\to\mathbb D$ with $\chi(z_*)=0$.
Continuity in phase gives $q=\sup_x\|W_+(z_*,x)\|<1$.
For $K\Subset D$, put $t=\sup_K|\chi|<1$.
By Schwarz--Pick applied to $u^*Wv$, $|u|=|v|=1$,
\[
 \sup_{z\in K}\|W(z,x)\|\le\frac{q+t}{1+qt}<1
\]
for a.e. $x$. For any phase $x$, choose $x_j\to x$ among the phases
covered by Step~1. Montel's theorem gives locally uniform subsequential
limits. Phase continuity identifies each with $W_+(\cdot,x)$ above and
$W_-^\natural(\cdot,x)$ below the real axis. Holomorphic uniqueness
therefore defines $W(\cdot,x)$ independently of the sequence and gives
\[
 \sup_{z\in K}\|W(z,x_j)-W(z,x)\|\longrightarrow0
 \qquad(K\Subset D).
\]
The same normal-family argument for arbitrary $x_j\to x$ gives
joint continuity on $D\times\T$. The bound persists at every phase.

\par\smallskip\noindent\emph{Step 3. A common holomorphic phase strip.}\par\nobreak
For each entry $f$ of $W$, put
\[
 a_k(z)=\int_0^1 f(z,x)e^{-2\pi ikx}\,dx,
 \qquad |a_k(z)|\le1.
\]
For every triangle $\Gamma\Subset D$, Fubini gives
\[
 \int_{\partial\Gamma}a_k(z)\,dz
 =\int_0^1e^{-2\pi ikx}
      \left(\int_{\partial\Gamma}f(z,x)\,dz\right)dx=0.
\]
Thus $a_k\in\operatorname{Hol}(D)$ by Morera's theorem.
By joint holomorphy above the real axis, there are a closed disk
$D_+\Subset D\cap\mathbb H$ and $h_1>0$, $C\ge1$, common to all entries,
such that
\[
 \sup_{D_+\times S_{h_1}}|f|\le C,
 \qquad\sup_{D_+}|a_k|\le Ce^{-2\pi h_1|k|}.
\]
Choose a closed disk $K\Subset D$ centered at $E_0$ and a smooth
simply connected $D_1\Subset D$ with $K\cup D_+\Subset D_1$. Let
\[
 \Delta u=0\text{ on }D_1\setminus D_+,\qquad
 u|_{\partial D_+}=1,\quad u|_{\partial D_1}=0,
 \qquad u=1\text{ on }D_+.
\]
By the maximum principle,
\[
 0<u\le1\text{ on }D_1,\qquad\delta:=\min_Ku>0.
\]
For $2\pi h_1|k|\ge\log C$, the subharmonic maximum principle gives
\begin{gather*}
 \log|a_k(z)|\le u(z)(\log C-2\pi h_1|k|)
 \qquad(z\in D_1\setminus D_+),\\
 \sup_K|a_k|\le Ce^{-2\pi\delta h_1|k|}.
\end{gather*}
The same bound holds on $D_+$. Enlarge $C$ for the finitely many
remaining $k$. For $0<h_2<\delta h_1$,
\[
 \sum_{k\in\Z}\sup_K|a_k|e^{2\pi h_2|k|}
 \le C\sum_{k\in\Z}e^{-2\pi(\delta h_1-h_2)|k|}<\infty.
\]
Hence
\[
 f(z,x)=\sum_{k\in\Z}a_k(z)e^{2\pi ikx}
\]
converges normally on $\operatorname{int}K\times S_{h_2}$.
Fourier uniqueness identifies the sum with $f$ on the real phase torus.
By holomorphic uniqueness, the graph of $W$ remains invariant under
$BNB^{-1}$. Take $\mathcal E_0=\operatorname{int}K$,
$h=h_2$, and use Step~2 for \eqref{eq:strict-interiority}.
\end{proof}

\begin{proof}[Proof of Theorem~\ref{thm:center-kotani}]
Restrict to the product domain $\mathcal E\times S_h$ of
Lemma~\ref{lem:analytic-continuation}. For real parameters, put
\[
 \mathcal P(W)=\begin{pmatrix}
 (I-WW^*)^{-1/2}&W(I-W^*W)^{-1/2}\\
 W^*(I-WW^*)^{-1/2}&(I-W^*W)^{-1/2}
 \end{pmatrix}.
\]
By $(WW^*)^nW=W(W^*W)^n$ and the binomial series,
\[
 (I-WW^*)^{-1/2}W=W(I-W^*W)^{-1/2}.
\]
Hence
\[
 \mathcal P(W)^*J_0\mathcal P(W)=J_0,\qquad
 \mathcal P(W)\binom0v=\binom{W}{I_m}(I-W^*W)^{-1/2}v.
\]
Therefore
\[
 \mathcal P(W^+)^{-1}BNB^{-1}\mathcal P(W)
\]
is $J_0$-unitary and preserves $\{0\}\times\C^m$.
Writing it as $\begin{psmallmatrix}a&0\\c&d\end{psmallmatrix}$ gives
\[
 d^*d=I,\qquad c^*d=0,\qquad a^*a-c^*c=I.
\]
Thus
\[
 d^{-1}=d^*,\qquad c=0,\qquad a^*a=d^*d=I.
\]

By \eqref{eq:strict-interiority}, a smaller product domain satisfies
\[
 \|WW^\natural\|<1,\qquad \|W^\natural W\|<1.
\]
Replace $W^*$ by $W^\natural$. The binomial series converges normally,
and holomorphic uniqueness gives
\[
 \mathcal P(W)^{-1}=J_0\mathcal P(W)^\natural J_0.
\]
Thus $\mathcal P(W)^{\pm1}$ is holomorphic. Put
\[
 C=V^{-1}Q_{\rm mid}B^{-1}\mathcal P(W)BV.
\]
Since
\[
 V^*J_{2m}V=\mathcal J_m,\quad
 Q_{\rm mid}^\natural J_{2m}Q_{\rm mid}=J_{2m},\quad
 B^*J_0B=-iJ_{2m},\quad
 BV=\begin{pmatrix}0&I_m\\I_m&0\end{pmatrix},
\]
we have
\begin{align*}
 C^*\mathcal J_mC
 &=i(BV)^*\mathcal P(W)^*J_0\mathcal P(W)BV=\mathcal J_m,\\
 (C^+)^{-1}MC
 &=(BV)^{-1}\mathcal P(W^+)^{-1}BNB^{-1}\mathcal P(W)BV
 \in\U(m)\times\U(m).
\end{align*}
Here $M$ is the original cocycle. The matrix $BV$ exchanges the two
signature blocks, and $C$ is periodic and holomorphic. Since $E_0\in I$
was arbitrary, the theorem follows.
\end{proof}

\subsection{The unitary representation and energy monotonicity}
\begin{proposition}\label{prop:compact-energy}
Let $J,\beta$ be as in Proposition~\ref{prop:actual-center}. For every
$E_0\in\operatorname{int}J$, there exist an open $\mathcal E\ni E_0$
with $\mathcal E\cap\R\subset\operatorname{int}J$, $h>0$,
$0<\kappa'\le K'<\infty$, and bounded periodic holomorphic maps
\[
 \widetilde Z:\mathcal E\times S_h\to\mathcal L(\C^{2m},X_\beta),
 \qquad U:\mathcal E\times S_h\to \mathrm{GL}(2m,\C)
\]
such that, for $(z,x)\in\mathcal E\times S_h$,
\[
 (L_{v,\alpha,x}-z)\widetilde Z_z(x)=0,\qquad
 \ker\widetilde Z_z(x)=\{0\},\qquad
 T\widetilde Z_z=\widetilde Z_z^+U_z.
\]
For $(E,x)\in(\mathcal E\cap\R)\times\T$,
\[
 \widetilde Z_E^*G_v\widetilde Z_E=\mathcal J_m,\qquad
 U_E=\diag(U_{+,E},U_{-,E})\in\U(m)\times\U(m),
\]
\begin{equation}\label{eq:compact-energy-signs}
 \begin{gathered}
 \kappa'I\le\frac1{2\pi i}U_{+,E}^*\partial_EU_{+,E}\le K'I,\\
 -K'I\le\frac1{2\pi i}U_{-,E}^*\partial_EU_{-,E}\le-\kappa'I,
 \end{gathered}
\end{equation}
\[
 L_i(\alpha,U_E(\cdot+iy))=0
 \quad(E\in\mathcal E\cap\R,\ |y|<h,\ 1\le i\le2m).
\]
\end{proposition}
\begin{proof}
\par\smallskip\noindent\emph{Step 1. The compact frame and its energy form.}\par\nobreak
Apply Theorem~\ref{thm:center-kotani} to the original frame $Z$
of Proposition~\ref{prop:actual-center}:
\[
 \widetilde Z_E=Z_EC_E,\qquad U_E=(C_E^+)^{-1}M_EC_E.
\]
By \eqref{eq:framegauge}, the solution and Green identities persist.
Shrink to a relatively compact product domain. Then $C_E^{\pm1}$
are bounded and the complexified exponents are unchanged.

For real $E,x$, put
\[
 H_E=\widetilde Z_E^*G_v\partial_E\widetilde Z_E,\qquad
 R_E=\widetilde Z_E(x;0)^*\widetilde Z_E(x;0),
\]
so that
\[
 \partial_E(\widetilde Z_E^*G_v\widetilde Z_E)=0
 \ \Longrightarrow\ H_E^*=H_E,\qquad R_E\ge0.
\]
Set
\[
 Q_E=U_E^*\mathcal J_m\partial_EU_E,\qquad
 \mathcal L_ES=U_E^*S^+U_E.
\]
By the differentiated Green identity,
\begin{align*}
 Q_E&=R_E+H_E-\mathcal L_EH_E,\\
 \sum_{\ell=0}^{N-1}\mathcal L_E^\ell Q_E
 &=\sum_{\ell=0}^{N-1}\mathcal L_E^\ell R_E
       +H_E-\mathcal L_E^NH_E.
\end{align*}
\par\smallskip\noindent\emph{Step 2. A uniformly positive finite average.}\par\nobreak
By covariance and Lemma~\ref{lem:finite-observation}, there are fixed
$\ell_0,c>0$ such that
\[
 \sum_{\ell=0}^{\ell_0-1}\mathcal L_E^\ell R_E\ge cI
\]
on a compact real neighborhood of $E_0$.
All energy suprema below are taken on this same neighborhood.
Since $\mathcal L_E$ preserves positivity, fixes $I$, and
$\|\mathcal L_E^NH_E\|_0=\|H_E\|_0$,
\[
 \begin{aligned}
 \sum_{\ell<N}\mathcal L_E^\ell R_E
 &\ge\sum_{j=0}^{\lfloor N/\ell_0\rfloor-1}
    \mathcal L_E^{j\ell_0}
    \left(\sum_{\ell=0}^{\ell_0-1}\mathcal L_E^\ell R_E\right)\\
 &\ge c\lfloor N/\ell_0\rfloor I.
 \end{aligned}
\]
Thus
\[
 \sum_{\ell<N}\mathcal L_E^\ell Q_E
 \ge\bigl(c\lfloor N/\ell_0\rfloor
      -2\|H_E\|_0\bigr)I.
\]
Choose an integer
\[
 N_0\ge\max\left\{2\ell_0,\frac{8\ell_0}{c}
 \sup_E\|H_E\|_0\right\}.
\]
Then
\[
 \begin{aligned}
 \frac1{N_0}\sum_{\ell<N_0}\mathcal L_E^\ell Q_E
 &\ge\left(\frac c{N_0}\left\lfloor\frac{N_0}{\ell_0}\right\rfloor
       -\frac{2\|H_E\|_0}{N_0}\right)I\\
 &\ge\left(\frac c{2\ell_0}-\frac c{4\ell_0}\right)I
 =\frac c{4\ell_0}I.
 \end{aligned}
\]

Put
\[
 S_E=\frac1{N_0}\sum_{\ell=0}^{N_0-2}(N_0-1-\ell)\mathcal L_E^\ell Q_E.
\]
Then
\[
 \mathcal L_ES_E-S_E
 =-\frac{N_0-1}{N_0}Q_E
   +\frac1{N_0}\sum_{\ell=1}^{N_0-1}\mathcal L_E^\ell Q_E,
\]
and hence
\[
 Q_E+\mathcal L_ES_E-S_E
 =\frac1{N_0}\sum_{\ell<N_0}\mathcal L_E^\ell Q_E.
\]
\par\smallskip\noindent\emph{Step 3. A change of frame depending on energy.}

\par\nobreak
Since $U_E\in\U(m)\times\U(m)$,
\[
 Q_E^*=Q_E,\quad S_E^*=S_E,\quad
 [Q_E,\mathcal J_m]=[S_E,\mathcal J_m]=0.
\]
Replacing adjoints by reflected adjoints extends $Q_E,S_E$
holomorphically. By~\cite[Section~4.1]{TeschlODE}, the initial value problem
\[
 \partial_EB_E=B_E\mathcal J_m^{-1}S_E,\qquad B_{E_0}=I,
\]
has a holomorphic solution on a sufficiently small disk
centered at $E_0$.

By analytic dependence and uniqueness, $B_E$ is periodic and holomorphic.
For real $E,x$,
\[
 (\mathcal J_m^{-1}S_E)^*=-\mathcal J_m^{-1}S_E,
\]
and
\[
 \partial_E[B_E,\mathcal J_m]
 =[B_E,\mathcal J_m]\mathcal J_m^{-1}S_E,\qquad
 [B_{E_0},\mathcal J_m]=0.
\]
Uniqueness gives $[B_E,\mathcal J_m]=0$. Also,
\[
 \partial_E(B_EB_E^*)
 =B_E\bigl(\mathcal J_m^{-1}S_E+(\mathcal J_m^{-1}S_E)^*\bigr)B_E^*=0.
\]
Since $B_{E_0}=I$,
\[
 B_E^*B_E=I,\qquad B_E^*\mathcal J_mB_E=\mathcal J_m,
\]
and the differential equation gives
\[
 B_E^*\mathcal J_m\partial_EB_E
 =B_E^*\mathcal J_mB_E\mathcal J_m^{-1}S_E=S_E
 \quad(E,x\text{ real}).
\]

Set $\widehat U_E=B_E^+U_EB_E^{-1}$. Since
\[
 \partial_E\widehat U_E
 =(\partial_EB_E)^+U_EB_E^{-1}
  +B_E^+(\partial_EU_E)B_E^{-1}
  -\widehat U_E(\partial_EB_E)B_E^{-1},
\]
we obtain
\begin{align*}
 \widehat U_E^*\mathcal J_m\partial_E\widehat U_E
 &=B_E^{-*}(Q_E+\mathcal L_ES_E-S_E)B_E^{-1}\\
 &=B_E^{-*}\left(\frac1{N_0}\sum_{\ell<N_0}\mathcal L_E^\ell Q_E\right)B_E^{-1}
 \ge\frac c{4\ell_0}I.
\end{align*}
The paired frame is $\widetilde Z_EB_E^{-1}$. Since
$\mathcal J_m=\diag(-iI_m,iI_m)$,
\[
 -i\widehat U_{+,E}^*\partial_E\widehat U_{+,E}\ge\frac c{4\ell_0}I,
 \qquad
 i\widehat U_{-,E}^*\partial_E\widehat U_{-,E}\ge\frac c{4\ell_0}I.
\]
Take $\kappa'=c/(8\pi\ell_0)$. Shrink to a relatively compact product
domain whose real-energy slice lies in the chosen compact neighborhood.
Bounded derivatives give $K'<\infty$; bounded $B_E^{\pm1}$ preserve
the zero complexified exponents on the final common strip.
Relabel $(\widetilde Z_EB_E^{-1},\widehat U_E)$ as
$(\widetilde Z_E,U_E)$.
\end{proof}

\section{Preparation of the compact center}\label{sec:global-to-local}
\begin{proposition}\label{prop:center-preparation}
Let $U_E=U_{+,E}\oplus U_{-,E}$ be as in
Proposition~\ref{prop:compact-energy}. For every $E_c\in\mathcal E\cap\R$,
there is $h_{\rm ARC}>0$ such that
\[
 \inf_{\substack{B\in C^\omega_{h_{\rm ARC}}(\T,\U(m)\times\U(m))\\
                  C\in\U(m)\times\U(m)}}
 \|B(\cdot+\alpha)U_{E_c}B^{-1}-C\|_{h_{\rm ARC}}=0.
\]
\end{proposition}

By the commuting returns of \cite[Section~4]{AK} and
\cite[Section~5.2]{HPZ}, we first obtain a cocycle close to a constant on a prescribed
strip. Apply the fixed-energy part of Lemma~\ref{lem:parameter-step},
whose proof is independent of this section, and then pull back by finite
inverse words. Throughout, let
$A\in C_h^\omega(\T,\U(m))$ satisfy
\begin{equation}\label{eq:subcritical}
 L_1(\alpha,A(\cdot+iy))=0\qquad(|y|<h).
\end{equation}

\subsection{Long products}
\begin{lemma}\label{lem:slow-variation}
For $0<r<h$, put $a_N(r)=\log\|A_N\|_r$. Then $a_N(r)=o(N)$.
If $N_j\to\infty$, $\sup_jN_j|t_j|<\infty$, and $K\subset\C$ is bounded,
then
\begin{equation}\label{eq:rescaled-products}
 \sup_{x\in\R,\,z\in K}
 \|A_{N_j}(x+t_jz)-A_{N_j}(x)\|\longrightarrow0.
\end{equation}
\end{lemma}
\begin{proof}
By 
uniform subadditivity, on $\Im z=\pm r$,
\eqref{eq:subcritical}, the maximum principle and real unitarity,
\[
 0\le\frac{a_N(r)}N
 \le\max_{\sigma=\pm1}\frac1N\sup_{x\in\R}
       \log\|A_N(x+i\sigma r)\|\longrightarrow0.
\]
Set $\delta_N=r/(1+a_N(r))$. Since $\|A_N\|_0=1$, the three-lines
theorem and Cauchy's estimate give
\[
 \|A_N\|_{\delta_N}
 \le\|A_N\|_r^{\delta_N/r}
 =e^{a_N(r)/(1+a_N(r))}\le e,\qquad
 \|\partial_xA_N\|_{\delta_N/2}\le\frac{2e}{\delta_N}.
\]
Moreover,
\[
 \frac{|t_j|}{\delta_{N_j}}
 =\frac{N_j|t_j|}{r}
   \left(\frac1{N_j}+\frac{a_{N_j}(r)}{N_j}\right)\longrightarrow0.
\]
For large $j$, $|t_j|\sup_K|z|<\delta_{N_j}/2$. Integrating along the segment
from $x$ to $x+t_jz$ in $S_{\delta_{N_j}/2}$ gives
\[
 \sup_{x\in\R,\,z\in K}
 \|A_{N_j}(x+t_jz)-A_{N_j}(x)\|
 \le2e\sup_K|z|\frac{|t_j|}{\delta_{N_j}}\longrightarrow0.
 \qedhere
\]
\end{proof}

\subsection{Equivariant normalization}
We regard $(\beta,F)$ as the map $(z,v)\mapsto(z+\beta,F(z)v)$.
With this convention,
\begin{equation}\label{eq:pair-product}
 (\beta,F)(\gamma,G)
 =(\beta+\gamma,F(\cdot+\gamma)G).
\end{equation}
When a matrix function is not periodic, we use the rectangle norm
$\|F\|_{r,J}=\sup_{z\in S_r,\,\Re z\in J}\|F(z)\|$.

\begin{lemma}\label{lem:unitary-logarithm}
For $U,V\in\U(m)$, there is $H\in\mathfrak u(m)$ such that
\[
 e^H=U,\qquad \|H\|\le2\pi,\qquad
 \|[H,V]\|\le C_m\|[U,V]\|.
\]
\end{lemma}
The proof is in Appendix~\ref{app:unitary-logarithm}.

\begin{lemma}\label{lem:equivariant-normalization}
Let $U:S_r\to\mathrm{GL}(m,\C)$ be holomorphic and unitary on $\R$.
If $\|U-I\|_{r,[-1,2]}\le\epsilon<c(m,r)$, there is a holomorphic
map $Q:S_r\to\mathrm{GL}(m,\C)$, unitary on $\R$, such that
\[
 Q(z+1)U(z)=Q(z),
\]
\[
 \|Q-I\|_{r,[-1,2]}+\|Q^{-1}-I\|_{r,[-1,2]}
 \le C_{m,r}\epsilon.
\]
The construction preserves orthogonal blocks.
\end{lemma}
\begin{proof}
\emph{Step 1. A smooth equivariant frame.}
Choose
$\chi\in C^\infty([0,1],[0,1])$, equal to zero near $0$ and one near
$1$. The principal logarithm is defined on the required rectangle, so
on $0\le\Re z\le1$ we may put
\[
 G(z)=e^{-\chi(\Re z)\log U(z-1)},\qquad
 \|\log U(z-1)\|\le\frac{\epsilon}{1-\epsilon}.
\]
Extend it to $S_r$ by
\begin{equation}\label{eq:smooth-normalizing-frame}
 G(z+1)=G(z)U(z)^{-1}.
\end{equation}
Since $\chi$ is constant near the endpoints, the extension is smooth and
\[
 \|G^{\pm1}-I\|_{r,[-1,2]}\le C\epsilon.
\]
Define $a=(\bar\partial G)G^{-1}$, where
$\bar\partial=(\partial_x+i\partial_y)/2$. On the initial cell,
holomorphy of the logarithm gives
\[
 a(z)=-\frac12\chi'(\Re z)\log U(z-1).
\]
By \eqref{eq:smooth-normalizing-frame} and holomorphy of $U$,
\[
 a(z+1)=a(z),\qquad \|a\|_r\le C\epsilon.
\]

\emph{Step 2. The holomorphic correction.}
For bounded periodic $f$, set
\[
 Tf(z)=\int_0^1\int_{-r}^r
          \cot\pi(z-s-it)f(s+it)\,dt\,ds.
\]
The kernel is periodic and satisfies, in distributions,
\[
 \bar\partial\cot\pi z=\sum_{k\in\Z}\delta_k.
\]
For $|u|\le1/2$,
\[
 |\cot\pi(u+iv)|\le C\left(1+
 \frac{\mathbf 1_{\{|v|\le1\}}}{\sqrt{u^2+v^2}}\right).
\]
By integration over one horizontal period,
\[
 Tf(z+1)=Tf(z),\qquad \bar\partial Tf=f,\qquad
 \|Tf\|_r\le C(1+r)\|f\|_r.
\]
In particular, $T1(z)=-2i\Im z$ and $\bar\partial T1=1$.
For sufficiently small $\epsilon$, $C(1+r)\|a\|_r\le1/2$. Then
\[
 \|T((R_1-R_2)a)\|_r
 \le C(1+r)\|a\|_r\|R_1-R_2\|_r
 \le\tfrac12\|R_1-R_2\|_r.
\]
Thus $R\mapsto I-T(Ra)$ has a unique bounded periodic fixed point,
with
\[
 R=I-T(Ra),\qquad
 \bar\partial R=-Ra,\qquad
 \|R^{\pm1}-I\|_r\le C_r\|a\|_r.
\]
Thus
\[
 \bar\partial(RG)=(-Ra)G+R(aG)=0.
\]
By Weyl's lemma, $Q_0=RG$ is holomorphic. Periodicity of $R$ gives
\[
 Q_0(z+1)U(z)=Q_0(z),\qquad
 \|Q_0^{\pm1}-I\|_{r,[-1,2]}\le C_{m,r}\epsilon.
\]
\emph{Step 3. Unitary normalization.}
Since $U^\natural=U^{-1}$ and $Q_0$ is equivariant,
$H=Q_0Q_0^\natural$ satisfies
\begin{gather*}
 Q_0^\natural(z+1)=U(z)Q_0^\natural(z),\\
 H(z+1)=Q_0(z)U(z)^{-1}U(z)Q_0^\natural(z)=H(z),\\
 H^\natural=H,\qquad \|H-I\|_r\le C_{m,r}\epsilon<\frac12.
\end{gather*}
By the binomial series, $H^{-1/2}$ is periodic and holomorphic, with
\[
 (H^{-1/2})^\natural=H^{-1/2},\qquad
 \|H^{-1/2}-I\|_r\le C_{m,r}\epsilon.
\]
Set $Q=H^{-1/2}Q_0$. Then
\[
 QQ^\natural=H^{-1/2}HH^{-1/2}=I,\qquad
 Q(z+1)U(z)=Q(z).
\]
The bounds for $Q_0^{\pm1}$ and $H^{\pm1/2}$ give those for $Q^{\pm1}$.
Every operation preserves the prescribed blocks.
\end{proof}

\begin{lemma}\label{lem:normalize-pair}
Let $\eta\in[-1,1]$. Suppose $U,V:S_r\to\mathrm{GL}(m,\C)$ are
holomorphic, unitary on $\R$, and satisfy
\[
 U(z+\eta)V(z)=V(z+1)U(z),\qquad
 \sup_{[-2,3]+i(-r,r)}(\|U-U_0\|+\|V-V_0\|)\le\epsilon,
\]
where $U_0=U(0)$ and $V_0=V(0)$. Choose $H$ by
Lemma~\ref{lem:unitary-logarithm} applied to $U_0,V_0$, and put
$Q_0(z)=e^{-zH}$. Then
\begin{gather*}
 Q_0(z+1)U(z)Q_0(z)^{-1}=I+E_1(z),\\
 Q_0(z+\eta)V(z)Q_0(z)^{-1}=e^{-\eta H}V_0+E_2(z),
\end{gather*}
with $\|E_1\|_{r,[-1,2]}+\|E_2\|_{r,[-1,2]}\le C_{m,r}\epsilon$.
\end{lemma}
\begin{proof}
At $z=0$, the commutation relation gives
\[
 [U_0,V_0]=(U_0-U(\eta))V_0+(V(1)-V_0)U_0,
\]
so Lemma~\ref{lem:unitary-logarithm} yields
\[
 \|[U_0,V_0]\|\le2\epsilon,\qquad
 \|[H,V_0]\|\le C_m\epsilon.
\]
Since $e^H=U_0$, the first constant part cancels. Also,
\[
 e^{-zH}V_0e^{zH}-V_0
 =-z\int_0^1e^{-tzH}[H,V_0]e^{tzH}\,dt.
\]
Hence
\begin{align*}
 E_1(z)&=e^{-(z+1)H}(U(z)-U_0)e^{zH},\\
 E_2(z)&=e^{-\eta H}e^{-zH}(V(z)-V_0)e^{zH}\\
 &\quad-z e^{-\eta H}\int_0^1
          e^{-tzH}[H,V_0]e^{tzH}\,dt.
\end{align*}
For real $t$, $H\in\mathfrak u(m)$ and $\|H\|\le2\pi$ give
$\|e^{tzH}\|\le e^{2\pi|t\Im z|}$. Also, $|z|\le2+r$ on the
stated rectangle and $e^{-\eta H}$ is unitary. Therefore
\[
 \|E_1\|_{r,[-1,2]}+\|E_2\|_{r,[-1,2]}
 \le e^{4\pi r}\bigl(2\epsilon+(2+r)\|[H,V_0]\|\bigr)
 \le C_{m,r}\epsilon.
\]
\end{proof}

\subsection{Continued-fraction renormalization}
Let $p_n/q_n$ be the convergents of $\alpha$, and put
\[
 s_n=q_{n-1}\alpha-p_{n-1},\qquad
 \eta_n=\frac{q_n\alpha-p_n}{s_n}.
\]
Then $|s_n|q_n<1$ and $\eta_n\in(-1,0)\setminus\Q$.

\begin{proposition}\label{prop:renormalization}
For every $R>0$ and all sufficiently large $n$, there are a holomorphic
map $W_n$ on $S_{2R}$, unitary on $\R$, and $C_n\in\U(m)$ such that
\begin{align}
 W_n(z+1)A_{q_{n-1}}(s_nz)W_n(z)^{-1}
 &=I,\label{eq:normalize-first}\\
 W_n(z+\eta_n)A_{q_n}(s_nz)W_n(z)^{-1}
 &=\widetilde A_n(z),\label{eq:renormalized-cocycle}
\end{align}
where $\widetilde A_n\in C_R^\omega(\T,\U(m))$ and
$\|\widetilde A_n-C_n\|_R\to0$.
\end{proposition}
\begin{proof}
Fix $R>0$. Since $F_1=(1,I)$ and $F_2=(\alpha,A)$ commute, so do
\[
 G_1=F_1^{-p_{n-1}}F_2^{q_{n-1}},\qquad
 G_2=F_1^{-p_n}F_2^{q_n}.
\]
Under $x=s_nz$, the pair becomes $(1,U_n)$, $(\eta_n,V_n)$, where
\[
 U_n(z)=A_{q_{n-1}}(s_nz),\qquad V_n(z)=A_{q_n}(s_nz).
\]
For large $n$, these functions are holomorphic on $S_{2R}$ because
$2R|s_n|<h$. Since $q_{n-1}|s_n|$ and $q_n|s_n|$ are bounded,
Lemma~\ref{lem:slow-variation} gives
\[
 \sup_{z\in[-3,4]+i[-2R,2R]}
 \bigl(\|A_{q_{n-1}}(s_nz)-A_{q_{n-1}}(0)\|
       +\|A_{q_n}(s_nz)-A_{q_n}(0)\|\bigr)\longrightarrow0.
\]
By Lemma~\ref{lem:normalize-pair}, the frame change
$e^{-zH_n}$ changes the two matrices to $I+E_{1,n}$ and
$C_n+E_{2,n}$, respectively, where
$C_n=e^{-\eta_nH_n}A_{q_n}(0)\in\U(m)$ and the rectangle errors tend
to zero on width $2R$. By Lemma~\ref{lem:equivariant-normalization},
there is a holomorphic $Q_n$ on $S_{2R}$ with
\begin{gather*}
 Q_n(z+1)(I+E_{1,n}(z))Q_n(z)^{-1}=I,\\
 \|Q_n^{\pm1}-I\|_{2R,[-1,2]}
 +\|E_{1,n}\|_{2R,[-1,2]}+\|E_{2,n}\|_{2R,[-1,2]}\longrightarrow0.
\end{gather*}
Set $W_n(z)=Q_n(z)e^{-zH_n}$. Then \eqref{eq:normalize-first} holds and
\[
 \widetilde A_n(z)
 =Q_n(z+\eta_n)(C_n+E_{2,n}(z))Q_n(z)^{-1}.
\]
Commutation with $(1,I)$ gives $\widetilde A_n(z+1)=\widetilde A_n(z)$.
The changes of frame, and hence $\widetilde A_n$, are unitary on the real axis.
For large $n$, $\|Q_n^{\pm1}\|_{2R,[-1,2]}\le2$. Since
$[0,1]+\eta_n\subset[-1,1]$, the formula for $\widetilde A_n$ gives
\begin{align*}
 \|\widetilde A_n-C_n\|_R
 &=\|\widetilde A_n-C_n\|_{R,[0,1]}\\
 &\le2\|Q_n-I\|_{2R,[-1,2]}
     +\|Q_n^{-1}-I\|_{2R,[-1,2]}
     +4\|E_{2,n}\|_{2R,[-1,2]}\longrightarrow0.
\end{align*}
If $A$ preserves prescribed orthogonal blocks, so does the construction.
\end{proof}

\subsection{Pullback}
\begin{proposition}\label{prop:pullback}
Suppose that, at one fixed return level $n$ and some $h>0$,
\[
 \inf_{B\in C_h^\omega(\T,\U(m)),\,C\in\U(m)}
 \|B(\cdot+\eta_n)\widetilde A_nB^{-1}-C\|_h=0.
\]
Then $(\alpha,A)$ is almost reducible on a fixed strip of positive width.
\end{proposition}
\begin{proof}
\emph{Step 1. Recover the original generators.}
Fix the return level $n$. Choose $0<r<h$ inside the holomorphy strip
of $W_n^{\pm1}$, and put $h_0=|s_n|r>0$. This width is fixed as
$j$ varies. Choose a sequence
\[
 B_j(\cdot+\eta_n)\widetilde A_nB_j^{-1}=D_j+R_j,\qquad
 D_j\in\U(m),\qquad \epsilon_j=\|R_j\|_r\longrightarrow0.
\]
Since $B_j$ is periodic, the transformed generators are $(1,I)$ and
$(\eta_n,D_j+R_j)$. Inverting the continued-fraction matrix gives
\begin{gather*}
 \begin{pmatrix}a&b\\c&d\end{pmatrix}
 =\begin{pmatrix}-p_{n-1}&q_{n-1}\\-p_n&q_n\end{pmatrix}^{-1}
 \in\mathrm{GL}(2,\Z),\\
 F_1=G_1^aG_2^b,\qquad F_2=G_1^cG_2^d,\\
 1=s_n(a+b\eta_n),\qquad \alpha=s_n(c+d\eta_n).
\end{gather*}
For $\epsilon_j<1/2$, the Neumann series gives
\[
 (D_j+R_j)^{-1}-D_j^{-1}
 =-(I+D_j^{-1}R_j)^{-1}D_j^{-1}R_jD_j^{-1},
\]
\[
 \|(D_j+R_j)^{\pm1}\|_r\le2,\qquad
 \|(D_j+R_j)^{-1}-D_j^{-1}\|_r\le2\epsilon_j.
\]
The word lengths are at most $L=L(n)$, independent of $j$. All their
shifts are real, so they preserve $S_r$. For the translated factors
$U_\nu$ and $\sigma_\nu\in\{-1,0,1\}$,
\[
 \|U_\nu\|_r\le2,\qquad
 \|U_\nu-D_j^{\sigma_\nu}\|_r\le2\epsilon_j.
\]
For $\ell\le L$, telescoping gives
\begin{align*}
 U_\ell\cdots U_1-D_j^{\sum_{\nu=1}^{\ell}\sigma_\nu}
 &=\sum_{\nu=1}^{\ell}U_\ell\cdots U_{\nu+1}
 (U_\nu-D_j^{\sigma_\nu})
 D_j^{\sum_{s<\nu}\sigma_s},\\
 \|U_\ell\cdots U_1-D_j^{\sum_{\nu=1}^{\ell}\sigma_\nu}\|_r
 &\le2\epsilon_j\sum_{\nu=1}^{\ell}2^{\ell-\nu}
 \le L2^L\epsilon_j.
\end{align*}
Thus, with
$T_j(x)=B_j(x/s_n)W_n(x/s_n)$, the recovered pair satisfies
\begin{gather*}
 T_j(x+1)T_j(x)^{-1}=D_j^b+E_{1,j}(x/s_n),\\
 T_j(x+\alpha)A(x)T_j(x)^{-1}=D_j^d+E_{2,j}(x/s_n),\\
 \|E_{1,j}\|_r+\|E_{2,j}\|_r\le C_{n,r}\epsilon_j.
\end{gather*}
\emph{Step 2. Make the change of frame periodic.}
In an eigenbasis of $D_j$, choose
\[
 H_j\in\mathfrak u(m),\qquad e^{H_j}=D_j^b,\qquad
 \|H_j\|\le\pi,\qquad [H_j,D_j]=0.
\]
Then
\begin{gather*}
 e^{-(x+1)H_j}D_j^be^{xH_j}=I,\qquad
 e^{-(x+\alpha)H_j}D_j^de^{xH_j}=e^{-\alpha H_j}D_j^d.
\end{gather*}
For $x\in S_{h_0}$, we have $x/s_n\in S_r$ and
\[
 \sup_{x\in S_{h_0}}
 \left\|e^{-(x+\beta)H_j}E_{i,j}(x/s_n)e^{xH_j}\right\|
 \le C_{n,r}e^{2\pi h_0}\epsilon_j
 \qquad(\beta\in\R,\ i=1,2).
\]
For large $j$, Lemma~\ref{lem:equivariant-normalization} normalizes the
first generator exactly by $Q_j$, with
\[
 \|Q_j^{\pm1}-I\|_{h_0,[-1,2]}\le C_{n,h_0}\epsilon_j.
\]
Consequently,
\[
 \widehat B_j=Q_je^{-(\cdot)H_j}T_j,\qquad
 \widehat B_j(x+1)=\widehat B_j(x).
\]
Since $x+\alpha\in[0,2]$ for $x\in[0,1]$, the bounds for
$Q_j^{\pm1}$ give
\[
 \|\widehat B_j(\cdot+\alpha)A\widehat B_j^{-1}
       -e^{-\alpha H_j}D_j^d\|_{h_0,[0,1]}
 \le C_{n,h_0}\epsilon_j\longrightarrow0.
\]
By periodicity, the estimate holds on $S_{h_0}$. For each $j$, the bounds
for $B_j^{\pm1}$ and $Q_j^{\pm1}$, and the compact-cell bounds for
$W_n^{\pm1}$, give
\[
 \sup_{\substack{x\in S_{h_0}\\0\le\Re x\le1}}
 \|\widehat B_j(x)^{\pm1}\|<\infty.
\]
Exact periodicity extends this bound to $S_{h_0}$. The change of frame is holomorphic
there and unitary on $\R$. The constants in the error estimate may depend
on $n,r$, but not on $j$.
\end{proof}

\begin{proof}[Proof of Proposition~\ref{prop:center-preparation}]
Fix $E_c\in\mathcal E\cap\R$. By Proposition~\ref{prop:compact-energy},
the complexified exponents of $U_{E_c}$ vanish on a strip. Both blocks
therefore satisfy \eqref{eq:subcritical}.

\emph{Step 1. Fix the tolerance before the return.}
Set
\[
 a=\frac1{128},\qquad
 h_j=\frac1{48}+\frac{2^{-j}}{48},\qquad
 d_j=\frac{2^{-j}}{96},\qquad K=4096(2m+1)^2.
\]
Then $h_0=1/24$, $0<d_j\le h_j/2$, and
$h_j-d_j=h_{j+1}$. Let $c_*$ be the fixed-energy threshold in
Lemma~\ref{lem:parameter-step} for dimension $2m$, and choose
\[
 0<\epsilon_0\le\min\{c_*(a/96)^K,\,2^{-K}\}.
\]
Then
\[
 \epsilon_0^{65^j}\le\epsilon_0^{j+1}
 \le c_*(a/96)^K2^{-Kj}=c_*(ad_j)^K
 \qquad(j\ge0).
\]

\emph{Step 2. Reduce one fixed return.}
The constants in Proposition~\ref{prop:embedding} and the fixed-energy
part of Lemma~\ref{lem:parameter-step} are frequency-independent.
By Proposition~\ref{prop:renormalization}, fix a common return $n$ for
both blocks on $S_{1/2}$ so that the suspension below has Fourier
error at most $\epsilon_0$.
By periodicity, the shifts $\eta_n$ and $1+\eta_n$ agree on $\T$.

Write $\widetilde A_n$ and $C_n$ for the two direct sums.
Choose a constant block-unitary $V_0$ diagonalizing $C_n$.
By Proposition~\ref{prop:embedding}, embed
$V_0\widetilde A_nV_0^{-1}$ on width $1/12$; restricting to width
$1/24$ and using \eqref{eq:fouriernorm} gives
\begin{gather*}
 \dot v=2\pi i(\Lambda_0+f_0)v,\qquad
 \dot\theta=\omega_n:=(1+\eta_n,1),\\
 \Lambda_0=\Lambda_0^*\text{ diagonal},\qquad
 |f_0|_{1/24}\le\epsilon_0.
\end{gather*}
Apply the fixed-energy step in dimension $2m$ within the two blocks.
With $\mathcal P_0=I$ and $\mathcal P_{j+1}=P_j\mathcal P_j$, induction gives
\begin{gather*}
 |f_j|_{h_j}\le\epsilon_j\le\epsilon_0^{65^j},\\
 \frac{(\partial_{\omega_n}\mathcal P_j)\mathcal P_j^{-1}}{2\pi i}
 +\mathcal P_j(\Lambda_0+f_0)\mathcal P_j^{-1}
 =\Lambda_j+f_j.
\end{gather*}
Since $h_j\downarrow1/48$, 
all changes of frame are defined on $S_{1/48}^2$.
Set $B_j(x)=\mathcal P_j(x,0)V_0$.
Lemma~\ref{lem:time-one}, applied to the remainder $2\pi i f_j$, gives
\[
 \|B_j(\cdot+\eta_n)\widetilde A_nB_j^{-1}
       -e^{2\pi i\Lambda_j}\|_{1/48}
 \le e^{2\pi\epsilon_j}-1\longrightarrow0.
\]
\emph{Step 3. Pull back.}
The return level $n$ has remained fixed throughout the iteration.
Apply Proposition~\ref{prop:pullback} to both blocks and take a common
positive width $h_{\rm ARC}$.
\end{proof}

\section{Local reduction for the scalar obstruction}\label{sec:local-marked}
\begin{proposition}\label{prop:marked-resonance}
Let $I\subset\R$ be open. Consider a family
$A_E=A_{+,E}\oplus A_{-,E}$ jointly analytic near $I\times\T$,
where $A_{\pm,E}(x)\in\U(m_\pm)$ for real $E,x$ and $m_\pm\ge1$.
Assume
\begin{equation}\label{eq:discrete-monotonicity}
 \kappa I\le\frac{A_{+,E}^{-1}\partial_EA_{+,E}}{2\pi i}\le K_pI,
 \qquad-K_pI\le\frac{A_{-,E}^{-1}\partial_EA_{-,E}}{2\pi i}\le-\kappa I
\end{equation}
for some $0<\kappa\le K_p<\infty$. Assume also that, for some $E_c\in I$
and $h>0$,
\[
 \inf_{\substack{B\in\coc h(\T,\U(m_+)\times\U(m_-))\\
 C\in\U(m_+)\times\U(m_-)}}
 \|B(\cdot+\alpha)A_{E_c}B^{-1}-C\|_h=0.
\]
Then, for every $\mu>0$, there are $E_*\in I$, $k\in\Z$,
constants $C,C_{\rm fixed}>0$ and real-analytic maps
$B_j:\T\to\U(m_+)\times\U(m_-)$, $j\ge0$, such that
\begin{equation}\label{eq:marked-smallness}
 \begin{gathered}
 B_j^+A_{E_*}B_j^{-1}=C_j+F_j,\qquad C_j\in\U(m_+)\times\U(m_-),\\
 S_j\longrightarrow\infty,\qquad \|F_j\|_0\le Ce^{-S_j},\\
 \|B_j^{\prime}\|_0+\|(B_j^{-1})^{\prime}\|_0
 \le C_{\rm fixed}+\mu S_j.
 \end{gathered}
\end{equation}
Writing $C_j=C_{+,j}\oplus C_{-,j}$, one can choose marked phases satisfying
\begin{equation}\label{eq:marked-gap}
 e^{2\pi i\theta_{\pm,j}}\in\operatorname{spec}C_{\pm,j},\qquad
 \theta_{+,j}-\theta_{-,j}=k\alpha\pmod{\mathbb Z}.
\end{equation}
The energy $E_*$, the integer $k$, and both constants are independent of $j$.
\end{proposition}

Throughout the section,
\[
 \omega=(\alpha,1),\qquad \theta=(x,t),\qquad
 \partial_\omega=\alpha\partial_x+\partial_t,
\]
and $|\cdot|_h$ denotes the Hilbert--Schmidt Fourier norm of
Section~\ref{subsec:analytic-notation}.

\subsection{Suspension and energy monotonicity}\label{subsec:embedding}\label{subsec:monotone-suspension}
We use \cite[Theorem~1.1 and Remarks~1.1,~1.3]{YZ} in the following form.
\begin{proposition}\label{prop:embedding}
For every $h>0$, there are constants $\epsilon_{m,h},C_{m,h}>0$ such
that the following holds. Given $\eta\in[-1,1]$, $C\in\U(m)$,
$A\in\coc h(\T,\U(m))$ with $\|A-C\|_h<\epsilon_{m,h}$, one can find
$D\in\uu(m)$ and $F\in\coc{h/6}(\T^2,\uu(m))$ satisfying
\begin{gather*}
 e^D=C,\qquad\|D\|\le\pi,\qquad
 \|F\|_{h/6}\le C_{m,h}\|A-C\|_h,\\
 \dot v=(D+F(x,t))v,\qquad\dot x=\eta,\qquad\dot t=1,
 \qquad\Phi(0,x)=I,\quad\Phi(1,x)=A(x).
\end{gather*}
If $A_E$ is a strip-holomorphic family near $E_c\in\R$, unitary for real
$E$ and satisfying $\|A_{E_c}-C\|_h<\epsilon_{m,h}$, the matrix $D$ can
be kept fixed while $F_E$ varies jointly holomorphically on a smaller energy
neighborhood. The construction can be carried out within each prescribed
orthogonal block.
\end{proposition}
\begin{proof}
\emph{Step 1. Lift the Fourier modes.}
By \cite[Sections~3.1--3.3]{YZ}, diagonalize within the prescribed blocks.
Choose $C=e^D$, $D=i\diag(\delta_a)$, $|\delta_a|\le\pi$, and set
\[
 r=h/2,\qquad G=\log(C^{-1}A),\qquad
 |G|_r\le C_{m,h}\|A-C\|_h.
\]
Choose paired integers $j_{ba}(-k)=-j_{ab}(k)$ so that
\[
 z_{ab}(k):=j_{ab}(k)+k\eta+\frac{\delta_b-\delta_a}{2\pi}
 \in[-1/2,1/2],\qquad |j_{ab}(k)|\le|k|+3/2.
\]
At endpoints, choose opposite signs in each pair; set $j_{aa}(0)=0$. For
$f(x)=\sum_{k,a,b}f_{ab}(k)e^{2\pi ikx}E_{ab}$, define its lift and its
interaction representation by
\[
 F_f(x,t)=\sum_{k,a,b}f_{ab}(k)e^{2\pi i(kx+j_{ab}(k)t)}E_{ab},\qquad
 F_{f,s}=e^{-sD}F_f(x+s\eta,s)e^{sD}.
\]
For real $s$, every entry is multiplied by a phase, so
$|F_{f,s}|_r=|f|_r$.

\emph{Step 2. Solve the endpoint equation.}
Let $\Phi_f$ be the normalized fundamental matrix for $D+F_f$.
Put $X_f(s)=e^{-sD}\Phi_f(s)$ and
$\mathcal L_Df=\int_0^1F_{f,s}\,ds$. Then
\begin{gather*}
 X_f'=F_{f,s}X_f,\qquad X_f(0)=I,\\
 X_f(1)-I-\mathcal L_Df
 =\sum_{n\ge2}\int_{0<s_n<\cdots<s_1<1}
 F_{f,s_1}\cdots F_{f,s_n}\,ds_n\cdots ds_1,\\
 |X_f(1)-I-\mathcal L_Df|_r
 \le e^{|f|_r}-1-|f|_r\le C|f|_r^2.
\end{gather*}
Expanding $\log X_f(1)$ gives
\begin{gather*}
 \log X_f(1)=\mathcal L_Df+\mathcal N_D(f),\\
 \widehat{(\mathcal L_Df)}_{ab}(k)=I(z_{ab}(k))f_{ab}(k),\qquad
 I(z)=\int_0^1e^{2\pi izs}\,ds
     =e^{\pi iz}\frac{\sin\pi z}{\pi z},\quad I(0)=1,\\
 \|\mathcal L_D^{-1}\|\le\pi/2,\qquad
 |\mathcal N_D(f)|_r\le C|f|_r^2,\qquad
 \operatorname{Lip}_{|f|_r\le R}\mathcal N_D\le CR.
\end{gather*}
Since $|I(z)|\ge2/\pi$ on $[-1/2,1/2]$, these constants are uniform
in $D,\eta$. For $|G|_r$ small,
$f\mapsto\mathcal L_D^{-1}(G-\mathcal N_D(f))$ maps the ball
$|f|_r\le\pi|G|_r$ into itself and is a contraction. Its fixed point
satisfies
\[
 \Phi_f(1)=e^De^G=A,\qquad |f|_r\le\pi|G|_r,\qquad
 \|F_f\|_{h/4}\le C_me^{3\pi h/4}|f|_r
 \le C_{m,h}\|A-C\|_h.
\]
Since $j_{ba}(-k)=-j_{ab}(k)$ and $I(-z)=\overline{I(z)}$, the
contraction preserves skew-Hermitian reality and blocks. Restrict to
width $h/6$. For $A_E$, fix $D$, $j_{ab}(k)$ and a common contraction
ball on a smaller energy disk. The fixed point is holomorphic in $E$.
\end{proof}

\begin{lemma}\label{lem:time-one}
Let $\eta\in\R$ and let $D\in\uu(m)$ be constant. Denote by $A$ the
time-one map of $\dot v=(D+F)v$, $\dot x=\eta$, $\dot t=1$, and assume
\[
 ((\partial_t+\eta\partial_x)P_j)P_j^{-1}
 +P_j(D+F)P_j^{-1}=D_j+F_j,
 \quad P_j\in\coc r(\T^2,\U(m)),
\]
where $D_j\in\uu(m)$ is constant. For $B_j(x)=P_j(x,0)$,
\begin{equation}\label{eq:time-one-error}
 \|B_j(\cdot+\eta)AB_j^{-1}-e^{D_j}\|_r
 \le e^{\|F_j\|_r}-1.
\end{equation}
\end{lemma}
\noindent See Appendix~\ref{app:time-one}.

\begin{lemma}\label{lem:monotone-suspension}
Let $E_c$ be an interior energy of the family
\begin{equation}\label{eq:suspension}
 \dot v=2\pi iH_E(\theta)v,\qquad\dot\theta=\omega.
\end{equation}
Let $A_E(x)$ be its time-one map with initial phase
$\theta(0)=(x,0)$, and assume \eqref{eq:discrete-monotonicity}.
Assume that $H_E=H_{+,E}\oplus H_{-,E}$ is jointly analytic on width $h$
and Hermitian for real $E,\theta$. For each prescribed $0<h^\prime<h$,
there is, on a smaller energy neighborhood, a periodic holomorphic block
change of frame $Q_E$ on $S_{h^\prime}^2$ satisfying
\begin{gather*}
 Q_E^*Q_E=I\quad(E,\theta\text{ real}),\qquad Q_{E_c}=I,\\
 2\pi i\widehat H_E=(\partial_\omega Q_E)Q_E^{-1}+2\pi iQ_EH_EQ_E^{-1},
 \qquad\widehat H_{E_c}=H_{E_c},
\end{gather*}
\begin{equation}\label{eq:flow-monotonicity}
 \frac\kappa2I\le\partial_E\widehat H_{+,E}\le (K_p+\kappa/2)I,
 \qquad -(K_p+\kappa/2)I\le\partial_E\widehat H_{-,E}\le-\frac\kappa2I.
\end{equation}
\end{lemma}
\begin{proof}
\emph{Step 1. Average the energy derivative.}
Fix a compact real energy interval about $E_c$, and define
\begin{gather*}
 \partial_s\Phi_E(s;\theta)=2\pi iH_E(\theta+s\omega)\Phi_E(s;\theta),
 \qquad \Phi_E(0;\theta)=I,\\
 V_E(s;\theta)=\Phi_E(s;\theta)^{-1}
       (\partial_EH_E)(\theta+s\omega)\Phi_E(s;\theta).
\end{gather*}
Since $\partial_s(\Phi_E^{-1}\partial_E\Phi_E)=2\pi iV_E$,
\begin{equation}\label{eq:flow-energy-integral}
 \frac{\Phi_E(N;\theta)^{-1}\partial_E\Phi_E(N;\theta)}{2\pi i}
 =\int_0^NV_E(s;\theta)\,ds.
\end{equation}
For $A_{E,N}(y)=A_E(y+(N-1)\alpha)\cdots A_E(y)$ and $A_{E,0}=I$,
\begin{gather*}
 \frac{A_{E,N}^{-1}\partial_EA_{E,N}}{2\pi i}
 =\sum_{j=0}^{N-1}A_{E,j}^{-1}
 \left[\frac{A_E^{-1}\partial_EA_E}{2\pi i}(y+j\alpha)\right]A_{E,j},\\
 N\kappa I\le\frac{A_{+,E,N}^{-1}\partial_EA_{+,E,N}}{2\pi i}\le NK_pI,
 \qquad -NK_pI\le\frac{A_{-,E,N}^{-1}\partial_EA_{-,E,N}}{2\pi i}\le-N\kappa I.
\end{gather*}
For real $E,x$, $0\le t\le1$, and $y=x-t\alpha$,
\begin{gather*}
 \Phi_E(N;(x,t))=
 \Phi_E(t;(y+N\alpha,0))A_{E,N}(y)\Phi_E(t;(y,0))^{-1},\\
 C_0:=\sup_{E,y,\,0\le t\le1}
 \left\|\frac{\Phi_E(t;(y,0))^{-1}\partial_E\Phi_E(t;(y,0))}{2\pi i}\right\|
 <\infty.
\end{gather*}
Differentiating the product and using real unitarity gives
\[
 \left\|\int_0^NV_E(s;\theta)\,ds
 -\Phi_E(t;(y,0))
 \frac{A_{E,N}^{-1}\partial_EA_{E,N}}{2\pi i}
 \Phi_E(t;(y,0))^{-1}\right\|\le2C_0.
\]
By \eqref{eq:flow-energy-integral},
\begin{gather*}
 (\kappa-2C_0/N)I\le\frac1N\int_0^NV_{+,E}(s;\theta)\,ds
                         \le(K_p+2C_0/N)I,\\
 (-K_p-2C_0/N)I\le\frac1N\int_0^NV_{-,E}(s;\theta)\,ds
                         \le(-\kappa+2C_0/N)I.
\end{gather*}
\emph{Step 2. Realize the average by a change of frame.}
Choose $N\in\N$ with $2C_0/N\le\kappa/2$, and set
\[
 S_E=\frac1N\int_0^N(N-s)V_E(s;\theta)\,ds.
\]
By real unitarity, $S_E^*=S_E$. By flow covariance,
\begin{gather*}
 V_E(s;\theta+t\omega)
 =\Phi_E(t;\theta)V_E(s+t;\theta)\Phi_E(t;\theta)^{-1},\\
 \partial_\omega V_E+[V_E,2\pi iH_E]=\partial_sV_E,\\
 \partial_\omega S_E+[S_E,2\pi iH_E]
 =\frac1N\int_0^N(N-s)\partial_sV_E\,ds
 =-\partial_EH_E+\frac1N\int_0^NV_E\,ds.
\end{gather*}
Fix $h^\prime<h^{\prime\prime}<h$ and a complex disk
$\mathcal E\ni E_c$ compactly contained in the energy domain.
By Gronwall's inequality, $\Phi_E^{\pm1}$ and $S_E$ are bounded on
$[0,N]\times\mathcal E\times S_{h^{\prime\prime}}^2$. Hence the equation
\begin{equation}\label{eq:monotone-gauge-ode}
 \partial_EQ_E=2\pi iQ_ES_E,\qquad Q_{E_c}=I,
\end{equation}
has a bounded jointly holomorphic block solution on a smaller disk.
By Cauchy's estimate, $\partial_\omega Q_E$ is bounded on $S_{h^\prime}^2$.
For real $E,\theta$,
\[
 \partial_E(Q_EQ_E^*)=2\pi iQ_E(S_E-S_E^*)Q_E^*=0,
 \qquad Q_EQ_E^*=I.
\]
By the energy equation and the transport identity,
\begin{gather*}
 \partial_E\widehat H_E
 =Q_E\left[\partial_EH_E+\partial_\omega S_E+[S_E,2\pi iH_E]\right]Q_E^{-1}
 =Q_E\left[\frac1N\int_0^NV_E\,ds\right]Q_E^{-1},\\
 \widehat H_{E_c}=H_{E_c}.
\end{gather*}
By uniqueness, the matrix $Q_E$ is periodic in $\theta$. By real unitarity, the bounds for
the average of $V_E$ give \eqref{eq:flow-monotonicity}.
\end{proof}

\subsection{One reduction step}\label{subsec:common-step}\label{subsec:parameter-step}

\begin{lemma}\label{lem:parameter-step}\label{lem:continued-fraction-local-step}\label{lem:uniform-local-smallness}
Let $m\in\N$, $0<a<1/64$, $0<d\le h/2\le1/2$, and
\[
 H=\Lambda+f,\qquad \Lambda=\diag(\lambda_1,\ldots,\lambda_m),\qquad
 |f|_h\le\epsilon,\qquad 0<\epsilon\le c_*(ad)^{4096(m+1)^2},
\]
where $\lambda_1,\ldots,\lambda_m\in\R$ and $f$ is analytic and
Hermitian on the real torus. Put $S=\log(1/\epsilon)$.
There are $T\ge128S$, a periodic holomorphic matrix $P$ on width
$h-d/2$ that is unitary on the real torus, and a constant real diagonal
matrix $\Lambda^+$ for which
\begin{gather}
 \frac{(\partial_\omega P)P^{-1}}{2\pi i}+PHP^{-1}
 =\Lambda^++f^+,\qquad
 \epsilon^+:=\epsilon e^{-T/2}\le\epsilon^{65},\qquad
 |f^+|_{h-d}\le\epsilon^+.\label{eq:quantitative-tail}
\end{gather}
For some $0\le\Gamma\le\frac a{64}\log(\epsilon/\epsilon^+)$ and
every analytic matrix $V$ with $|V|_h<\infty$,
\begin{gather}
 \max\{\|P\|_{h-d/2},\|P^{-1}\|_{h-d/2}\}\le e^{\Gamma/2},\qquad
 |PVP^{-1}|_{h-d}\le e^\Gamma|V|_h,\label{eq:product-cost}\\
 \|\partial_xP\|_0+\|\partial_x(P^{-1})\|_0
 \le\frac\Gamma{h-d/2}.\label{eq:local-derivative-cost}
\end{gather}
If $\Lambda,f$ preserve a prescribed orthogonal block decomposition,
then $P,\Lambda^+$ preserve the same decomposition.

For the parameter assertion, suppose $H=H_{E_0}$ belongs to a family
$H_E=\Lambda+f_E$ jointly analytic near
$I_\epsilon\times\overline{\T_h^2}$ and Hermitian for real parameters,
where $I_\epsilon=[E_0-W\epsilon,E_0+W\epsilon]$.
Assume $H_E=H_{+,E}\oplus H_{-,E}$, $\Lambda=\Lambda_+\oplus\Lambda_-$ and
\begin{equation}\label{eq:normalized-flow-signs}
 \partial_EH_{+,E}\ge\kappa I,\qquad
 \partial_EH_{-,E}\le-\kappa I
 \quad(E\in I_\epsilon,\ \theta\in\T^2).
\end{equation}
Assume $\kappa>0$, $W\ge\max\{1,4/\kappa\}$ and
$\sup_{I_\epsilon}|\partial_EH_E|_h\le\epsilon^{-a}$.
Given marked values $\lambda_\pm\in\spec\Lambda_\pm$, the preceding
conclusions hold with $H,f^+$ replaced by $H_{E^+},f_{E^+}^+$ for an
energy $E^+\in I_\epsilon$. The new constant has marked values
$\lambda_\pm^+\in\spec\Lambda_\pm^+$ satisfying
\begin{equation}\label{eq:local-resonance-output}
 |E^+-E_0|\le\frac8{3\kappa}\epsilon,\qquad
 \lambda_+^+-\lambda_-^+=\lambda_+-\lambda_-.
\end{equation}
After $P,\Lambda^+$ have been fixed at $E^+$, the transformed family
satisfies
\begin{equation}\label{eq:local-step-output}
 \widehat H_E=\frac{(\partial_\omega P)P^{-1}}{2\pi i}+PH_EP^{-1}
 =\Lambda^++f_E^+,\qquad
 \partial_E\widehat H_E=P(\partial_EH_E)P^{-1}.
\end{equation}
The constant $c_*>0$ depends only on $m$ in the fixed-energy assertion,
and may also depend on $\kappa,W$ in the parameter assertion. In both
cases it is independent of $\Lambda,a,d,\alpha$.
\end{lemma}

\begin{proof}[Proof of Lemma~\ref{lem:parameter-step}]
We use the elimination of \cite[Lemma~3.1]{HY}. Set
\[
 L=\sup_{E\in I_\epsilon}|\partial_EH_E|_h\le\epsilon^{-a}.
\]
For the fixed-energy assertion, suppress $E$ and set $L=0$.
All Fourier projections are chosen from $\Lambda$ and fixed in energy.

\emph{Step 1. Resonant components.}
Choose $C_0=C_0(m)$ large, and then choose $C_A=C_A(m,C_0)$ large.
Select consecutive continued-fraction denominators $q,q_+$, with
convergent $p/q$, satisfying
\begin{equation}\label{eq:cf-eligible-level}
 q\le C_A(ad)^{-(m+2)}(S+\log(1/d)+1)\le q_+.
\end{equation}
Put $\tau=q\alpha-p$, so $(q+q_+)^{-1}<|\tau|<q_+^{-1}$, and set
$\eta=\epsilon^{1/8}$ and
\[
 T_0=\lceil C_0/(ad)\rceil,\qquad
 K_0=\left\lfloor\frac{q_+}{64mT_0^m}\right\rfloor,\qquad
 K_i=T_0^iK_0\quad(0\le i\le m).
\]
Increase $C_A$ so that $q_+/(64mT_0^m)\ge2$.
At scale $K_i$, join indices $b\ne c$ in the same prescribed block
whenever some $|k|_1\le K_i$ satisfies
$|\lambda_b-\lambda_c-\langle k,\omega\rangle|<\eta$.
There are $m$ adjacent pairs of partitions and at most $m-1$ strict
mergers. Choose $i$ with equal consecutive partitions and set $K=K_i$,
$N=K_{i+1}=T_0K$. Then
\begin{equation}\label{eq:cutoff-lower}
 N\ge\frac{q_+}{128mT_0^{m-1}},\qquad
 T:=\pi dN\ge cC_Aa^{-3}d^{-2}(S+\log(1/d)+1)\ge128S.
\end{equation}
In each component $B$, choose a spanning tree at scale $K$, rooted at
the marked index if present. Sum its oriented edge vectors from the root
$b_0$ to obtain $\ell_b\in\Z^2$, $\ell_{b_0}=0$. With
$\Lambda_B=\lambda_{b_0}$,
\begin{gather*}
 |\ell_b|_1,\ |\ell_b-\ell_c|_1\le(m-1)K,\\
 |\lambda_b-\lambda_c-\langle\ell_b-\ell_c,\omega\rangle|
 \le(m-1)\eta,\qquad
 |\lambda_b-\langle\ell_b,\omega\rangle-\Lambda_B|\le(m-1)\eta.
\end{gather*}
By stability, every divisor of modulus $<\eta$ with $|k|_1<N$ inside
a prescribed block joins indices in one component. For $v=k-\ell_b+\ell_c$,
\[
 |v|_1\le2K_m\le q_+/(32m),\qquad
 |\langle v,\omega\rangle|<m\eta<1/(7q).
\]
Here $(2m-1)q\eta<1/7$ follows from
\eqref{eq:uniform-master-smallness} below. Since $|\tau|<1/q_+$,
\[
 |pv_1+qv_2|=|q\langle v,\omega\rangle-v_1\tau|
 <1/7+1/6<1.
\]
Hence $pv_1+qv_2=0$. By coprimality, $v\in\Z(q,-p)$.

Define $P_{\rm line}$ to retain entries
$b,c\in B$, $k=\ell_b-\ell_c+j(q,-p)$, and define $P_{\rm re}$ to
retain these and the remaining entries in the prescribed blocks with
divisor of modulus $<\eta$. Both projections contract $|\cdot|_h$,
preserve reality and blocks, and commute with $\partial_\omega$ and
$[\Lambda,\cdot]$. Let $T_N$ retain $|k|_1<N$, and put $R_N=I-T_N$.
Then
\[
 T_NP_{\rm re}=T_NP_{\rm line},\qquad
 Q(\theta)=\diag_b e^{-2\pi i\langle\ell_b,\theta\rangle}.
\]

\emph{Step 2. Elimination.}
Work in the algebra preserving the prescribed blocks, with $I$ its
identity operator. Set
$P_{\rm nre}=I-P_{\rm re}$ and
\[
 \mathcal L Y=\partial_\omega Y-[2\pi i\Lambda,Y],\qquad
 \mathcal J_YV=\int_0^1e^{-tY}Ve^{tY}\,dt.
\]
The Fourier multiplier is
\[
 \widehat{(\mathcal LY)}_{bc}(k)
 =2\pi i\bigl(\langle k,\omega\rangle-\lambda_b+\lambda_c\bigr)
       \widehat Y_{bc}(k).
\]
On the range of $P_{\rm nre}$, every divisor has modulus at least
$\eta$. Hence
\[
 |2\pi i\mathcal L^{-1}Z|_h\le\eta^{-1}|Z|_h.
\]
For $Y=2\pi i\mathcal L^{-1}Z$,
\begin{align*}
 &(\partial_\omega e^{-Y})e^Y+e^{-Y}(2\pi i\Lambda)e^Y-2\pi i\Lambda
 =-\mathcal J_Y(\mathcal LY)=-2\pi i\mathcal J_YZ,\\
 &\frac{(\partial_\omega e^{-Y})e^Y}{2\pi i}
       +e^{-Y}(\Lambda+f)e^Y
 =\Lambda+e^{-Y}fe^Y-\mathcal J_YZ.
\end{align*}
Solve for $Z$ in the range of $P_{\rm nre}$:
\[
 Z=P_{\rm nre}\bigl(e^{-Y}fe^Y-(\mathcal J_Y-I)Z\bigr),
 \qquad Y=2\pi i\mathcal L^{-1}Z.
\]
If $|f|_h\le r$, $|Z|_h\le2r$ and $C_mr/\eta\le1/4$, the map
preserves this ball with Lipschitz constant at most $C_mr/\eta$, since
\[
 |Y|_h\le2r/\eta,\qquad
 |e^{-Y}fe^Y-f|_h+|(\mathcal J_Y-I)Z|_h
 \le C_mr^2/\eta.
\]
At its fixed point,
\begin{gather}
 \frac{(\partial_\omega e^{-Y_E})e^{Y_E}}{2\pi i}
 +e^{-Y_E}H_Ee^{Y_E}=\Lambda+\widetilde f_E,
 \qquad P_{\rm re}\widetilde f_E=\widetilde f_E,
 \label{eq:hy-elimination-conjugacy}\\
 |Y_E|_h\le C_m|f_E|_h/\eta,\qquad
 |\widetilde f_E|_h\le2|f_E|_h,\qquad
 |\widetilde f_E-P_{\rm re}f_E|_h\le C_m|f_E|_h^2/\eta.
 \notag
\end{gather}
For Hermitian $Z$, paired divisors give $Y^*=-Y$ on the real torus.
The contraction preserves reality and blocks; its fixed point is holomorphic in $f$.
Only $\mathcal LY$ enters the estimates, without separate bounds for
$\partial_\omega Y$ and $[\Lambda,Y]$; hence the constants are independent
of $\Lambda$.

For the family assertion, we have
\[
 |f_E|_h\le\epsilon+W\epsilon L
 \le\epsilon^{1/2}/1024.
\]
The remainder map $f\mapsto\widetilde f-P_{\rm re}f$ is holomorphic
on the ball of radius $\epsilon^{1/2}/64$;
each $f_E$ lies in its concentric quarter ball. The quadratic remainder
bound and Cauchy's estimate give
\[
 |D_f(\widetilde f-P_{\rm re}f)[V]|_h
 \le C_m\epsilon^{1/2}\eta^{-1}|V|_h
 =C_m\epsilon^{3/8}|V|_h.
\]
Since $\partial_Ef_E=\partial_EH_E$, the chain rule yields
\begin{equation}\label{eq:retained-derivative}
 |\partial_E\widetilde f_E-P_{\rm re}\partial_EH_E|_h
 \le C_m\epsilon^{3/8}L.
\end{equation}
At $E_0$,
$|\widetilde f_{E_0}|_h\le2\epsilon$.

\emph{Step 3. Periodic reduction.}
The identity $T_NP_{\rm re}=T_NP_{\rm line}$ shows that rotation by
$Q$ makes every retained mode depend on $\phi=q\theta_1-p\theta_2$.
Thus
\[
 \frac{(\partial_\omega Q)Q^{-1}}{2\pi i}
 +Q(\Lambda+T_N\widetilde f_E)Q^{-1}
 =\bigoplus_B(\Lambda_BI+G_{B,E}(\phi)),\qquad
 \phi=q\theta_1-p\theta_2,
\]
\[
 G_{B,E}=\diag_{b\in B}
 (\lambda_b-\langle\ell_b,\omega\rangle-\Lambda_B)
 +(QT_N\widetilde f_EQ^{-1})_B,\qquad
 \|G_{B,E}\|_0\le C_m\eta.
\]
Here the truncation is performed before the rotation.
Apply Floquet theory to each component, including the singleton components:
\begin{equation}\label{eq:retained-flow}
 \tau\Phi_{B,E}'=2\pi iG_{B,E}\Phi_{B,E},\qquad
 \Phi_{B,E}(0)=I,\qquad M_B=\Phi_{B,E}(1)=e^{L_B}.
\end{equation}
Initially choose $L_B^*=-L_B$ with spectrum in $i[-\pi,\pi]$.
The logarithms and diagonalizing bases may be chosen separately at each
real energy. Step~5 will fix the marked lifts at the selected energy.
On the real interval, $\|\Phi_{B,E}\|=1$ and
\[
 M_B-I=\frac{2\pi i}{\tau}\int_0^1
       G_{B,E}(\phi)\Phi_{B,E}(\phi)\,d\phi.
\]
The spectral theorem therefore gives
\[
 |\tau|\|L_B\|\le\tfrac\pi2|\tau|\|M_B-I\|
 \le\pi^2\|G_{B,E}\|_0\le C_m\eta.
\]
Define $B_B(\phi)=e^{\phi L_B}\Phi_{B,E}(\phi)^{-1}$.
The monodromy identity $\Phi_{B,E}(\phi+1)=\Phi_{B,E}(\phi)M_B$
gives $B_B(\phi+1)=B_B(\phi)$, and differentiation gives
\[
 \tau B_B'B_B^{-1}+2\pi iB_BG_{B,E}B_B^{-1}=\tau L_B.
\]
Choose a unitary $U=\bigoplus_BU_B$ for which each
$U_BL_BU_B^{-1}$ is diagonal, and set $B=U\bigoplus_BB_B$.
Then
\begin{gather*}
 P=BQe^{-Y_E},\qquad
 \Lambda^+=U\bigoplus_B(\Lambda_BI+\tau L_B/(2\pi i))U^{-1},\\
 \frac{(\partial_\omega P)P^{-1}}{2\pi i}+PH_EP^{-1}
 =\Lambda^++f^+,\qquad
 f^+=(BQ)R_N\widetilde f_E(BQ)^{-1}.
\end{gather*}
On singleton components use scalar integration. If all components are
singletons, $Q=I$, $B$ is diagonal, and the remainder is
$BR_N\widetilde f_EB^{-1}$.

\emph{Step 4. Cost and remainder.}
Define
\begin{equation}\label{eq:gamma-definition}
 e^\Gamma=C_m(1+d^{-2})\max\{1,
 \|P\|_{h-d/2}\|P^{-1}\|_{h-d/2},
 \|BQ\|_{h-d/2}\|(BQ)^{-1}\|_{h-d/2}\}.
\end{equation}
By reflected unitarity, each matrix and its inverse have equal strip norms.
Put
\[
 w=\frac{(1,-\operatorname{sgn}p)}{q+|p|},\qquad
 \Psi(\phi)=Q_B(w\phi)^{-1}\Phi_{B,E}(\phi).
\]
The equation $\tau\Psi'=2\pi i\widetilde G\Psi$ has coefficient
\[
 \widetilde G=\diag_{b\in B}
 (\lambda_b-\langle\ell_b,\omega\rangle-\Lambda_B
                  +\tau\langle\ell_b,w\rangle)
 +(T_N\widetilde f_E)_B(w\phi),
\]
\[
 \|\widetilde G\|_{h(q+|p|)}
 \le C_m(\eta+|\tau|K/(q+|p|)).
\]
By real unitarity and integration along a vertical segment,
\[
 \|\Psi(x+iy)^{\pm1}\|\le
 \exp\left(\frac{C_m|y|}{|\tau|}
 [\eta+|\tau|K/(q+|p|)]\right).
\]
We have $|\Im\phi|\le2hq$, $|\Im(\theta-w\phi)|_\infty\le2h$, and
\[
 (BQ)_B=U_Be^{\phi L_B}\Psi(\phi)^{-1}Q_B(w\phi)^{-1}Q_B(\theta).
\]
Using $P=BQe^{-Y_E}$, $T=\pi dT_0K$, and
$|\tau|^{-1}<q+q_+\le2q_+$, we obtain
\begin{gather*}
 \Gamma\le C_mK+C_mq\eta/|\tau|
       +C_m(1+\log(1/d))+C_m\epsilon^{3/8},\\
 \frac\Gamma T\le\frac{C_m}{dT_0}
 +\frac{C_mqT_0^{m-1}\eta}d
 +\frac{C_mT_0^{m-1}}{dq_+}
       (1+\log(1/d)+\epsilon^{3/8})\le a/128.
\end{gather*}
For the last inequality, see Step~6. Converting to the Fourier norm on
the remaining half-strip gives
\begin{gather*}
 |PVP^{-1}|_{h-d}\le e^\Gamma|V|_h,\qquad
 |R_N\widetilde f_E|_{h-d/2}\le e^{-T}|\widetilde f_E|_h,\\
 |f^+|_{h-d}\le C_m\epsilon^{1/2}e^{-T+\Gamma}
 \le\epsilon e^{-T/2}=\epsilon^+.
\end{gather*}
For the last inequality, the ratio of the two sides is at most
$C_m\epsilon^{-1/2}e^{-(1/2-a/128)T}
\le C_m\epsilon^{127/2-a}\le1$.
Fix the second coordinate on the real torus and set $r=h-d/2$.
Since $P$ is unitary there and $\|P^{\pm1}\|_r\le e^{\Gamma/2}$,
the three-lines theorem gives
\[
 \|P(x+iy)\|\le e^{|y|\Gamma/(2r)}.
\]
Since $P(x)^*\partial_xP(x)$ is skew-Hermitian, expansion at the real
axis with both signs of the imaginary displacement gives
\[
 P(x+iy)^*P(x+iy)=I+2iyP(x)^*\partial_xP(x)+O(y^2),\qquad
 -\frac\Gamma{2r}I\le iP(x)^*\partial_xP(x)\le\frac\Gamma{2r}I.
\]
Finally,
\[
 \partial_x(P^{-1})=-P^{-1}(\partial_xP)P^{-1},\qquad
 \|\partial_x(P^{-1})\|_0=\|\partial_xP\|_0.
\]
The Hermitian bound above therefore gives
\[
 \|\partial_xP\|_0+\|\partial_x(P^{-1})\|_0\le\Gamma/r,
\]
which proves \eqref{eq:local-derivative-cost}.

\emph{Step 5. Energy selection.}
For the parameter assertion, the full resonant-line projection satisfies
\[
 [QP_{\rm line}(V)Q^{-1}]_B
 =\int_0^1[Q(\theta+s(p,q))V(\theta+s(p,q))
             Q(\theta+s(p,q))^{-1}]_B\,ds.
\]
The full average preserves the signs; estimate the cutoff error afterward.
Since
\[
 P_{\rm line}-T_NP_{\rm re}=R_NP_{\rm line},\qquad
 |R_NP_{\rm line}V|_0\le e^{-2\pi hN}|V|_h,
\]
and $Q$ is unitary on the real torus, \eqref{eq:retained-derivative}
gives
\[
 \left\|\partial_EG_{B,E}
       -[QP_{\rm line}(\partial_EH_E)Q^{-1}]_B\right\|_0
 \le C_m(\epsilon^{3/8}+e^{-2\pi hN})L\le\kappa/4.
\]
Consequently,
\begin{gather*}
 \partial_EG_{+,B,E}\ge\tfrac34\kappa I,\qquad
 \partial_EG_{-,B,E}\le-\tfrac34\kappa I,\\
 \|\partial_EG_{B,E}\|_0\le L+\kappa/4\le2L.
\end{gather*}
Here $L\ge\kappa$ follows from \eqref{eq:normalized-flow-signs}.
Let $G_s$ be a real-analytic Hermitian family in \eqref{eq:retained-flow},
and choose $|\xi|=1$ outside $\spec M_{s_0}$. A neighborhood of $s_0$
then admits the Hermitian Cayley chart
\[
 C_s=i(\xi I+M_s)(\xi I-M_s)^{-1}=C_s^*,\qquad
 \beta_a(s)=\frac{\arg\xi}{2\pi}+\frac12+
                 \frac{\arctan r_a(s)}\pi,
\]
where $r_a(s)$ are the ordered eigenvalues of $C_s$.
By min--max, $\beta_a$ is locally Lipschitz and
$e^{2\pi i\beta_a(s)}\in\spec M_s$. At a differentiability point,
choose normalized eigenvectors of the chosen branch at $s+t$ along a
subsequence $u_t\to u$.
Normality gives
\[
 \begin{gathered}
 M_su=e^{2\pi i\beta_a(s)}u,\\
 \bigl(e^{2\pi i\beta_a(s+t)}-e^{2\pi i\beta_a(s)}\bigr)u^*u_t
 =u^*(M_{s+t}-M_s)u_t.
 \end{gathered}
\]
Dividing by $t$ and passing to the limit gives
\[
 \beta_a'(s)=\frac{u^*M_s^*(\partial_sM_s)u}{2\pi i}.
\]
The differentiated flow satisfies
\begin{gather*}
 \partial_\phi(\Phi_s^{-1}\partial_s\Phi_s)
 =\frac{2\pi i}{\tau}\Phi_s^{-1}(\partial_sG_s)\Phi_s,\qquad
 (\Phi_s^{-1}\partial_s\Phi_s)(0)=0,\\
 M_s^{-1}\partial_sM_s
 =\frac{2\pi i}{\tau}\int_0^1
 \Phi_s^{-1}(\partial_sG_s)\Phi_s\,d\phi.
\end{gather*}
Since $\Phi_s^{-1}=\Phi_s^*$ on the real interval,
\begin{equation}\label{eq:floquet-phase-derivative}
 \tau\beta_a'(s)=\int_0^1
 \langle\Phi_su,(\partial_sG_s)\Phi_su\rangle\,d\phi
 \quad\text{for a.e. }s.
\end{equation}
On a compact interval, use a finite chart cover. On overlaps, match the
eigenvalue and adjust its phase by an integer to keep the lift continuous.
The locally Lipschitz pieces give absolutely continuous branches satisfying
\eqref{eq:floquet-phase-derivative} almost everywhere, including at multiplicities.

For each marked root, first follow the homotopy
\[
 G_s=\diag_{b\in B}(\lambda_b-\langle\ell_b,\omega\rangle-\Lambda_B)
       +s(QT_N\widetilde f_{E_0}Q^{-1})_B,\qquad 0\le s\le1.
\]
Along this homotopy, choose the marked Floquet eigenphase lift with
$\beta_\pm(0)=0$. Since
$\|\partial_sG_s\|_0\le2\epsilon$, the frequencies
$b_\pm(E_0)=\Lambda_B+\tau\beta_\pm$ satisfy
$|b_\pm(E_0)-\lambda_\pm|\le2\epsilon$.
Continue in $E$. The identity $b_\pm'=\tau\beta_\pm'$ applies for
both signs of $\tau$. By \eqref{eq:floquet-phase-derivative} and
monotonicity,
\begin{equation}\label{eq:parameter-mark-signs}
 b_+'\ge3\kappa/4,\qquad b_-'\le-3\kappa/4,\qquad
 |b_\pm'|\le2L\quad\text{a.e.}
\end{equation}
Thus $g=b_+-b_-$ satisfies
\[
 |g(E_0)-(\lambda_+-\lambda_-)|\le4\epsilon,\qquad
 g(E')-g(E)\ge\tfrac32\kappa(E'-E)\quad(E<E').
\]
Since $W\ge4/\kappa$,
\[
 \begin{aligned}
 g(E_0-W\epsilon)&\le g(E_0)-\tfrac32\kappa W\epsilon
 \le\lambda_+-\lambda_--2\epsilon,\\
 g(E_0+W\epsilon)&\ge g(E_0)+\tfrac32\kappa W\epsilon
 \ge\lambda_+-\lambda_-+2\epsilon.
 \end{aligned}
\]
By the intermediate value theorem,
\[
 \exists!E^+\in I_\epsilon:\quad
 g(E^+)=\lambda_+-\lambda_-,\qquad |E^+-E_0|\le8\epsilon/(3\kappa),
\]
for the chosen branches. Moreover,
\[
 |b_\pm(E^+)-\lambda_\pm|\le C\epsilon(1+L)\le C\epsilon^{1-a}.
\]
At $E^+$, choose in each marked component an orthonormal eigenbasis
of $M_B$ containing the marked eigenvector. Assign that vector the
logarithm eigenvalue $2\pi i\beta_\pm(E^+)$ and give all other vectors
principal lifts. There is at most one mark per component; unmarked
components keep principal logarithms. This defines $L_B^*=-L_B$ with
$e^{L_B}=M_B$, including on repeated eigenspaces. Since
$\Lambda_B=\lambda_\pm$ in a marked component and $C\epsilon^{1-a}\le\eta$,
\[
 |\tau|\|L_B\|
 \le\max\{C_m\eta,\,2\pi|b_\pm(E^+)-\Lambda_B|\}
 \le C_m\eta.
\]
The cost bounds remain valid, and the new marked values satisfy
\[
 \lambda_\pm^+=b_\pm(E^+),\qquad
 \lambda_+^+-\lambda_-^+=\lambda_+-\lambda_-.
\]
Now freeze $P$ and $\Lambda^+$ at $E^+$. The identity
\[
 f_E^+=f_{E^+}^++P(H_E-H_{E^+})P^{-1}
\]
gives \eqref{eq:local-step-output} on the original parameter domain.
Since $P$ is energy-independent and unitary on the real torus,
the signed energy inequalities are preserved.

\emph{Step 6. Uniform choice of constants.}
First choose $C_0$ so that $C_m/(dT_0)\le a/512$, and then $C_A$ so that
\[
 \frac{C_mT_0^{m-1}}{dq_+}(1+\log(1/d)+\epsilon^{3/8})\le a/512
\]
and so that \eqref{eq:cutoff-lower} gives $T\ge128S$.
The entry condition yields $\log(1/d)\le S/[4096(m+1)^2]$, and hence
\begin{align}\label{eq:uniform-master-smallness}
 \frac{qT_0^{m-1}\eta}d
 &\le C_m a^{-(2m+1)}d^{-(2m+2)}(S+1)\epsilon^{1/8}\notag\\
 &\le C_m(ad)^{-2(m+1)}\epsilon^{1/16}\notag\\
 &\le C_m c_*^{1/16}(ad)^{256(m+1)^2-2(m+1)}
 \le\frac{a}{512C_m}
\end{align}
after decreasing $c_*$. The same estimate gives
$(2m-1)q\eta<1/7$, as used in the resonant-line argument.
For the parameter assertion, decreasing $c_*$ in terms of $m,\kappa,W$ gives
\[
 |f_E|_h\le\epsilon+W\epsilon L
 \le\epsilon+W\epsilon^{1-a}\le\epsilon^{1/2}/1024.
\]
Since $h\ge2d$, $T\ge128S$, and $a<1/64$,
\begin{gather*}
 e^{-2\pi hN}\le e^{-4T}\le\epsilon^{512},\\
 \begin{aligned}
 C_m(\epsilon^{3/8}+e^{-2\pi hN})L
 &\le C_m(\epsilon^{3/8-a}+\epsilon^{512-a})\\
 &\le2C_m\epsilon^{23/64}\le\kappa/4.
 \end{aligned}
\end{gather*}
For fixed energy, use $|f|_h\le\epsilon$ and omit energy selection; the constant
depends only on the dimension.
\end{proof}

\subsection{Preparation and iteration}\label{subsec:energy-selection}
\begin{proof}[Proof of Proposition~\ref{prop:marked-resonance}]
\emph{Step 1. Initial suspension.}
Decrease $h$ so that the family is jointly holomorphic near
$\{E_c\}\times\overline{\T_h}$, and fix
\[
 \mu>0,\qquad m=m_++m_-,\qquad
 \varrho=\min\{1/128,\mu\},\qquad
 0<h_{\rm in}<\min\{h/12,1/8\}.
\]
Choose $W\ge\max\{1,8/\kappa\}$, and fix the parameter constant $c_*$
of Lemma~\ref{lem:parameter-step} for the lower monotonicity bound
$\kappa/2$.
Choose an almost-reduction $B_{\rm in}$ at $E_c$. Absorb the constant
block-unitary diagonalization into $B_{\rm in}$, and freeze it in $E$.
Proposition~\ref{prop:embedding}, followed by
Lemma~\ref{lem:monotone-suspension} on width $2h_{\rm in}<h/6$
and the norm conversion \eqref{eq:fouriernorm}, gives on a smaller energy
neighborhood
\[
 \dot v=2\pi iH_Ev,\qquad H_E=\Lambda+f_E,\qquad
 \Lambda=\Lambda^*\text{ diagonal},\qquad Q_{E_c}=I,
\]
with $|f_{E_c}|_{h_{\rm in}}$ arbitrarily small, since $Q_{E_c}=I$ and thus the Hamiltonian at $E_c$ is unchanged.

Relabel the bounds in \eqref{eq:flow-monotonicity} as $\kappa,K_p$.
Fix a compact interval $I_0$ in this neighborhood with
$E_c\in\operatorname{int}I_0$. The interval and initial analytic
derivative bound may depend on $B_{\rm in}$.

\emph{Step 2. Finite preparation.}
Apply the fixed-energy assertion at $E_c$, keeping each resulting
matrix $P$ fixed as $E$ varies.

Set
\begin{gather*}
 H_{\ell,E}=\Lambda_\ell+f_{\ell,E},\qquad
 h_\ell=4^{-\ell}h_{\rm in},\quad d_\ell=h_\ell/2,\quad a_\ell=\varrho h_\ell,\\
 S_\ell=\log(1/\epsilon_\ell),\qquad
 L_\ell=\max\{1,\sup_{I_0}|\partial_EH_{\ell,E}|_{h_\ell}\}.
\end{gather*}
Choose the initial error below the fixed-energy threshold of
Lemma~\ref{lem:parameter-step}, which depends only on data fixed before
the almost-reduction. Each step gives
\begin{gather*}
 |f_{\ell,E_c}|_{h_\ell}\le\epsilon_\ell,\qquad
 S_{\ell+1}\ge65S_\ell,\qquad
 \log L_{\ell+1}\le\log L_\ell+\frac{a_\ell}{64}(S_{\ell+1}-S_\ell),\\
 a_{\ell+1}d_{\ell+1}=a_\ell d_\ell/16,\qquad
 S_{\ell+1}-S_\ell\ge64S_\ell\ge4096(m+1)^2\log16.
\end{gather*}
The last two relations preserve the entry condition. Since
\[
 a_{\ell+1}S_{\ell+1}\ge\tfrac{65}4a_\ell S_\ell,
\]
we have
\begin{align*}
 \log L_\ell
 &\le\log L_0+\sum_{i=1}^{\ell}
       \frac{a_{i-1}}{64}(S_i-S_{i-1})\\
 &\le\log L_0+\frac1{16}\sum_{i=1}^\ell a_iS_i\\
 &\le\log L_0+\frac{a_\ell S_\ell}{16}
       \sum_{r=0}^{\ell-1}(4/65)^r
 \le\log L_0+\frac{65}{976}a_\ell S_\ell.
\end{align*}
\begin{gather*}
 a_\ell S_\ell\ge(65/4)^\ell a_0S_0\longrightarrow\infty,
 \qquad
 \limsup_{\ell\to\infty}\frac{\log L_\ell}{a_\ell S_\ell}
 \le\frac{65}{976}<\frac14.
\end{gather*}
Since
\[
 \log\frac1{a_\ell d_\ell}
 =\log\frac1{a_0d_0}+\ell\log16,\qquad
 S_\ell\ge65^\ell S_0,
\]
we have
\begin{gather*}
 \frac{S_\ell-\log2}{1+\log(1/(a_\ell d_\ell))}\longrightarrow\infty,\\
 \limsup_{\ell\to\infty}
 \frac{\log L_\ell}{a_\ell(S_\ell-\log2)}
 \le\frac{65}{976}<\frac14.
\end{gather*}
Also $\epsilon_\ell\to0$. Choose one finite stage satisfying the parameter
entry and energy-window bounds below. Double its error bound and renumber:
\begin{gather*}
 |f_{0,E_c}|_{h_0}\le\epsilon_0/2,\qquad
 \log L_0\le a_0S_0,\qquad a_0=\varrho h_0,\quad d_0=h_0/2,\\
 0<\epsilon_0<1/2,\qquad
 \epsilon_0\le c_*(a_0d_0)^{4096(m+1)^2},\qquad
 64S_0\ge4096(m+1)^2\log16,\\
 \left(\frac{16}{3\kappa}+W\right)\epsilon_0
 <\operatorname{dist}(E_c,\partial I_0).
\end{gather*}
Fix this finite product $\mathcal P_{\rm pre}$. Choose marked entries
$\lambda_{\pm,0}$ in the two blocks. By density of $\Z\alpha+\Z$, change
$\lambda_{+,0}$ by at most $\epsilon_0/4$ and subtract the same change
from the corresponding diagonal entry of $f_{+,0,E}$.
Still denote the adjusted entry by $\lambda_{+,0}$. Then
\[
 H_{0,E}\text{ is unchanged},\qquad
 |f_{0,E_c}|_{h_0}\le3\epsilon_0/4,\qquad
 \lambda_{+,0}-\lambda_{-,0}=k\alpha+l\quad(k,l\in\Z).
\]

\emph{Step 3. Iteration with a fixed marked gap.}
Set
\begin{gather*}
 E_0=E_c,\quad\mathcal P_0=I,\quad
 h_j=4^{-j}h_0,\quad d_j=h_j/2,\quad a_j=\varrho h_j,\\
 S_j=\log(1/\epsilon_j),\qquad
 L_j=\max\{1,\sup_{E\in I_0}|\partial_EH_{j,E}|_{h_j}\}.
\end{gather*}

The case $j=0$ follows from Step~2. Apply Lemma~\ref{lem:parameter-step}
at $E_j$ and keep $P_j$ fixed at its value at the selected energy $E_{j+1}$  and restrict to
$h_{j+1}=h_j/4$. Then
\begin{gather*}
 |E_{j+1}-E_j|\le\frac8{3\kappa}\epsilon_j,\qquad
 |f_{j+1,E_{j+1}}|_{h_{j+1}}\le\epsilon_{j+1},\qquad
 S_{j+1}\ge65S_j,\\
 \lambda_{+,j+1}-\lambda_{-,j+1}=k\alpha+l,\qquad
 \mathcal P_{j+1}=P_j\mathcal P_j,\qquad
 \partial_EH_{j+1,E}=P_j(\partial_EH_{j,E})P_j^{-1}.
\end{gather*}
The $P_j$ are independent of $E$, so $H_{j,E}$ is defined for every
$E\in I_0$. Since $P_j^*P_j=I$, the real energy bounds persist on
$I_0$. If $L_j\le e^{a_jS_j}$, then by \eqref{eq:product-cost},
\begin{align*}
 \log L_{j+1}
 &\le a_jS_j+\frac{a_j}{64}(S_{j+1}-S_j)
 =\frac{a_j}{64}(63S_j+S_{j+1})\\
 &\le\frac{2a_j}{65}S_{j+1}\le a_{j+1}S_{j+1}.
\end{align*}
The entry condition persists as in preparation. Since
$\epsilon_{j+1}\le\epsilon_j^{65}$,
\begin{gather*}
 \epsilon_{j+n}\le\epsilon_j^{65^n}\le\epsilon_j^{1+64n}
 \quad(n\ge0),\\
 \sum_{t\ge j}\epsilon_t
 \le\frac{\epsilon_j}{1-\epsilon_j^{64}}\le2\epsilon_j,\\
 |E_j-E_c|+W\epsilon_j
 \le\left(\frac{16}{3\kappa}+W\right)\epsilon_0
 <\operatorname{dist}(E_c,\partial I_0).
\end{gather*}
Thus $[E_j-W\epsilon_j,E_j+W\epsilon_j]$ lies in $I_0$ at every step,
and the induction continues.

\emph{Step 4. Real derivative budget and limiting energy.}
By \eqref{eq:local-derivative-cost},
\[
 \|\partial_xP_j\|_0+\|\partial_x(P_j^{-1})\|_0
 \le\frac{a_j(S_{j+1}-S_j)}{64(h_j-d_j/2)}
 =\frac\varrho{48}(S_{j+1}-S_j).
\]
Real unitarity and the product rule therefore give
\begin{equation}\label{eq:cf-global-budget}
 \begin{aligned}
 \|\partial_x\mathcal P_j\|_0+\|\partial_x(\mathcal P_j^{-1})\|_0
 &\le\sum_{t<j}\bigl(\|\partial_xP_t\|_0+
                         \|\partial_x(P_t^{-1})\|_0\bigr)\\
 &\le\frac\varrho{48}(S_j-S_0).
 \end{aligned}
\end{equation}
Also,
\[
 |E_n-E_j|\le\frac8{3\kappa}\sum_{t=j}^{n-1}\epsilon_t
 \le\frac{16}{3\kappa}\epsilon_j\quad(n>j),\qquad
 E_j\longrightarrow E_*\in\operatorname{int}I_0.
\]
Since $\mathcal P_j$ is independent of $E$ and unitary on the real
torus, we have for every $E\in I_0$
\[
 \partial_Ef_{j,E}=\partial_EH_{j,E}
 =\mathcal P_j(\partial_EH_{0,E})\mathcal P_j^{-1},\qquad
 \|\partial_EH_{j,E}\|_0\le K_p.
\]
Use this real bound to compare the two energies:
\[
 f_{j,E_*}-f_{j,E_j}
 =\int_{E_j}^{E_*}\partial_EH_{j,E}\,dE.
\]
\begin{equation}\label{eq:cf-global-family}
 |E_*-E_j|\le\frac{16}{3\kappa}\epsilon_j,\qquad
 \|f_{j,E_*}\|_0\le\epsilon_j+K_p|E_*-E_j|
 \le\left(1+\frac{16K_p}{3\kappa}\right)\epsilon_j.
\end{equation}
Set
\[
 B_j(x)=\mathcal P_j(x,0)\mathcal P_{\rm pre}(x,0)
 Q_{E_*}(x,0)B_{\rm in}(x),\qquad C_j=e^{2\pi i\Lambda_j}.
\]
The product of the last three factors and its inverse have real derivative
norms summing to at most a fixed $C_{\rm fixed}>0$.
By Duhamel's formula, real unitarity and \eqref{eq:cf-global-budget},
\begin{gather*}
 \|B_j^+A_{E_*}B_j^{-1}-C_j\|_0
 \le2\pi\|f_{j,E_*}\|_0\le Ce^{-S_j},\\
 \|B_j'\|_0+\|(B_j^{-1})'\|_0
 \le C_{\rm fixed}+\frac\varrho{48}(S_j-S_0)
 \le C_{\rm fixed}+\mu S_j.
\end{gather*}

Finally, 
$\lambda_{+,j}-\lambda_{-,j}=k\alpha+l$ implies\eqref{eq:marked-gap}.
\end{proof}

\section{Aubry duality and the scalar Wronskian}\label{sec:duality-wronskian}
We use the scalar Wronskian and apply Proposition~\ref{prop:marked-resonance} to bound $\|F\|_0$ from below in
$
 B^+U_EB^{-1}=C+F,
$
where $C$ is constant and has resonant eigenphases in opposite blocks.

\begin{proposition}\label{prop:duality-obstruction}
Let $\widetilde Z_E,U_E$ be as in Proposition~\ref{prop:compact-energy}
near a compact $J\subset\R$. Suppose $E\in J$, $B$ is periodic,
$C^1$ and block-unitary, and
\[
 B^+U_EB^{-1}=C+F,\qquad C\in\U(m)\times\U(m),
\]
\[
 e^{2\pi i\theta_\pm}\in\spec C_\pm,\qquad
 \theta_+-\theta_-=k\alpha\pmod\Z,\qquad k\in\Z.
\]
Set $K=2+\sup_{E\in J}|E|+\|v\|_0$ and $\delta=\|F\|_0$.
There are $A\ge1$ and $c>0$, depending only on the fixed frame
neighborhood, $J$ and $v$, such that
\begin{equation}\label{eq:duality-lower}
 \begin{gathered}
 R=\left\lceil A\bigl(1+\|B'\|_0+\|(B^{-1})'\|_0+|k|\bigr)\right\rceil,\\
 \delta R K^{2R}\ge c.
 \end{gathered}
\end{equation}
\end{proposition}
\begin{proof}
All constants are uniform for $E\in J$ and independent of $B,k$.

\smallskip
\noindent\textit{Step 1. Scalar solutions and the Gram bound.}
Choose unit eigenvectors $e_\pm$ in the indicated opposite blocks. Put
\begin{gather*}
 Q_k(x)=(e_+,e^{-2\pi ikx}e_-),\qquad
 (u\ w)(x)=\widetilde Z_E(x)B(x)^{-1}Q_k(x),\\
 h=(h_1,h_2)=(u_0,w_0),\qquad
 Q_k^*Q_k=(B^{-1}Q_k)^*(B^{-1}Q_k)=I_2.
\end{gather*}
By \eqref{eq:rows} and Proposition~\ref{prop:compact-energy}
(cf.\ \cite[Corollary~10.1 and Section~10.2]{GJY}),
\[
 \widetilde Z_E(x;n)=\widetilde Z_E(x+n\alpha;0)U_{E,n}(x),\qquad
 \|U_{E,n}(x)\|=1\quad(x\in\T,\ n\in\Z),
\]
\[
 (L_{v,\alpha,x}-E)u=(L_{v,\alpha,x}-E)w=0,\qquad
 \sup_{n\in\Z}(\|u_n\|_0+\|w_n\|_0)\le C.
\]
Since $CQ_k=e^{2\pi i\theta_+}Q_k^+$, the reduction gives
\[
 U_EB^{-1}Q_k
 =e^{2\pi i\theta_+}(B^+)^{-1}Q_k^++(B^+)^{-1}FQ_k.
\]
Hence
\begin{gather*}
 (u_{n+1}\ w_{n+1})-e^{2\pi i\theta_+}(u_n\ w_n)^+
 =\widetilde Z_E(x+\alpha;n)(B^+(x))^{-1}F(x)Q_k(x),\\
 \|(u_{n+1}\ w_{n+1})-e^{2\pi i\theta_+}(u_n\ w_n)^+\|_0
 \le C\delta.
\end{gather*}
Iteration in both directions gives
\begin{equation}\label{eq:gjy-reconstruct}
 \|(u_n\ w_n)-e^{2\pi in\theta_+}h(\cdot+n\alpha)\|_0
 \le C|n|\delta,\qquad n\in\Z.
\end{equation}

The change of frame from $Z_E$ to $\widetilde Z_E$ has
uniformly bounded inverse. Thus
Lemma~\ref{lem:finite-observation}, together with
$(B^{-1}Q_k)^*(B^{-1}Q_k)=I_2$, gives fixed $p\in\N$
and $c_1>0$ such that

\[
 \sum_{n=0}^{p-1}(u_n\ w_n)^*(u_n\ w_n)\ge c_1I_2.
\]
Set
\[
 G_h=\int_\T h(x)^*h(x)\,dx
     =(\langle h_i,h_j\rangle)_{i,j=1}^2.
\]
The identity
$X^*X-Y^*Y=(X-Y)^*X+Y^*(X-Y)$,
\eqref{eq:gjy-reconstruct} and translation invariance give
\begin{gather*}
 \left\|\int_\T(u_n\ w_n)^*(u_n\ w_n)\,dx-G_h\right\|
 \le C|n|\delta\qquad(0\le n<p),\\
 pG_h\ge(c_1-C\delta)I_2.
\end{gather*}
Fix $\delta_0>0$ with $C\delta_0\le c_1/2$, and put $c_0=c_1/(2p)$.
Then
\begin{equation}\label{eq:closure-Gram}
 \begin{gathered}
 c_0I_2\le G_h\le CI_2,\qquad\delta\le\delta_0,\\
 c_0|z|^2\le\|z_1h_1+z_2h_2\|_2^2\le C|z|^2,
 \qquad z\in\C^2.
 \end{gathered}
\end{equation}
At site zero,
\begin{align*}
 (e_1,e_2)&:=(\mathcal L_{\theta_+}-E)h\\
 &=\sum_s\widehat v_s
    \bigl[e^{2\pi is\theta_+}h(\cdot+s\alpha)-(u_s,w_s)\bigr].
\end{align*}
By \eqref{eq:gjy-reconstruct} and \eqref{eq:potential},
\[
 \|e_1\|_0+\|e_2\|_0
 \le C\delta\sum_s|s\widehat v_s|\le C\delta.
\]
For $a_j=\widehat h_j$, by Fourier duality we have
\begin{equation}\label{eq:closure-residual}
 (H_{v,\alpha,\theta_+}-E)a_j=\widehat e_j,\qquad
 \|a_j\|_2\le C,\qquad\|\widehat e_j\|_2\le C\delta
 \quad(j=1,2).
\end{equation}

\smallskip
\noindent\textit{Step 2. Localization.}
Let $R$ be as in \eqref{eq:duality-lower}, and write
$Z_0=\widetilde Z_E(\cdot;0)$. Since $h=Z_0B^{-1}Q_k$,
\begin{gather*}
 h'=Z_0'B^{-1}Q_k+Z_0(B^{-1})'Q_k+Z_0B^{-1}Q_k',\\
 \|h'\|_0
 \le\|Z_0'\|_0+\|Z_0\|_0\bigl(\|(B^{-1})'\|_0+2\pi|k|\bigr)
 \le CR/A.
\end{gather*}
By Parseval,
\begin{align*}
 \sum_m m^2\bigl(|a_1(m)|^2+|a_2(m)|^2\bigr)
 &=\frac{\|h_1'\|_2^2+\|h_2'\|_2^2}{4\pi^2}\\
 &=\frac1{4\pi^2}\int_\T\|h'(x)\|^2\,dx\\
 &\le\frac{\|h'\|_0^2}{4\pi^2}
 \le CR^2/A^2,
\end{align*}
where $C$ is independent of $A\ge1$. Fix $A$ sufficiently large that
\begin{equation}\label{eq:closure-tail}
 \sum_{|m|>R}\bigl(|a_1(m)|^2+|a_2(m)|^2\bigr)
 \le C/A^2\le c_0/32.
\end{equation}
First let $\delta\le\delta_0$. By \eqref{eq:closure-Gram} and
\eqref{eq:closure-tail},
\[
 \sum_{|m|\le R}\bigl(|a_1(m)|^2+|a_2(m)|^2\bigr)
 \ge2c_0-c_0/32\ge c_0.
\]
Thus some $|m_0|\le R$ satisfies
\[
 |a_1(m_0)|^2+|a_2(m_0)|^2\ge\frac{c_0}{2R+1}.
\]

\smallskip
\noindent\textit{Step 3. The scalar lower bound.}
Set
\[
 W_m=a_1(m)a_2(m-1)-a_1(m-1)a_2(m).
\]
By \eqref{eq:closure-residual},
\[
 W_{m+1}-W_m
 =\widehat e_1(m)a_2(m)-a_1(m)\widehat e_2(m).
\]
Since $a_1,a_2\in\ell^2(\Z)$,
\[
 \begin{gathered}
 W_m\longrightarrow0\quad(m\to\pm\infty),\\
 W_m=\sum_{n<m}
 \bigl(\widehat e_1(n)a_2(n)-a_1(n)\widehat e_2(n)\bigr).
 \end{gathered}
\]
Hence
\[
 |W_m|
 \le\|\widehat e_1\|_2\|a_2\|_2+\|a_1\|_2\|\widehat e_2\|_2
 \le C\delta.
\]
Choose
\[
 z=\frac{(a_2(m_0),-a_1(m_0))^{\mathsf T}}
          {\sqrt{|a_1(m_0)|^2+|a_2(m_0)|^2}},
 \qquad d=z_1a_1+z_2a_2.
\]
Then
\[
 |z|=1,\qquad d_{m_0}=0,\qquad
 |d_{m_0-1}|
 =\frac{|W_{m_0}|}{\sqrt{|a_1(m_0)|^2+|a_2(m_0)|^2}}
 \le C\delta\sqrt R.
\]
Put $r_m=z_1\widehat e_1(m)+z_2\widehat e_2(m)$. By
\eqref{eq:closure-residual},
\[
 \|r\|_\infty\le\|r\|_2\le C\delta,
\]
\begin{align*}
 \binom{d_{m+1}}{d_m}
 &=\begin{pmatrix}E-v(\theta_++m\alpha)&-1\\1&0\end{pmatrix}
       \binom{d_m}{d_{m-1}}+\binom{r_m}{0},\\
 \binom{d_m}{d_{m-1}}
 &=\begin{pmatrix}0&1\\-1&E-v(\theta_++m\alpha)\end{pmatrix}
       \binom{d_{m+1}}{d_m}+\binom{0}{r_m}.
\end{align*}
Since
\[
 \left\|\begin{pmatrix}
 E-v(\theta_++m\alpha)&-1\\1&0
 \end{pmatrix}^{\!\pm1}\right\| 
 \le K,
\]
iteration in both directions gives, for $|m|\le R$,
\[
 |d_m|
 \le K^{2R}|d_{m_0-1}|+C\delta\sum_{\ell=0}^{2R}K^\ell
 \le C\delta\sqrt R K^{2R}.
\]
Since $|z|=1$,
\[
 |d_m|^2\le|a_1(m)|^2+|a_2(m)|^2.
\]
By \eqref{eq:closure-Gram} and \eqref{eq:closure-tail},
\[
 \sum_{|m|\le R}|d_m|^2
 =\|d\|_2^2-\sum_{|m|>R}|d_m|^2
 \ge c_0-c_0/32=31c_0/32.
\]
Thus
\[
 0<\sqrt{31c_0/32}
 \le\|d\|_{\ell^2([-R,R])}
 \le\sqrt{2R+1}\,C\delta\sqrt R K^{2R}
 \le C\delta RK^{2R}.
\]
For $\delta>\delta_0$,
\[
 R\ge1,\qquad K\ge2,\qquad
 \delta RK^{2R}\ge4\delta_0.
\]
Decreasing $c$ proves \eqref{eq:duality-lower} in both cases.
\end{proof}

\subsection{Finishing the proof.}
\begin{theorem}\label{thm:no-interval}
Let $v\in C^\omega(\T,\R)$ be nonconstant and $\alpha$ irrational.
For $a\le b$,
\[
 [a,b]\subset\Sigma_{v,\alpha}\cap\{L>0\}
 \quad\Longrightarrow\quad a=b.
\]
\end{theorem}
\begin{proof}
Suppose $I\subset\Sigma_{v,\alpha}\cap\mathcal O$ is nondegenerate.
By Propositions~\ref{prop:actual-center} and \ref{prop:compact-energy},
choose a nondegenerate compact interval
$J\Subset\operatorname{int}I$ in the common energy domain and
$E_c\in\operatorname{int}J$.
Proposition~\ref{prop:center-preparation} gives fixed-strip
almost reducibility at $E_c$.
Fix $A,K$ from Proposition~\ref{prop:duality-obstruction}, and then choose
\[
 0<\mu<(4A\log K)^{-1}.
\]
Proposition~\ref{prop:marked-resonance} gives a fixed
$E_*\in\operatorname{int}J$, a fixed $k\in\Z$, and reductions
\[
 B_j^+U_{E_*}B_j^{-1}=C_j+F_j,\qquad C_j=C_{+,j}\oplus C_{-,j},
\]
\[
 e^{2\pi i\theta_{\pm,j}}\in\spec C_{\pm,j},\qquad
 \theta_{+,j}-\theta_{-,j}=k\alpha\pmod\Z,
\]
\[
 \begin{gathered}
 \delta_j=\|F_j\|_0\le Ce^{-S_j},\qquad S_j\longrightarrow\infty,
 \\
 \|B_j'\|_0+\|(B_j^{-1})'\|_0\le C_{\rm fixed}+\mu S_j.
 \end{gathered}
\]
Since $E_*,k$ are independent of $j$,
\[
 \begin{aligned}
 R_j&=\left\lceil A\bigl(1+\|B_j'\|_0
                +\|(B_j^{-1})'\|_0+|k|\bigr)\right\rceil\\
 &\le C+A\mu S_j.
 \end{aligned}
\]
Thus \eqref{eq:duality-lower} implies
\[
 0<c\le\delta_jR_jK^{2R_j}
 \le C(1+S_j)e^{-(1-2A\mu\log K)S_j}
 \le C(1+S_j)e^{-S_j/2}\longrightarrow0,
\]
a contradiction.
\end{proof}

\begin{proof}[Proof of Theorem~\ref{thm:main}]
It follows directly from Theorem \ref{thm:no-interval}.

\end{proof}

\appendix

\section{Elementary calculations}\label{app:elementary}
\subsection{The matrix identity}\label{app:matrix-identity}
\begin{proof}[Proof of Lemma~\ref{lem:matrix-defect}]
Set $S=M+N$, $X=\Im M$ and $Y=\Im N$. Since
\[
 Sa=0\ \Longrightarrow\
 0=\Im(a^*Sa)=a^*(X+Y)a\ \Longrightarrow\ a=0,
\]
$S$ is invertible. Moreover,
\[
 M^*+N=S^*+2iY,\qquad M+N^*=S-2iY,
\]
\begin{align*}
 (M^*+N)Y^{-1}(M+N^*)
 &=S^*Y^{-1}S+2i(S-S^*)+4Y\\
 &=S^*Y^{-1}S-4X.
\end{align*}
Hence
\[
 S^{-*}(M^*+N)Y^{-1}(M+N^*)S^{-1}
 =Y^{-1}-4S^{-*}XS^{-1}.
\]
By trace cyclicity,
\begin{align*}
 &2\Re\tr\left[
 (AX^{-1/2})^*(2iC^*S^{-*}X^{1/2})\right]\\
 &\qquad=4\Re\left(i\,\overline{\tr(S^{-1}CA)}\right)
 =4\Im\tr(S^{-1}CA).
\end{align*}
Thus
\begin{align*}
 \|AX^{-1/2}+2iC^*S^{-*}X^{1/2}\|_{\HS}^2
 &=\tr(A^*AX^{-1})+4\tr(CC^*S^{-*}XS^{-1})\\
 &\quad+4\Im\tr(S^{-1}CA),\\
 \|C^*S^{-*}(M^*+N)Y^{-1/2}\|_{\HS}^2
 &=\tr\bigl(CC^*(Y^{-1}-4S^{-*}XS^{-1})\bigr).
\end{align*}
Adding gives the asserted identity.
\end{proof}

\subsection{A unitary logarithm}\label{app:unitary-logarithm}
\begin{proof}[Proof of Lemma~\ref{lem:unitary-logarithm}]
For $m=1$, take
\[
 H=i\theta,\qquad e^{i\theta}=U,\qquad|\theta|\le\pi,\qquad
 [H,V]=[U,V]=0.
\]
Let $m>1$.
Cut the unit circle at the midpoint of a largest eigenvalue gap,
whose length is at least $2\pi/m$. Lift equal eigenvalues equally.
After a common $2\pi\Z$-translation, in a unitary eigenbasis,
\[
 U=\diag(e^{i\theta_a}),\qquad H=\diag(i\theta_a),
\]
\[
 \max_a\theta_a-\min_a\theta_a\le2\pi-\frac{2\pi}{m},\qquad
 \left|\frac{\max_a\theta_a+\min_a\theta_a}{2}\right|\le\pi.
\]
Consequently,
\[
 e^H=U,\qquad\|H\|=\max_a|\theta_a|
 \le2\pi-\frac\pi m<2\pi.
\]
For $0<u<\pi$,
\[
 \left(\frac{u}{\sin u}\right)'
 =\frac{\int_0^u t\sin t\,dt}{\sin^2u}\ge0.
\]
Thus, for $0<|t|\le2\pi-2\pi/m$,
\[
 \frac{|t|}{|e^{it}-1|}
 =\frac{|t|/2}{\sin(|t|/2)}
 \le\frac{\pi(1-1/m)}{\sin(\pi/m)}=:c_m.
\]
Including $\theta_a=\theta_b$, this gives
\[
 |\theta_a-\theta_b|\le c_m|e^{i\theta_a}-e^{i\theta_b}|.
\]
By
\[
 [H,V]_{ab}=i(\theta_a-\theta_b)V_{ab},\qquad
 [U,V]_{ab}=(e^{i\theta_a}-e^{i\theta_b})V_{ab},
\]
\[
 \|[H,V]\|
 \le\|[H,V]\|_{\mathrm{HS}}
 \le c_m\|[U,V]\|_{\mathrm{HS}}
 \le c_m\sqrt m\,\|[U,V]\|.\qedhere
\]
\end{proof}

\subsection{The time-one estimate}\label{app:time-one}
\begin{proof}[Proof of Lemma~\ref{lem:time-one}]
Let $\Phi_t$ be the original fundamental solution, $\Phi_0=I$, and set
\[
 \Phi_t^j=P_j(x+t\eta,t)\Phi_tP_j(x,0)^{-1}.
\]
Differentiating and using the defining differential equation, we obtain,
\begin{align*}
 \partial_t\Phi_t^j
 &=\left[((\partial_t+\eta\partial_x)P_j)P_j^{-1}
       +P_j(D+F)P_j^{-1}\right]_{(x+t\eta,t)}\Phi_t^j\\
 &=(D_j+F_j(x+t\eta,t))\Phi_t^j.
\end{align*}
Also,
\[
 \Phi_0^j=P_j(x,0)IP_j(x,0)^{-1}=I,\qquad
 \|e^{\pm tD_j}\|=1\quad(t\in\R).
\]
Duhamel's formula gives
\[
 \Phi_t^j=e^{tD_j}
   +\int_0^te^{(t-s)D_j}F_j(x+s\eta,s)\Phi_s^j\,ds.
\]
Since
\[
 x\in S_r,\quad0\le s\le1\quad\Longrightarrow\quad
 (x+s\eta,s)\in S_r^2,
\]
Gronwall's inequality yields
\[
 \|\Phi_t^j\|_r
 \le1+\|F_j\|_r\int_0^t\|\Phi_s^j\|_r\,ds
 \le e^{t\|F_j\|_r},\qquad0\le t\le1.
\]
Hence
\[
 \|\Phi_1^j-e^{D_j}\|_r
 \le\int_0^1\|F_j\|_re^{s\|F_j\|_r}\,ds
 =e^{\|F_j\|_r}-1.
\]
By periodicity,
\[
 \begin{gathered}
 P_j(x+\eta,1)=P_j(x+\eta,0)=B_j(x+\eta),\\
 \Phi_1^j=B_j(x+\eta)A(x)B_j(x)^{-1}.
 \end{gathered}
\]
This proves \eqref{eq:time-one-error}.
\end{proof}

\section{Acknowledgement and AI-use} 
L. Ge was partially supported by NSFC grant (12371185) and the Fundamental Research Funds for the Central Universities (the start-up fund), Peking University. J. You   was partially supported by NSFC grant (12531006,12526201) and
Nankai Zhide Foundation. Research of SJ was supported by NSF DMS-2052899, DMS-2155211, and Simons 896624.

The authors conceived the entire proof strategy  in  2025, without AI involvement. Our plan was to use the intrinsic symplectic center of \cite{GJmanuscript} to extend Kotani theory from a two-dimensional center to a center of arbitrary finite dimension, and then use renormalization to bring the compact unitary center cocycle into the local regime. Our idea for the final step was to isolate a single pair of resonant eigenphases in a local KAM argument with parameters and use the Wronskian to derive a contradiction.

Our first use of AI in this work came much later, after we had formulated this strategy, completed the construction of the intrinsic symplectic center, and extended Kotani theory to higher-dimensional centers. We then used an LLM to help carry out the renormalization and local parameter KAM steps. In particular, it helped us derive the renormalization estimate in Lemma 5.1 and discover the monotonicity property in the Lemma 3.2. Its assistance was particularly valuable in finding how to use this property to simplify our original local KAM argument with parameters.

We also used  LLM to speed up preparation of the manuscript, stress-test the arguments, identify overlooked cases and points requiring clarification, and improve the exposition. We take full responsibility for the mathematical arguments and the final text.

\raggedbottom
\begingroup
\linespread{0.98}\selectfont

\endgroup


\begin{thebibliography}{99}
\bibitem{ArgentieriAvila} F.~Argentieri and A.~Avila.
In preparation.

\bibitem{AubryAndre} S.~Aubry and G.~Andr\'e.
Analyticity breaking and Anderson localization in incommensurate lattices.
Ann. Israel Phys. Soc. \textbf{3} (1980), 133--164.

\bibitem{AvilaGlobal} A.~Avila. Global theory of one-frequency Schr\"odinger operators.
 Acta Math. \textbf{215} (2015), 1--54.

\bibitem{ABD09} A.~Avila, J.~Bochi and D.~Damanik.
Cantor spectrum for Schr\"odinger operators with potentials arising from
generalized skew-shifts. Duke Math. J. \textbf{146} (2009), 253--280.

\bibitem{ABD12} A.~Avila, J.~Bochi and D.~Damanik.
Opening gaps in the spectrum of strictly ergodic Schr\"odinger operators.
J. Eur. Math. Soc. \textbf{14} (2012), 61--106.

\bibitem{AJTen} A.~Avila and S.~Jitomirskaya. The Ten Martini Problem.
 Ann. of Math. \textbf{170} (2009), 303--342.

\bibitem{AJS} A.~Avila, S.~Jitomirskaya and C.~Sadel. Complex one-frequency cocycles.
 J. Eur. Math. Soc. \textbf{16} (2014), 1915--1935.

\bibitem{AK} A.~Avila and R.~Krikorian.
 Reducibility or nonuniform hyperbolicity for quasiperiodic Schr\"odinger cocycles.
 Ann. of Math. \textbf{164} (2006), 911--940.

\bibitem{AKMonotonic} A.~Avila and R.~Krikorian.
Monotonic cocycles. Invent. Math. \textbf{202} (2015), 271--331.

\bibitem{avronsimon} J. Avron and B. Simon. Almost periodic operators, II. The density of states. {\it Duke Math J.} {\bf 50} (1983), 369–391.


\bibitem{AYZ} A.~Avila, J.~You and Q.~Zhou.
 Dry Ten Martini Problem in the non-critical case.
 arXiv:2306.16254v2 (2024); first version, 2023.

 \bibitem{AYZc} A.~Avila, J.~You and Q.~Zhou. In preparation.

\bibitem{AvronSimon} J.~Avron and B.~Simon.
Almost periodic Schr\"odinger operators. II. The integrated density of states.
Duke Math. J. \textbf{50} (1983), 369--391.

\bibitem{BBL} R.~Band, S.~Beckus, and R.~Loewy.
 The Dry Ten Martini Problem for Sturmian Hamiltonians.
 arXiv:2402.16703 (2024).

\bibitem{BF} D.~S.~Borgnia and M.~Faust.
 The Dry Ten Martini Problem at Criticality.
 arXiv:2112.06869v3 [math-ph] (2026).

 \bibitem{BandSofer} R.~Band and G.~Sofer.
 Johnson-Schwartzman Gap Labelling for Metric and Discrete Decorated Graphs.
 arXiv:2604.08496 (2026).

\bibitem{BJK} S.~Becker, S.~Jitomirskaya, and I.~Krasovsky.
 Critical almost Mathieu operator: hidden singularity, gap continuity, and the Hausdorff dimension of the spectrum.
 arXiv:1909.04429v2 (2026).

\bibitem{BS82} J.~Bellissard and B.~Simon.
Cantor spectrum for the almost Mathieu equation.
J. Funct. Anal. \textbf{48} (1982), 408--419.

\bibitem{bourg} J. Bourgain. Positivity and continuity of the Lyapunov exponent for shifts on ${\mathbb T}^d$ with arbitrary frequency vector and real analytic potential. {\it J. Anal. Math.} {\bf 96} (2005), 313–355.

\bibitem{BJContinuity} J.~Bourgain and S.~Jitomirskaya.
Continuity of the Lyapunov exponent for quasiperiodic operators with analytic potential.
J. Stat. Phys. \textbf{108} (2002), 1203--1218.

\bibitem{CEY90} M.-D.~Choi, G.~A.~Elliott and N.~Yui.
Gauss polynomials and the rotation algebra.
Invent. Math. \textbf{99} (1990), 225--246.
\bibitem{dl}
D.~Damanik and D.~Lenz,
\emph{Zero-measure Cantor spectrum for Schr\"odinger operators
with low-complexity potentials},
J. Math. Pures Appl. (9) \textbf{85} (2006), no.~5, 671--686.
\bibitem{DS75} E.~I.~Dinaburg and Ya.~G.~Sinai.
The one-dimensional Schr\"odinger equation with a quasiperiodic potential.
Funct. Anal. Appl. \textbf{9} (1975), 279--289.

\bibitem{Eliasson} L.~H.~Eliasson.
 Floquet solutions for the 1-dimensional quasi-periodic Schr\"odinger equation.
 Comm. Math. Phys. \textbf{146} (1992), 447--482.

\bibitem{FormanVandenBoom} Y.~Forman and T.~VandenBoom.
Localization and Cantor spectrum for quasiperiodic discrete Schr\"odinger operators
with asymmetric, smooth, cosine-like sampling functions.
Mem. Amer. Math. Soc. \textbf{312} (2025), no.~1583, v+86 pp.

\bibitem{OkaGrauert} F.~Forstneri\v{c}.
 The homotopy principle in complex analysis: a survey.
 Contemp. Math. \textbf{332} (2003), 73--99.

%\bibitem{Furman} A.~Furman.
 %On the multiplicative ergodic theorem for uniquely ergodic systems.
 %Ann. Inst. H. Poincar\'e Probab. Statist. \textbf{33} (1997), 797--815.

\bibitem{GeDry} L.~Ge.
The Dry Ten Martini Problem for subcritical Type I operators.
arXiv:2609.37438v1 (2026).



\bibitem{GJmanuscript} L.~Ge and S.~Jitomirskaya.
Intrinsic Symplectic Structure and Sharp Arithmetic Universality.
 arXiv:2407.08866v3 (2026).

\bibitem{GJY2023} L.~Ge, S.~Jitomirskaya and J.~You.
 Kotani theory, Puig's argument, and stability of The Ten Martini Problem.
 arXiv:2308.09321 (2023).

\bibitem{GJY} L.~Ge, S.~Jitomirskaya and J.~You.
The Robust Ten Martini Problem. arXiv:2609.15812v1 (2026).

\bibitem{GJYZ} L.~Ge, S.~Jitomirskaya, J.~You and Q.~Zhou.
 Multiplicative Jensen's formula and quantitative global theory of one-frequency
 Schr\"odinger operators. Forum Math. Pi \textbf{14} (2026), e1.
\url{https://doi.org/10.1017/fmp.2025.10019}.

\bibitem{GeXu} L.~Ge and D.~Xu.
In preparation.
\bibitem{GXWZ} L.~Ge, Y.~Wang, J.~Xu and X. ZHao. In preparation.
\bibitem{GWXDry} L.~Ge, Y.~Wang and J.~Xu.
The Dry Ten Martini Problem for $C^2$ cosine-type quasiperiodic Schr\"odinger operators.
arXiv:2503.06918 (2025).

\bibitem{GS} M.~Goldstein and W.~Schlag.
 On resonances and the formation of gaps in the spectrum of quasi-periodic
 Schr\"odinger equations. Ann. of Math. \textbf{173} (2011), 337--475.

\bibitem{HelfferSjostrand} B.~Helffer and J.~Sj\"ostrand.
Semi-classical analysis for Harper's equation. III. Cantor structure of the spectrum.
M\'em. Soc. Math. France (N.S.) \textbf{39} (1989), 1--124.

\bibitem{HL} X.~Hou and J.~Li.
Almost reducibility of quasi-periodic linear system on unitary group.
J. Differential Equations \textbf{448} (2025), 113674.

\bibitem{HPZ} X.~Hou, Y.~Pan and Q.~Zhou.
 Dynamical classification of analytic one-frequency quasi-periodic
 $\mathrm{SO}(3,\R)$-cocycles.
 Adv. Math. \textbf{457} (2024), 109943.

\bibitem{HY} X.~Hou and J.~You.
 Almost reducibility and non-perturbative reducibility of quasi-periodic linear systems.
 Invent. Math. \textbf{190} (2012), 209--260.
\url{https://doi.org/10.1007/s00222-012-0379-2}.
\bibitem{jk} S.~Jitomirskaya and I.~Kachkovskiy.
 All couplings localization for quasiperiodic operators with monotone potentials.
 J. Eur. Math. Soc. \textbf{21} (2019), no.~3, 777--795.
\bibitem{jls}
S.~Jitomirskaya, N.~Levi, and T.~Stoiber,
(Sharp arithmetic) localization beyond transversality,
preprint, arXiv:2610.17139 (2026).
\bibitem{JS}
S.~Jitomirskaya and T.~Stoiber,
Large gaps and Ten Martini problem for monotone
quasiperiodic Schr\"odinger operators,
Preprint, 2026.
\bibitem{JM} R.~Johnson and J.~Moser.
 The rotation number for almost periodic potentials.
 Comm. Math. Phys. \textbf{84} (1982), 403--438.

 \bibitem{kps} I.~Kachkovskiy, L.~Parnovski, and R.~Shterenberg.
 On gaps in the spectra of quasiperiodic Schr\"odinger operators with discontinuous monotone potentials.
 Int. Math. Res. Not. IMRN \textbf{2025}, no.~7, 21 pp.

\bibitem{KotaniFloquet} S.~Kotani.
Generalized Floquet theory for stationary Schr\"odinger operators in one dimension.
Chaos Solitons Fractals \textbf{8} (1997), 1817--1854.

\bibitem{KS} S.~Kotani and B.~Simon.
 Stochastic Schr\"odinger operators and Jacobi matrices on the strip.
 Comm. Math. Phys. \textbf{119} (1988), 403--429.

\bibitem{KrCompact} R.~Krikorian.
R\'eductibilit\'e des syst\`emes produits-crois\'es \`a valeurs dans des groupes compacts.
Ast\'erisque \textbf{259} (1999), 222 pp.

\bibitem{Kr99} R.~Krikorian.
 $C^0$-densit\'e globale des syst\`emes produits-crois\'es sur le cercle r\'eductibles.
 Ergodic Theory Dynam. Systems \textbf{19} (1999), 61--100.

\bibitem{Kr01} R.~Krikorian.
 Global density of reducible quasi-periodic cocycles on
 $\T^1\times\mathrm{SU}(2)$.
 Ann. of Math. \textbf{154} (2001), 269--326.

\bibitem{Last94} Y.~Last.
Zero measure spectrum for the almost Mathieu operator.
Comm. Math. Phys. \textbf{164} (1994), 421--432.

\bibitem{LiDry} X.~Li.
Dry Ten Martini Problem in the Subcritical Type I Regime.
arXiv:2609.23467v2 (2026).

\bibitem{LXZ} X.~Li, D.~Xu and Q.~Zhou.
 Monotonicity, global symplectification and the stability of Dry Ten Martini Problem.
 arXiv:2601.02222v4 (2026).

\bibitem{MarxJitomirskaya2017} C.~A.~Marx and S.~Jitomirskaya.
Dynamics and spectral theory of quasi-periodic Schr\"odinger-type operators.
Ergodic Theory Dynam. Systems \textbf{37} (2017), no.~8, 2353--2393.

\bibitem{Peierls} R. Peierls. Zur theorie des diamagnetismus von leitungselektronen. {\it Zeitschrift f\"ur Physik} {\bf 80(11-12)} (1933), 763-791.


\bibitem{MP} J.~Moser and J.~P\"oschel.
 An extension of a result by Dinaburg and Sinai on quasi-periodic potentials.
 Comment. Math. Helv. \textbf{59} (1984), 39--85.

 \bibitem{ntw} Q. Niu, D.J. Thouless and Y.S. Wu. Quantized Hall conductance as a topological invariant. {\it Phys. Rev. B} {\bf 31} (1985), 3372.

\bibitem{Novikov98}S. P. Novikov, Discrete Schrödinger operators and topology, Asian Journal of Mathematics 2 (1998), no. 4, 921–934.
\bibitem{Puig} J.~Puig. Cantor spectrum for the almost Mathieu operator.
 Comm. Math. Phys. \textbf{244} (2004), 297--309.

 \bibitem{sim15} B. Simon. Fifteen problems in mathematical physics. {\it Perspectives in Mathematics: Anniversary of Oberwolfach} (1984), 423–454.

\bibitem{simXXI} B. Simon. Schr\"odinger operators in the twenty-first century. {\it Mathematical Physics} Imp. Coll. Press, London (2000), 283–288.


\bibitem{Sinai1987} Ya.~G.~Sinai.
Anderson localization for one-dimensional difference Schr\"odinger operator
with quasiperiodic potential.
J. Stat. Phys. \textbf{46} (1987), 861--909.

\bibitem{SoretsSpencer} E.~Sorets and T.~Spencer.
Positive Lyapunov exponents for Schr\"odinger operators with quasi-periodic potentials.
Comm. Math. Phys. \textbf{142} (1991), 543--566.

\bibitem{TeschlJacobi} G.~Teschl.
 \emph{Jacobi Operators and Completely Integrable Nonlinear Lattices}.
 Mathematical Surveys and Monographs, vol.~72, American Mathematical Society,
 Providence, RI, 2000.

\bibitem{TeschlODE} G.~Teschl.
 \emph{Ordinary Differential Equations and Dynamical Systems}.
 Graduate Studies in Mathematics, vol.~140, American Mathematical Society,
 Providence, RI, 2012.

\bibitem{WangZhang} Y.~Wang and Z.~Zhang.
Cantor spectrum for a class of $C^2$ quasiperiodic Schr\"odinger operators.
Int. Math. Res. Not. IMRN (2017), no.~8, 2300--2336.

\bibitem{XuJEMS} D.~Xu.
Density of positive Lyapunov exponents for symplectic cocycles.
J. Eur. Math. Soc. \textbf{21} (2019), 3143--3190.

\bibitem{YouKAM2024} J.~You.
The KAM method for the spectral theory of quasi-periodic Schr\"odinger operators
(in Chinese). Sci. Sin. Math. \textbf{54} (2024), no.~6, 863--870.
\url{https://doi.org/10.1360/SSM-2024-0013}.

\bibitem{YZ} J.~You and Q.~Zhou.
 Embedding of analytic quasi-periodic cocycles into analytic quasi-periodic
 linear systems and its applications.
 Comm. Math. Phys. \textbf{323} (2013), 975--1005.
\end{thebibliography}
\end{document}